\documentclass[aps,amsmath,amssymb, prc,twocolumn,superscriptaddress]{revtex4-1}
\usepackage{graphicx}% Include figure files
\usepackage{dcolumn}% Align table columns on decimal point
\usepackage{bm}% bold math
\usepackage[utf8]{inputenc}
\usepackage[ngerman,english]{babel}
\usepackage{float}
\usepackage{multirow}
\usepackage{longtable}
\usepackage{dcolumn}
\newcolumntype{d}{D{.}{.}{-1}}
\usepackage{setspace}
\usepackage{threeparttable}
\usepackage{tabularx}
\usepackage[symbol*]{footmisc}
\DefineFNsymbolsTM{myfnsymbols}{% def. from footmisc.sty "bringhurst" symbols
  \textasteriskcentered *
  \textdagger    \dagger
  \textdaggerdbl \ddagger
  \textsection   \mathsection
  \textbardbl    \|%
  \textparagraph \mathparagraph
}%
\setfnsymbol{myfnsymbols}

\begin{document}

%\preprint{}

\title{High-resolution $(p,t)$ study of low-spin states in $^{240}$Pu: Octupole excitations, $\alpha$\,clustering, and other structure features}


\author{M. Spieker}
\email[]{spieker@nscl.msu.edu ; present address: NSCL, 640 South Shaw Lane, East Lansing, MI 48824, USA}
%\homepage[Member of the Bonn-Cologne Graduate School of Physics and Astronomy]{}
%\thanks{}
%\altaffiliation{test}
\affiliation{Institut f\"{u}r Kernphysik, Universit\"{a}t zu K\"{o}ln, Z\"{u}lpicher Straße 77, D-50937 K\"{o}ln, Germany}

\author{S. Pascu}
\affiliation{Institut f\"{u}r Kernphysik, Universit\"{a}t zu K\"{o}ln, Z\"{u}lpicher Straße 77, D-50937 K\"{o}ln, Germany}
\affiliation{Horia Hulubei National Institute of Physics and Nuclear Engineering, Bucharest, Romania}

\author{D. Bucurescu}
\affiliation{Horia Hulubei National Institute of Physics and Nuclear Engineering, Bucharest, Romania}

\author{T. M. Shneidman}
\affiliation{Bogoliubov Laboratory of Theoretical Physics, Joint Institute for Nuclear Research, Dubna 141980, Russia}
\affiliation{Kazan Federal University, Kazan 420008, Russia}

\author{T. Faestermann}
\affiliation{Physik Department, Technische Universität München, München, Germany}

\author{R. Hertenberger}
\affiliation{Fakultät für Physik, Ludwig-Maximilians-Universität München, München, Germany}

\author{H.-F. Wirth}
\affiliation{Fakultät für Physik, Ludwig-Maximilians-Universität München, München, Germany}

\author{N.-V. Zamfir}
\affiliation{Horia Hulubei National Institute of Physics and Nuclear Engineering, Bucharest, Romania}

\author{A. Zilges}
\affiliation{Institut f\"{u}r Kernphysik, Universit\"{a}t zu K\"{o}ln, Z\"{u}lpicher Straße 77, D-50937 K\"{o}ln, Germany}


\date{\today}

\begin{abstract}

\noindent\textbf{Background:} Many nuclear-structure features have been observed in the actinides during the last decades. Especially the octupole degree of freedom has been discussed lately after the successful measurement of the B$\left( E3; 0^+_1 \rightarrow 3^-_1 \right)$ reduced transition strength in $^{224}$Ra. Recent results stemming from $\gamma$-spectroscopy experiments and high-resolution $(p,t)$ experiments suggested, that strong octupole correlations might be observed for some positive-parity states of actinide nuclei.

\noindent\textbf{Purpose:} This work completes a series of $(p,t)$ experiments on actinide nuclei by adding the data on $^{240}$Pu. The $(p,t)$ experiments allow to study low-spin states up to $J^{\pi} = 6^+$. Besides two-nucleon transfer cross sections, spin and parity can be assigned to excited states by measuring angular distributions, and several rotational bands are recognized based on these assignments.  

\noindent\textbf{Methods:} A high-resolution {\it (p,t)} experiment at $E_{p}$=~24~MeV was performed to populate low-spin states in the actinide nucleus ${}^{240}$Pu. The Q3D magnetic spectrograph of the Maier-Leibnitz Laboratory (MLL) in Munich (Germany) was used to identify the ejected tritons via $dE/E$ particle identification with its focal-plane detection system. Angular distributions were measured at nine different Q3D angles to assign spin and parity to the excited states based on a comparison with coupled-channels DWBA calculations.

\noindent\textbf{Results:} In total, 209 states have been excited in $^{240}$Pu up to an excitation energy of 3\,MeV. Many previously known states have also been observed and their spin-parity assignments were confirmed. However, many of the populated states have been seen for the first time, {\it e.g.}, 15 new and firmly assigned $J^{\pi} = 0^+$ states. In addition, all low-spin one-octupole phonon excitations, i.e. $K^{\pi} = 0^-,1^-,2^-,3^-$, could be observed and a new candidate for the $K = 3$ projection is proposed. Furthermore, the double-octupole or $\alpha$-cluster structure of the $0^+_2$ state in $^{240}$Pu has been studied in more detail. It is shown that the $0^+_2$ state in $^{230}$Th has a distinctly different structure. In addition, strongly excited $1^-$ states have been observed at 1.5\,MeV and 1.8\,MeV in $^{240}$Pu. The present study suggests that similar states might be observed in $^{230}$Th. 

\noindent\textbf{Conclusions:} At least two different and distinct structures for $J^{\pi} = 0^+$ states are present in the actinides. These are pairing states and states with enhanced octupole correlations. We have shown that it is crucial to consider negative-parity single-particle states being admixed to some $K^{\pi} = 0^+_2$ rotational bands to understand the $\alpha$-decay hindrance factors and enhanced $E1$-decay rates. Based on our analysis, we have identified the double-octupole or $\alpha$-cluster $K^{\pi} = 0^+$ candidates from $^{224}$Ra to $^{240}$Pu.

\end{abstract}


\maketitle

%\section*{}
%\subsection{}
%\subsubsection{}

\section{Introduction}

During the last decade renewed interest to study the octupole degree of freedom in atomic nuclei and, especially, in the actinides has grown, see, {\it e.g.}, Refs.\,\cite{Robl11a, Gaff13, Nomu14a, Birk15a, Zimba16a, Butl16a, Agbe16a, Buch16a, Maq17a, Buch17a} and references therein. Many of these experimental and theoretical studies were triggered by the observation of the enhanced B$\left( E3; 3^-_1 \rightarrow 0^+_1 \right)$ value of 42(3)\,W.u. in $^{224}$Ra which, in combination with an alternating-parity band at low energies, was interpreted as clear signature of static octupole deformation in the ground state of this nucleus\,\cite{Gaff13}. Strong octupole correlations are expected and observed in many actinide nuclei owing the fact that the Fermi surface for both protons and neutrons lies between single-particle orbitals differing by $\Delta j = \Delta l =3$, see, {\it e.g.}, the review article\,\cite{Butl96}. However, only a few Ra and Th nuclei are considered to show signs of static octupole deformation already in their ground state. For instance, in $^{240}$Pu strong octupole correlations were observed by means of an alternating-parity band at high spins, i.e. $J \sim 20$\,\cite{Wied99}. Using two-center octupole wave functions in the framework of supersymmetric quantum mechanics, Refs. \cite{Jolo11, Jolo12} explained the experimental data as a second-order phase transition from an octupole-nondeformed to an octupole-deformed shape at higher spins. In a consecutive high-statistics ``unsafe'' Coulomb excitation experiment\,\cite{Wang09}, the $K^{\pi}=$~0$^+_2$ rotational band of $^{240}$Pu was investigated up to highest spins ($J^{\pi}=~30^+$). Enhanced $E1$ transitions were observed which deexcited its high-spin members exclusively to the $K^{\pi}= 0^-_1$ one-octupole phonon band. Following the concept of multiphonon condensation proposed in Ref.\,\cite{Frauend08}, the experimental observations in ${}^{240}$Pu were explained in terms of the condensation of rotation-aligned octupole phonons\,\cite{Wang09}. As a consequence, the $K^{\pi}=$~0$^+_2$ rotational band has been proposed as a candidate for the double-octupole band. This hypothesis was later on supported by the work of Refs.\,\cite{Jolo13, Spiek13a}. The new $(p,t)$ data on $^{228}$Th, $^{232}$U and $^{240}$Pu helped to clearly identify the double-octupole $J^{\pi} = 0^+$ candidates in combination with enhanced $E1$ transitions measured in previously performed $\gamma$-ray spectroscopy experiments\,\cite{Levon13, Spiek13a, Levon15}.
\\
However, the nature of the $0^+_2$ states in the even-even actinides has been controversially discussed for decades\,\cite{Maher72, Cast72, Rij72, Fried73, Fried74, Ragn76}. Extensive experimental studies had shown an asymmetry between the population in $(p,t)$ and $(t,p)$ reactions for some actinides\,\cite{Maher72, Fried74, Cast72}. Ragnarsson and Broglia introduced the concept of pairing isomers which should have a smaller neutron pairing gap $\Delta_{n}$ than the ground state itself\,\cite{Ragn76, Rij72}. These isomers would be present in the case of an inhomogeneity of weakly coupled prolate and oblate levels around the Fermi surface for comparable monopole and quadrupole pairing strengths. The experimental signature of pairing isomers would indeed be large $(p,t)$ cross sections and almost vanishing $(t,p)$ cross sections. We note that these have been recently discussed in $^{154}$Gd\,\cite{Allm17a}. However, Rij and Kahana predicted a negligible population of pairing isomers in single-neutron transfer reactions~\cite{Rij72} which was not observed in $^{240}$Pu, i.e. $\sigma_{0^+_2}/\sigma_{0^+_1} \approx 18\%$\,\cite{Fried73}. It might, thus, be possible that several configurations coexist at energies around the neutron-2QP energy, i.e. $2 \Delta_n$ in the actinides. In our previous publication\,\cite{Spiek13a}, we have already shown that two different and very distinct structures are close in energy in $^{240}$Pu, i.e. $\Delta E_x = 230$\,keV. Besides the double-octupole phonon candidate, we identified a quadrupole-type excitation built upon the ground state which did not show the common signatures of the classical $\beta$ vibration.
\\
\\
In his recent topical review\,\cite{Butl16a}, P.\,A.\,Butler pointed out the importance to identify the possible double-octupole phonon bands and to clarify the nature of the $K^{\pi} = 0^+_2$ bands in the actinides. He stated that the existence of low-lying double-octupole phonon bands in the context of multiphonon condensation might be hard to reconcile with the picture of rigid octupole deformation in the ground state of, {\it e.g.}, $^{226}$Ra. We already stressed that two-neutron transfer reactions, i.e. $(p,t)$ experiments can provide important information on the pairing character of these states. In a recent global analysis of octupole deformation within the covariant density functional theory (CDFT)\,\cite{Agbe16a}, the authors have shown that enhanced pairing correlations can weaken the octupole correlations in the actinides since more spherical shapes are favored. It is, thus, also instructive to study pairing correlations in nuclei with enhanced octupole correlations.
\\
\\ 
This publication features all the data obtained from the $^{242}$Pu$(p,t){}^{240}$Pu experiment performed at the Maier-Leibnitz Laboratory (MLL) in Munich which we will present in Sec.\,\ref{sec:results}. In Sec.\,\ref{sec:disc} we will mainly discuss possible origins of $0^+$ states in the actinides and provide strong evidence for the coexistence of at least two different structures. Since it has been recently shown that also $\alpha$-clustering in the actinides could possibly explain the signatures which are usually attributed to octupole-type excitations\,\cite{Shneid15a}, we will comment on these two mechanisms causing reflection asymmetry in the atomic nucleus by studying the negative-parity states in $^{240}$Pu in more detail.

\begin{figure*}[!t]
\centering
\includegraphics[width=0.9\linewidth]{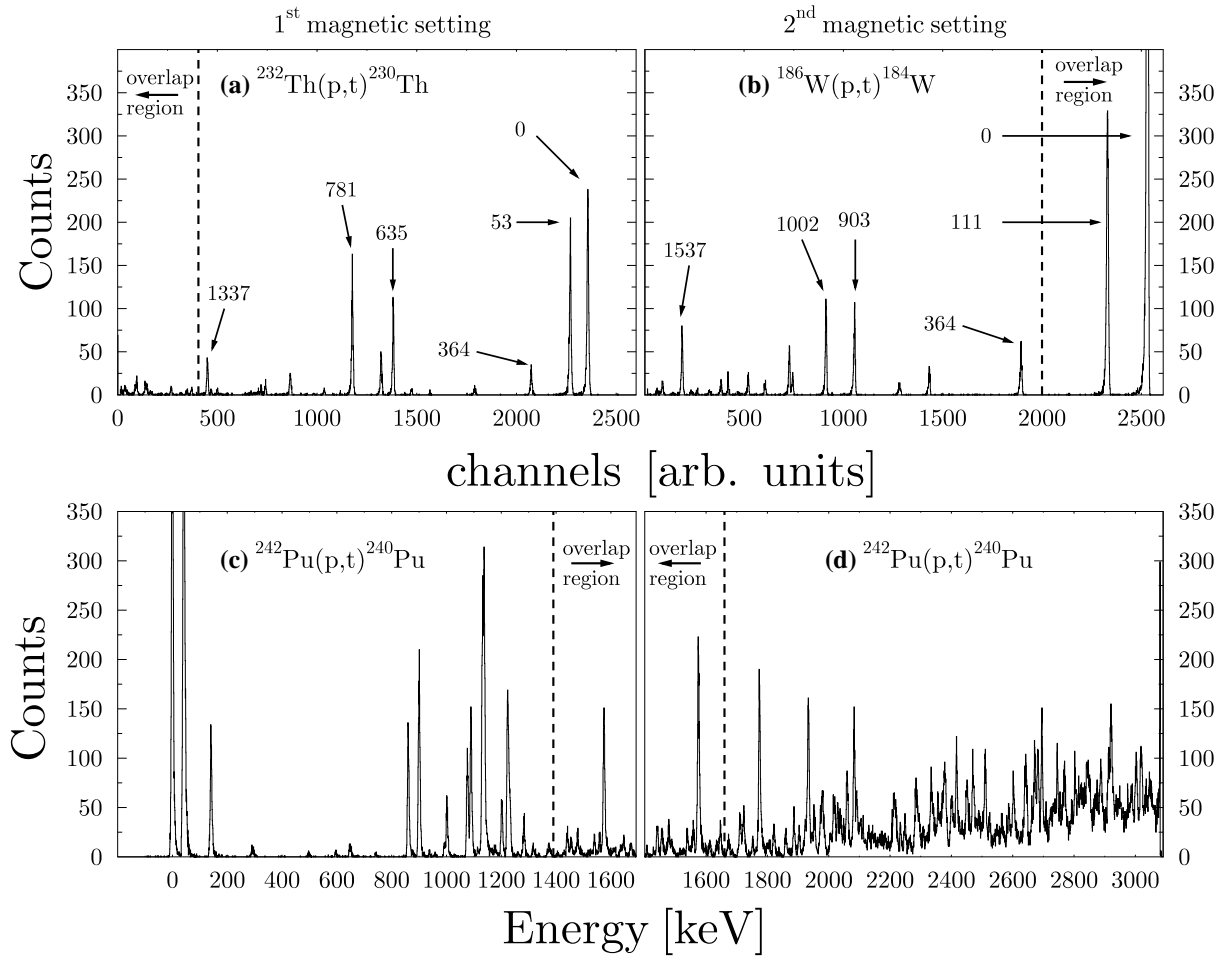}
\caption{\label{fig:spectrum}{\bf (a)}${}^{232}$Th$(p,t){}^{230}$Th, {\bf (b)}${}^{186}$W$(p,t){}^{184}$W and {\bf (c)}, {\bf (d)} ${}^{242}$Pu$(p,t){}^{240}$Pu spectra at 10$^{\circ}$ for the two magnetic settings which were used for the energy calibration of the focal plane detection system, see text. The overlap regions are marked by dashed lines. Some prominent peaks in $^{230}$Th and $^{184}$W are highlighted with their excitation energy in keV.}
\end{figure*}

\section{experimental details}

To study 0$^+$ states, see Ref.\,\cite{Spiek13a}, and other low-spin excitations in ${}^{240}$Pu a high-resolution $(p,t)$ study was performed at the Q3D magnetic spectrograph of the MLL in Munich~\cite{loef73}. A 120~$\mathrm{\mu g/cm^2}$ thick and highly-enriched ${}^{242}$Pu target (99.93$\%$, T$_{1/2}$=~3.75~$\times$~10$^5$ years) was provided by Oak Ridge National Laboratory, and was evaporated onto a 25~$\mathrm{\mu g/cm^2}$ carbon backing. Possible target contaminations were excluded using the measured triton spectra, see the supplemental material\,\cite{Sup18a}. The $E_{p}$=~24~MeV proton beam impinged onto the ${}^{242}$Pu target with an average beam current of 1~$\mu$A. The ejected tritons were bent to the focal plane detection system of the Q3D magnetic spectrograph, where they were unambiguously selected via {\it dE/E} particle identification \cite{Wirt00}. The energy calibration of the detection system was done in terms of well-known $(p,t)$ reactions as presented in, {\it e.g.}, Ref.~\cite{Levon09}. Figure~\ref{fig:spectrum} shows the excitation spectrum of ${}^{240}$Pu for the two magnetic settings at 10$^{\circ}$, which have been used to cover excitation energies up to 3~MeV. To unambiguously assign spin and parity to excited states, angular distributions were measured at nine laboratory angles ranging from 5$^{\circ}$ to 40$^{\circ}$ and compared to the distributions calculated by the CHUCK3 code \citep{CHUCK}. Except for the measurements at 5$^{\circ}$ (9.3 msr), the maximum Q3D solid angle of 13.9 msr was chosen. This procedure has already been successfully applied to the ${}^{232}$Th$(p,t){}^{230}$Th\,\cite{Levon09}, ${}^{230}$Th$(p,t){}^{228}$Th\,\cite{Levon13}, and ${}^{234}$U$(p,t){}^{232}$U\,\cite{Levon15} reactions. In total, 209 states in ${}^{240}$Pu have been identified in the present $(p,t)$ study. Many previously known low-spin states have also been observed and their spin-parity assignments were confirmed. However, most of the populated states have been seen for the first time.

\section{data analysis}

\subsection{Energy calibration}

The energy calibration of the focal-plane detection system was done in terms of well-known $(p,t)$ reactions. The reactions ${}^{232}$Th$(p,t){}^{230}$Th\,\cite{Levon09} and ${}^{186}$W$(p,t){}^{184}$W\,\cite{Mort80} were chosen and measured at laboratory angles of 10$^{\circ}$ for both magnetic settings, see Fig.\,\ref{fig:spectrum}\,{\bf (a)} and {\bf (b)}. The individual channels were identified with their respective level energies using a second-order polynomial. Once calibrated, the triton energies for the respective level energies were calculated using the reaction kinematic program CatKin\,\cite{catkin}. In this way, a reaction-independent relation between triton energies and channels was found. This procedure allowed the calculation of the triton energies for the reaction ${}^{242}$Pu$(p,t){}^{240}$Pu, which could then be converted into excitation energies in $^{240}$Pu for both magnetic settings, see Fig.\,\ref{fig:spectrum}\,{\bf (c)} and {\bf (d)}. Due to the reaction $Q$-value difference of about 1.4\,MeV between the ${}^{232}$Th$(p,t){}^{230}$Th ($Q(p,t) = -3076.5(11)$\,keV) and ${}^{186}$W$(p,t){}^{184}$W ($Q(p,t) = -4463.1(16)$\,keV) reaction\,\cite{Audi03}, all relevant triton energies were measured. Each magnetic setting typically covers an excitation-energy range of 1.7\,MeV, see also Fig.\,\ref{fig:spectrum}. The accuracy of the energy calibration was cross-checked with well-known excited states in $^{240}$Pu and has a precision of at least 1\,keV. Discrepancies arise mainly due to the uncertainties of the $Q(p,t)$ values.

\subsection{Cross sections and angular distributions}

The differential cross sections of the $^{242}$Pu$(p,t) ^{240}$Pu reaction were measured at nine angles between 5$^{\circ}$ and 40$^{\circ}$. Some examples are shown in Figs.~\ref{fig:0+_angdist}, \ref{fig:others_angdist} and \ref{fig:octu_angdist}, respectively. The differential cross sections were calculated according to equation\,(\ref{eq:cs_01}) and corrected for the deadtime of the data-acquisition system, which is well below 10$\%$.

\begin{align}
\frac{\mathrm{d}\sigma}{\mathrm{d}\Omega}\left(\theta\right) = \frac{\mathrm{N}\left(\theta\right)}{\mathrm{d}\Omega\times\mathrm{N}_{\mathrm{total}}\times\mathrm{F}_{\mathrm{target}}}
\label{eq:cs_01}
\end{align}

In equation\,(\ref{eq:cs_01}), N$\left(\theta\right)$ corresponds to the number of tritons measured at a Q3D angle $\theta$, d$\Omega$ to the solid angle covered by the spectrograph, and N$_\mathrm{total}$ to the total number of protons, which were impinging onto the $^{242}$Pu target. The latter was on the order of 10$^{16}$ to 10$^{17}$ protons for one measurement at a given Q3D angle. This is equivalent to a measurement of about three hours per angle for an average beam current of 1\,$\mu$A on target. The target thickness F$_{\mathrm{target}}$ is calculated with respect to $\vartheta$, which is the angle between the target and the beam axis. This tilting angle is used to minimize straggling effects in the target. 
\\
\\
The nearly background-free detection of tritons at the focal plane allowed the determination of differential cross sections as low as 0.1\,$\mu$b/sr. Combined with the superior energy resolution of the Munich Q3D spectrograph of less than 10\,keV, also weakly excited states could be identified in the dense excitation spectrum of $^{240}$Pu, see Fig.\,\ref{fig:spectrum}. All excited states are given in Table\,\ref{longtable}. To determine the integrated $(p,t)$ excitation cross section, the differential cross sections were integrated over the covered angular range.

\subsection{DWBA calculations}
\label{sec:dwba}

\begin{figure}[!t]
\centering
\includegraphics[width=0.8\linewidth]{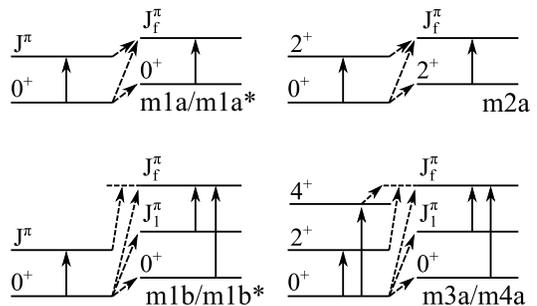}
\caption{\label{fig:multistep}Excitation schemes used in the DWBA calculations. A ``$\ast$'' indicates that the intermediate states were the $2^+_1$ states in $^{242}$Pu and $^{240}$Pu, respectively. m3a corresponds to $J_1^{\pi} = 2^+_1$ and m4a to $J_1^{\pi} = 4^+_1$. In most cases, it was not necessary to include all intermediate states when m3a or m4a had to be used.}
\end{figure}

Direct reactions are expected to take place on a time scale of 10$^{-22}$ s. The reaction process, {\it e.g.}, $(p,t)$ reactions can be described by the Distorted Wave Born Approximation (DWBA) and the optical model. To calculate the differential cross sections, the computer code {\small CHUCK}3 of P.D. Kunz \cite{CHUCK} was used. The calculations were performed by solving an appropriate set of coupled equations within the program code. The optical potentials as well as particle masses, binding energies and the respective Q values are reaction specific. While performing the calculation the binding energies of the two neutrons are calculated such that they match the respective energies of the outgoing tritons for every considered excited state \cite{Wirt04}. It is also possible to calculate multi-step processes within the code while the normal DWBA only considers one-step processes. Already in Ref.\,\cite{Baer73} it was pointed out that multi-step processes may alter the shape of angular distributions. In  Ref.\,\cite{Levon09} it was found that these processes had indeed to be included in the ${}^{232}$Th$(p,t){}^{230}$Th reaction already for the description of the ground-state rotational band members. In general, it is possible to include eight channels into the program's calculations and define their individual coupling to the other channels. Different coupling schemes were used in the analysis. The population of excited states in ${}^{240}$Pu has been possible by a coupling of inelastic and direct-transfer channels, i.e. $(p,p') \rightarrow (p,t) \rightarrow (t,t')$, see Fig.\,\ref{fig:multistep}. In Ref.\,\cite{Mahg08} even sequential, i.e. $(p,d) \rightarrow (d,t)$ transfers had to be implemented for low-lying even- and odd-parity states.

Following the common notation of C.M.~Perey and F.G.~Perey \cite{Pere76}, the optical model parameters are the potentials V$_{\mathrm{r}}$ and W$_0$ for the Volume Woods-Saxon, and W$_{\mathrm{D}}$ for the Surface Woods-Saxon parts, as well as V$_{\mathrm{so}}$ for the spin-orbit interaction. The subscript ``c'' indicates an additional Coulomb potential contribution. For a specific realization of the optical-model potential see, {\it e.g.}, Refs.\,\cite{Mahg08, Pere76}. The global optical-model parameters used in this work are given in Table\,\ref{optmodtable}. They are taken from Ref.\,\cite{Becch69} for the protons and from Ref.\,\cite{Flynn69} for the tritons. The neutron parameters are adopted from Ref.\,\cite{Wirt04,Levon09}.

\begin{table}[!t]
\caption{\label{optmodtable}The optical-model parameters for the $^{242}$Pu$(p,t){}^{240}$Pu reaction used for the DWBA calculations. See text for more information.}
\begin{ruledtabular}
\begin{tabular}{lddd}
Parameters & p{\footnote{Ref.\,\cite{Becch69}}} & t{\footnote{Ref.\,\cite{Flynn69}}} & n{\footnote{Ref.\,\cite{Wirt04,Levon09}}} \\ 
\hline
V$_\mathrm{r}$ [MeV] & 57.71 & 166.70 &  \\
4W{\small $_\mathrm{D}$} [MeV] & 33.91 &  &  \\
W{\small$_0$} [MeV] & 2.58 & 10.28 &  \\
4V$_{\mathrm{so}}$ [MeV] & 24.80 &  &  \\
r$_{\mathrm{r}}$ [fm] & 1.17 & 1.16 & 1.17 \\
r{\small$_{\mathrm{D}}$} [fm] & 1.32 &  &  \\
r{\small $_{0}$} [fm] & 1.32 & 1.50 &  \\
r$_{\mathrm{so}}$ [fm] & 1.01 &  &  \\
r$_{\mathrm{c}}$ [fm] & 1.30 & 1.30 &  \\
a$_{\mathrm{r}}$ [fm] & 0.75 & 0.75 & 0.75 \\
a{\small $_{\mathrm{D}}$} [fm] & 0.67 &  &  \\
a{\small $_{0}$} [fm] & 0.67 & 0.82 &  \\
a$_{\mathrm{so}}$ [fm] & 0.75 &  &  \\
nlc & 0.85 & 0.25 &  \\
$\lambda$ &  &  & 25 \\
\end{tabular}
\end{ruledtabular}
\end{table}

An important aspect of the two-neutron transfer calculations are the chosen transfer configurations. For a ground-state deformation of $\beta_2$= 0.224 \cite{Moel95} one finds several Nilsson orbitals close to the fermi surface of ${}^{242}$Pu \cite{Bohr75}. The spherical analoges to these orbitals are:
\begin{center}
2g$_{9/2}$, 3d$_{5/2}$, 1j$_{15/2}$, 1i$_{11/2}$, and 3p$_{1/2}$ 
\end{center}
In addition, the following orbitals are fairly close:
\begin{center}
2g$_{7/2}$, 2f$_{5/2}$, and 1i$_{13/2}$
\end{center}
The latter two are especially important to generate negative-parity states as transfer configurations like (3d$_{5/2}$)(2f$_{5/2}$) are needed for this. In addition, one needs to break with the convention of $\Delta s = 0$ in the $(p,t)$ reaction to describe the excitation of unnatural parity states, i.e. $J^{\pi} = 2^-, 3^+$, and $5^+$. For these unnatural parity states at least two $L$ transfers are possible. However, depending on the transfer configuration, one might be dominant.

\section{ Experimental Results}
\label{sec:results}

\begin{figure}[!t]
\centering
\includegraphics[width=0.65\linewidth]{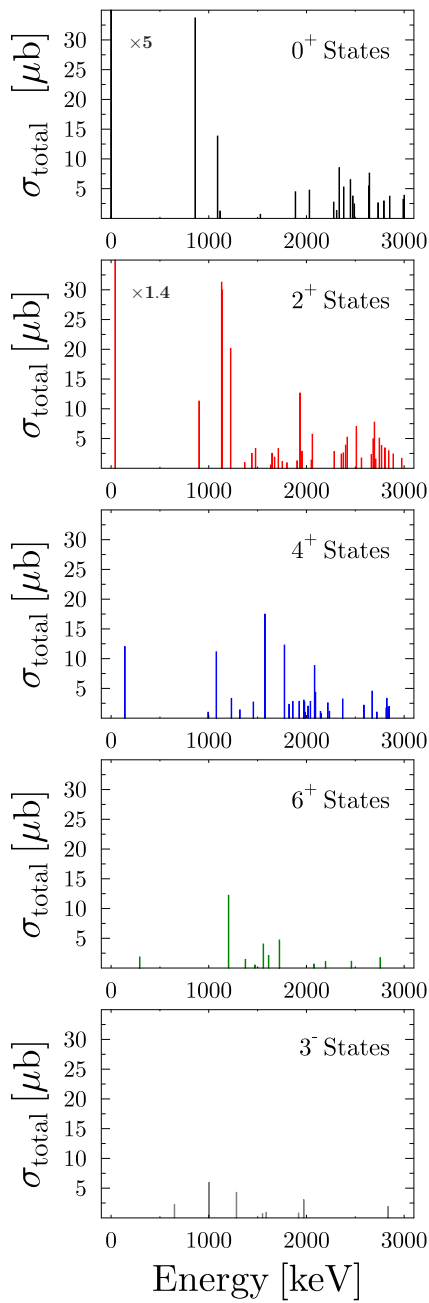}
\caption{\label{fig:crosssections}(Color online) Integrated $(p,t)$ cross sections $\sigma_{\mathrm{total}}$ for $0^+, 2^+, 4^+$, $6^+$ , and $3^-$ states in $^{240}$Pu.}
\end{figure}

All experimental information, which has been obtained, can be found in Tables\,\ref{longtable}-\ref{tab:negbands} as well as in Figs.\,\ref{fig:crosssections}-\ref{fig:octu_angdist}. Before discussing some specific states in detail, we will shortly comment on the general strength distribution observed for $0^+$, $2^+$, $4^+$, $6^+$, and $3^-$ states in $^{240}$Pu. As mentioned earlier, the spin-parity assignment is based on a stringent comparison of the experimentally measured angular distributions and the DWBA predictions. In section \ref{sec:moi}, the rotational bands will be discussed, i.e. those which were previously known and those which have been recognized in this work.
\\

\renewcommand*{\arraystretch}{1.2}
\begin{longtable*}{ddccddc}
%\captionsetup[longtable]{\LTcapwidth=\linewidth}
%\begin{ruledtabular}
%\begin{tabular}
\caption{\label{longtable}Experimental data and two-neutron transfer configurations for the $^{242}$Pu$(p,t)^{240}$Pu reaction. The level energies and spin-parity assignments of states in ${}^{240}$Pu as observed in the $(p,t)$ experiment, and as listed in Ref. \cite{Sing08} are given in the first four columns. Additionally, the integrated $(p,t)$ cross section and the ratio $\sigma_{\mathrm{exp}}/\sigma_{\mathrm{DWBA}}$ is shown. The last column highlights the two-neutron transfer configuration and the excitation scheme if multi-step processes had to be added to the DWBA calculations. Tentative spin-parity assignments are given in parentheses. The stated uncertainties are statistical only.}
\\
\hline
\hline
\multicolumn{2}{c}{ Level Energy [keV] } & \multicolumn{2}{c}{ $J^{\pi}$ } & \multicolumn{1}{c}{$\sigma_{\mathrm{total}}$} &  \multicolumn{1}{c}{\multirow{2}{*}{$\frac{\sigma_{\mathrm{exp}}}{\sigma_{\mathrm{DWBA}}}$}} & Transfer \\
\multicolumn{1}{c}{This Work} & \multicolumn{1}{c}{Ref: \cite{Sing08}} & This Work & Ref. \cite{Sing08} & \multicolumn{1}{c}{[$\mu$b]} &  & configuration\\ 
\hline
\endfirsthead
\multicolumn{7}{c}{Table \ref{table_energy}: ({\it Continued.)}}
\\
\\
\hline
\hline
\multicolumn{2}{c}{ Level energy [keV] } & \multicolumn{2}{c}{ $J^{\pi}$ } & \multicolumn{1}{c}{$\sigma_{\mathrm{total}}$} & \multicolumn{1}{c}{\multirow{2}{*}{$\frac{\sigma_{\mathrm{exp}}}{\sigma_{\mathrm{DWBA}}}$}} & Transfer \\
\multicolumn{1}{c}{This Work} & \multicolumn{1}{c}{Ref: \cite{Sing08}} & This Work & Ref. \cite{Sing08} & \multicolumn{1}{c}{[$\mu$b]} &  & configuration\\ 
\hline
\endhead
\hline
\endfoot
\hline
\hline
\endlastfoot
0.0(1) 	&	 0.0 	&	 0$^+$ 	&	 0$^+$ 	&	 173.75(7) 	&	 10.9 	&	 (2g$_{9/2}$)$^2$  \\
42.0(1) 	&	 42.824(8) 	&	 2$^+$ 	&	 2$^+$ 	&	 49.41(3) 	&	 \multicolumn{1}{c}{22} 	&	 m1a - (2g$_{9/2}$)$^2$ \\
141.5(1) 	&	 141.690(15) 	&	 4$^+$ 	&	 4$^+$ 	&	 12.01(2) 	&	 \multicolumn{1}{c}{70} 	&	 m1b - (1i$_{11/2}$)$^2$ \\
293.5(6) 	&	 294.319(24) 	&	 6$^+$ 	&	 6$^+$ 	&	 1.855(7) 	&	 0.028 	&	 m1b - (1i$_{11/2}$)$^2$ + (3d$_{5/2}$)$^2$ \\
499.1(14) 	&	 497.37(20) 	&	  	&	 8$^+$ 	&	 1.026(5) 	&	 	&  \\
597.2(4) 	&	 597.34(4) 	&	 1$^-$ 	&	 1$^-$ 	&	 0.667(4) 	&	 0.45 	&	 m1a$^{\ast}$ - (1i$_{13/2}$)(1j$_{15/2}$) \\
648.8(4) 	&	 648.86(4) 	&	 3$^-$ 	&	 3$^-$ 	&	 2.209(7) 	&	 2.5 	&	 m1a - (2g$_{9/2}$)(2f$_{5/2}$) \\
745.3(8) 	&	 742.33(4) 	&	 5$^-$ 	&	 5$^-$ 	&	 0.642(4) 	&	 0.118 	&	 (3d$_{5/2}$)(1j$_{15/2}$) \\
861.2(1) 	&	 860.71(7) 	&	 0$^+$ 	&	 0$^+$ 	&	 33.69(3) 	&	 1.4 	&	 (2g$_{9/2}$)$^2$ \\
878.8(5) 	&	 878.1(4) 	&	  	&	 (7$^-$) 	&	 0.774(12)	&	  	&	  \\
901.1(1) 	&	 900.32(4) 	&	 2$^+$ 	&	 2$^+$ 	&	 11.29(2) 	&	 5.8 	&	 m1a - (2g$_{9/2}$)$^2$ \\
938.2(3) 	&	 938.06(6) 	&	 (1$^-$) 	&	 (1$^-$) 	&	 0.838(6) 	&	 0.08 	&	 m1a$^{\ast}$ - (3d$_{5/2}$)(2f$_{5/2}$) \\
959.4(5) 	&	 958.85(6) 	&	 2$^-$ 	&	 (2$^-$) 	&	 0.435(4) 	&	 0.035 	&	 m2a - (3d$_{5/2}$)(3p$_{1/2}$) \\  & & & & & & +(2g$_{9/2}$)(2f$_{5/2}$)\\
993.2(4) 	&	 992.4(5) 	&	 4$^+$ 	&	 4$^+$ 	&	 0.987(5) 	&	 0.072 	&	 m4a - (2g$_{9/2}$)$^2$ \\
1002.3(3) 	&	 1001.94(8) 	&	 3$^-$ 	&	 (3$^-$) 	&	 5.970(12) 	&	 1.858 	&	 (2g$_{9/2}$)(2f$_{5/2}$) \\
1033.3(5) 	&	 1030.55(4)	&	 3$^+$ 	&	 (3$^+$) 	&	 0.636(5) 	&	 0.54 	&	 m1b* - (1j$_{15/2}$)$^2$ + (2g$_{9/2}$)$^2$ \\
1077.2(1) 	&	 1076.22(9) 	&	 4$^+$ 	&	 (4$^+$) 	&	 11.14(2) 	&	 3.6 	&	 m2a - (2g$_{9/2}$)$^2$ \\
1090.3(1) 	&	 1089.45(10) 	&	 0$^+$ 	&	 0$^+$ 	&	 13.83(2) 	&	 0.8 	&	 (2g$_{7/2}$)$^2$  \\
1115.7(5) 	&	 1115.53(6) 	&	 0$^+$ 	&	 (5$^-$) 	&	 1.230(7) 	&	 0.07 	&	 (2g$_{7/2}$)$^2$ \\
1131.9(1) 	&	 1130.95(9) 	&	 2$^+$ 	&	 (2$^+$) 	&	 31.27(5) 	&	 7.4 	&	 (2g$_{7/2}$)$^2$ \\
1138.1(1) 	&	 1136.97(13) 	&	 2$^+$ 	&	 (2$^+$) 	&	 29.97(5) 	&	 8.0 	&	 (2g$_{7/2}$)$^2$ \\
1179.9(4)	&	 1177.63(8) 	&	 ($3^+$) 	&	 (3$^+$) 	&	1.386(6)	&	1.1  	&	m1b* - (1j$_{15/2}$)$^2$ + (2g$_{9/2}$)$^2$  \\
	&	 1180.5(4) 	&	  	&	 (2$^+$) 	&		&	  	&	 \\
1202.8(2) 	&	 \multicolumn{1}{c}{1199(2)} 	&	 (6$^+$) 	&	  	&	 12.23(2) 	&	 0.085 	&	 m1b - (2g$_{7/2}$)$^2$ + (2g$_{9/2}$)$^2$ \\
1224.3(2) 	&	 1222.99(13) 	&	 2$^+$ 	&	 (2$^+$) 	&	 20.14(3) 	&	 7.4 	&	 (2g$_{7/2}$)$^2$ \\
1232.0(5) 	&	 1232.46(10) 	&	 4$^+$ 	&	 (4$^+$) 	&	 3.31(2) 	&	 1.15 	&	 m2a - (2g$_{9/2}$)$^2$ \\
1261.6(6) 	&	 1262.08(24) 	&	 (3$^+$) 	&	 (3$^+$) 	&	 0.597(5) 	&	 0.44 	&	m1b* - (1j$_{15/2}$)$^2$ + (2g$_{9/2}$)$^2$ \\
1283.6(2) 	&	 \multicolumn{1}{c}{1282(2)} 	&	 3$^-$ 	&	 (3$^-$) 	&	 4.286(13) 	&	 0.065 	&	 m1a - (2g$_{9/2}$)(2f$_{5/2}$) \\
1318.7(1) 	&	  	&	 4$^+$ 	&	  	&	 1.393(7) 	&	 0.42 	&	 m2a - (2g$_{9/2}$)$^2$ \\
1325.6(8) 	&	 1323.4(4) 	&	  	&	 (8$^+$) 	&	 0.426(8) 	&	  	&	  \\
1340.5(6) 	&	 1337.02(2) 	&	  	&	 (2$^+$,3,4$^+$) 	&	 0.375(4) 	&	 - 	&	 - \\
1368.8(11) 	&	  	&	 (2$^+$) 	&	  	&	 0.95(2) 	&	 0.19 	&	 (2g$_{7/2}$)$^2$ \\
1375.0(6) 	&	 \multicolumn{1}{c}{1379(4)} 	&	 (6$^+$) 	&	  	&	 1.44(2) 	&	 0.035 	&	 m1b - (2g$_{9/2}$)$^2$ + (3d$_{5/2}$)$^2$ \\
1407.5(6) 	&	 \multicolumn{1}{c}{1407(3)} 	&	 (5$^-$) 	&	  	&	 0.685(6) 	&	 0.0265 	&	 m2a - (3d$_{5/2}$)(2f$_{5/2}$) \\
 	&	 1410.75(11) 	&	  	&	 0$^{(-)}$ 	&	  	&	  	&	  \\
1441.4(1) 	&	 1438.45(8) 	&	 2$^+$ 	&	 2$^{(-)}$ 	&	 2.489(9) 	&	 \multicolumn{1}{c}{30} 	&	 (1i$_{11/2}$)$^2$ \\
1456.5(1) 	&	  	&	 (4$^+$) 	&	  	&	 2.698(11) 	&	 3.6 	&	 m4a - (2g$_{9/2}$)$^2$ \\
1464.1(7) 	&	 	&	  	&	  	&	 0.685(9) 	&	 	&	  \\
1473.0(5) 	&	  	&	 (6$^+$) 	&	  	&	 0.497(10) 	&	 0.027 	&	 m1b - (2g$_{7/2}$)$^2$ \\
1479.2(3) 	&	  	&	 (2$^+$) 	&	  	&	 3.292(11) 	&	 0.47 	&	 m2a - (2g$_{9/2}$)$^2$ \\
1488.4(4) 	&	 1488.17(7) 	&	  	&	 (1,2$^+$)	&	 0.717(14)	&	 - 	&	 - \\
1515.1(4) 	&	  	&	  	&	  	&	 0.539(6) 	&		&	 \\
1528.6(6) 	&	 1525.86(8) 	&	 (5$^+$) 	&	 (0$^+$) 	&	 0.983(6) 	&	 34 	&	 m1b* - (2g$_{9/2}$)$^2$  \\
1540.1(1)	&	1539.67(6)	&	1$^-$	&	(1$^-$)	&	 2.584(9)	&	0.15	&	 m1a$^*$ -  (3d$_{5/2}$)(2f$_{5/2}$) \\
1550.3(6) 	&	  	&	 (3$^-$) 	&	  	&	 0.680(10) 	&	 0.13 	&	 m1a - (2g$_{9/2}$)(2f$_{5/2}$) \\
1559.0(1) 	&	 1558.87(5) 	&	 (6$^+$) 	&	 (2$^+$) 	&	 4.033(11) 	&	 0.0178 	&	 m4a - (2g$_{9/2}$)$^2$ + (3d$_{5/2}$)$^2$ \\
1575.5(1)	&	 \multicolumn{1}{c}{1574} 	&	 4$^+$ 	&	  	&	 17.48(2) 	&	 4.8 	&	 m2a - (2g$_{9/2}$)$^2$ \\
 	&	 \multicolumn{1}{c}{1580(5)} 	&	  	&	  	&	  	&	  	&	  \\
1588.0(6)	&		&	(3$^-$)	&		&	0.941(10)	&	 0.11	&	m1a$^*$ - (1i$_{13/2}$)(1j$_{15/2}$) \\
1612.6(2) 	&	 1607.72(13) 	&	 (6$^+$) 	&	 (1$^-$) 	&	 2.101(9) 	&	 0.12 	&	 m4a - (2g$_{7/2}$)$^2$ + (3d$_{5/2}$)$^2$ \\
 	&	\multicolumn{1}{c}{1609(6)} 	&	  	&	  	&	  	&	  	&	  \\
1626.6(9) 	&	 1626.77(15) 	&	 	&	 (1,2$^+$) 	&	 0.306(9)	&	  	&	  \\
1633.6(3) 	&	 1633.37(7) 	&	 (2$^+$) 	&	 (1,2$^+$) 	&	 0.55(4) 	&	 0.05 	&	 (2g$_{9/2}$)$^2$ \\
1638.6(6) 	&	 \multicolumn{1}{c}{1641(5)} 	&	 (5$^+$) 	&	  	&	 2.072(15) 	&	-  	&	  m1b - (2g$_{9/2}$)$^2$ + (2f$_{5/2}$)$^2$\\
1647.6(4) 	&	 \multicolumn{1}{c}{1641(5)} 	&	 2$^+$ 	&	  	&	 2.511(9) 	&	 0.2 	&	 (2g$_{9/2}$)$^2$ \\
1669.5(9) 	&	  	&	 (5$^-$) 	&	  	&	 0.89(2) 	&	 0.03 	&	 m2a - (3d$_{5/2}$)(2f$_{5/2}$) \\
1674.1(4) 	&	 \multicolumn{1}{c}{1675(2)} 	&	 2$^+$ 	&	  	&	 1.86(2) 	&	 0.03 	&	 (3d$_{5/2}$)$^2$ \\
1686.2(4)	&		&	(5$^-$)	&		&	0.432(6)	&	0.0225	&	m2a – (3d$_{5/2}$)(1j$_{15/2}$) \\
1712.1(2) 	&	 1710.43(8) 	&	 2$^+$ 	&	 (2$^+$) 	&	 3.309(11) 	&	 0.01 	&	 m1b - (3d$_{5/2}$)$^2$ \\
1723.5(1) 	&	  	&	 (6$^+$) 	&	  	&	 4.710(14) 	&	 0.18 	&	 m1b - (2g$_{7/2}$)$^2$ \\
1752.7(2) 	&	 \multicolumn{1}{c}{1752(3)} 	&	 (2$^+$) 	&	  	&	 1.121(8) 	&	 0.004 	&	 m1b - (3d$_{5/2}$)$^2$ \\
1774.8(1) 	&	\multicolumn{1}{c}{1784(3)} 	&	 4$^+$ 	&	  	&	 12.28(2) 	&	 \multicolumn{1}{c}{350} 	&	 m2a - (1i$_{11/2}$)$^2$ \\
1800.2(2) 	&	 1796.34(13) 	&	 (2$^+$) 	&	 (1,2$^+$) 	&	 0.916(7) 	&	 0.004 	&	 m1b - (3d$_{5/2}$)$^2$ \\
1807.4(2)	&	1808.02(13)	&	1$^-$	&	 (1$^-$,2$^+$) 	&	1.298(8)	&	0.11	&	m1a$^*$ - (3d$_{5/2}$)(2f$_{5/2}$) \\
1821.9(1)	&	  	&	 4$^+$ 	&	  	&	 2.31(2) 	&	 0.85 	&	 m2a - (2g$_{9/2}$)$^2$ \\
1860.8(1) 	&	 \multicolumn{1}{c}{1861(3)} 	&	 (4$^+$) 	&	  	&	 2.802(9) 	&	 0.175 	&	 m4a - (2g$_{9/2}$)$^2$ \\
1887.3(1) 	&	 1881.1 	&	 0$^+$ 	&	 (0,1,2) 	&	 4.469(11) 	&	 0.018 	&	 (3d$_{5/2}$)$^2$  \\
1904.1(1)	&	 \multicolumn{1}{c}{1902(3)} 	&	 (2$^+$) 	&	  	&	 1.245(7) 	&	 6.5 	&	 m1b - (1j$_{15/2}$)$^2$ \\
1919.5(6)	&	1917.8(3)	&	(3$^-$)	&	(1$^-$)	&	0.808(13)	&	0.006	&	(3d$_{5/2}$)(3p$_{1/2}$) \\
1925.4(3) 	&	 \multicolumn{1}{c}{1923(3)} 	&	 (4$^+$) 	&	  	&	 2.829(12) 	&	 0.2 	&	 m4a - (2g$_{9/2}$)$^2$ \\
1934.2(1) 	&	  	&	 2$^+$ 	&	  	&	 12.65(2) 	&	 2.6 	&	 (2g$_{7/2}$)$^2$ \\
1946.4(3) 	&	  	&	 (2$^+$) 	&	  	&	 1.292(12) 	&	 6.5 	&	 (2g$_{9/2}$)$^2$ \\
1954.2(3) 	&	 1954.51(8) 	&	 2$^+$ 	&	 2$^+$ 	&	 2.855(11) 	&	 0.6 	&	 (2g$_{7/2}$)$^2$ \\
1967.2(13)	&		&	(5$^-$)	&		&	0.581(13)	&	0.011	&	(3d$_{5/2}$)(1j$_{15/2}$) \\
1973.5(1) 	&	  	&	 (3$^-$, 4$^+$) 	&	 	&	 3.03(2) 	&	 0.33 	&	m1a$^*$ - (1i$_{13/2}$)(1j$_{15/2}$), \\
	&		&		&		&		&	\multicolumn{1}{c}{31}	&	 m3a - (1i$_{11/2}$)$^2$ \\
1980.3(1) 	&	  	&	 (4$^+$) 	&	  	&	 2.74(2) 	&	 \multicolumn{1}{c}{75} 	&	 m2a - (1i$_{11/2}$)$^2$ \\
1987.1(4) 	&	  	&	 (4$^+$) 	&	  	&	 0.944(10) 	&	 0.03 	&	 m2a - (3d$_{5/2}$)$^2$ \\
2016.3(2) 	&	  	&	 4$^+$ 	&	  	&	 2.01(2) 	&	 \multicolumn{1}{c}{44} 	&	 m2a - (1i$_{11/2}$)$^2$ \\
2020.8(3)	&	  	&	  	&	  	&	 4.78(2) 	&	  	&	  \\
2030.4(1) 	&	  	&	 0$^+$ 	&	  	&	 4.744(13) 	&	0.017 	&	 (3d$_{5/2}$)$^2$ \\
2040.6(5) 	&	  	&	 (4$^+$) 	&	  	&	 2.803(13) 	&	 0.21 	&	 m4a - (2g$_{9/2}$)$^2$ \\
2050.0(2) 	&	  	&	 2$^+$ 	&	  	&	 1.368(9) 	&	 0.2 	&	 (2g$_{9/2}$)$^2$ \\
2060.4(1) 	&	  	&	 2$^+$ 	&	  	&	 5.686(15) 	&	 \multicolumn{1}{c}{55} 	&	 (1i$_{11/2}$)$^2$ \\
2076.7(6) 	&	  	&	 (6$^+$) 	&	  	&	 0.664(13) 	&	 0.055 	&	 m1b - (2g$_{7/2}$)$^2$ + (3d$_{5/2}$)$^2$ \\
2083.4(1) 	&	  	&	 4$^+$ 	&	  	&	 8.82(2) 	&	 0.38 	&	 m2a - (3d$_{5/2}$)$^2$ \\
2092.7(1) 	&	  	&	 (4$^+$) 	&	  	&	 4.34(2) 	&	 1.14 	&	 m2a - (2g$_{9/2}$)$^2$ \\
2105.3(4) 	&	  	&	 	&	  	&	 1.209(11) 	&	  	&	  \\
2112.9(2) 	&	  	&	 	&	  	&	 0.676(7) 	&	  	&	 \\
2143.4(3) 	&	  	&	 (4$^+$) 	&	  	&	 1.121(11) 	&	 0.325 	&	 m2a - (2g$_{9/2}$)$^2$ \\
2151.2(5) 	&	  	&	 (4$^+$) 	&	  	&	 0.754(9) 	&	 0.378 	&	 (2g$_{7/2}$)$^2$ \\
2184.9(4) 	&	  	&	($5^+$)  	&	  	&	0.809(7) 	&	  34	&	m1b* - (2g$_{9/2}$)$^2$\\
2195.4(6) 	&	  	&	 (6$^+$) 	&	  	&	 1.054(9) 	&	 0.014 	&	 m1b - (2g$_{7/2}$)$^2$ + (2g$_{9/2}$)$^2$ \\
2209.2(3)	&	  	&	  	&	  	&	 2.11(2) 	&	  	&	  \\
2219.4(3) 	&	  	&	 (4$^+$) 	&	  	&	 2.57(2) 	&	0.115 	&	 m2a - (3d$_{5/2}$)$^2$ \\
2235.2(1) 	&	  	&	 4$^+$ 	&	  	&	 1.134(8) 	&	 0.052 	&	 m2a - (3d$_{5/2}$)$^2$ \\
2279.1(6) 	&	  	&	 0$^+$ 	&	  	&	 2.73(3) 	&	 0.06 	&	 (2g$_{9/2}$)$^2$ \\
2284.2(3) 	&	  	&	 2$^+$ 	&	  	&	 2.81(4) 	&	 \multicolumn{1}{c}{27} 	&	 (1i$_{11/2}$)$^2$ \\
2289.1(4) 	&	  	&	  	&	  	&	 2.02(3) 	&	  	&	  \\
2309.3(2) 	&	  	&	 (0$^+$) 	&	  	&	 1.308(11) 	&	 0.043 	&	 (2g$_{7/2}$)$^2$  \\
2335.7(4) 	&	  	&	 0$^+$ 	&	  	&	 8.52(3) 	&	 0.034 	&	 (3d$_{5/2}$)$^2$  \\
2357.3(1) 	&	  	&	 2$^+$ 	&	  	&	 2.39(3) 	&	 0.19 	&	 (2g$_{9/2}$)$^2$ \\
2371.6(3) 	&	  	&	 (4$^+$) 	&	  	&	 3.21(2) 	&	 0.15 	&	 m2a - (3d$_{5/2}$)$^2$ \\
2377.5(4) 	&	  	&	 2$^+$ 	&	  	&	 2.59(3) 	&	 0.06 	&	 m1b - (3d$_{5/2}$)$^2$ \\
2381.8(4) 	&	  	&	 0$^+$ 	&	  	&	 5.27(3) 	&	 0.016 	&	 (3d$_{5/2}$)$^2$  \\
2401.3(6) 	&	  	&	 2$^+$ 	&	  	&	 3.89(2) 	&	 0.55 	&	 (2g$_{7/2}$)$^2$ \\
2417.5(1) 	&	  	&	 2$^+$ 	&	  	&	 5.22(2) 	&	 0.8 	&	 (2g$_{7/2}$)$^2$ \\
2450.3(7) 	&	  	&	 0$^+$ 	&	  	&	 6.50(4) 	&	 0.018 	&	 (3d$_{5/2}$)$^2$ \\
2460.5(5) 	&	  	&	 (6$^+$) 	&	&	 1.13(2) 	&	 0.016 	&	 m1b - (2g$_{7/2}$)$^2$ + (2g$_{9/2}$)$^2$ \\
2470.4(2) 	&	  	&	  	&	 	&	 3.58(4) 	&	  	&	 \\
2474.9(5) 	&	  	&	 0$^+$ 	&	  	&	 3.75(4) 	&	 0.011 	&	 (3d$_{5/2}$)$^2$  \\
2490.1(7) 	&	  	&	 0$^+$ 	&	  	&	 2.43(2) 	&	 0.007 	&	 (3d$_{5/2}$)$^2$  \\
2504.5(8) 	&	  	&	  	&	  	&	 1.820(13) 	&	 	&	  \\
2510.9(1) 	&	  	&	 2$^+$ 	&	  	&	 6.99(2) 	&	 3.9 	&	 m2a - (2g$_{7/2}$)$^2$ \\
2524.9(2) 	&	  	&	  	&	  	&	 0.983(15) 	&	  	&	  \\
2563.2(4) 	&	  	&	 (2$^+$) 	&	  	&	 1.73(2) 	&	 \multicolumn{1}{c}{16} 	&	 (1i$_{11/2}$)$^2$ \\
2588.8(2) 	&	  	&	 4$^+$ 	&	  	&	 2.179(13) 	&	 0.115 	&	 m2a - (3d$_{5/2}$)$^2$ \\
2640.4(1) 	&	  	&	 0$^+$ 	&	  	&	 5.50(3) 	&	 0.016 	&	 (3d$_{5/2}$)$^2$  \\
2644.6(2) 	&	  	&	 0$^+$ 	&	  	&	 7.61(4) 	&	 0.023 	&	 (3d$_{5/2}$)$^2$  \\
2664.3(5) 	&	  	&	 2$^+$ 	&	  	&	 2.288(13) 	&	 \multicolumn{1}{c}{21} 	&	 (1i$_{11/2}$)$^2$ \\
2672.6(2) 	&	  	&	 (4$^+$) 	&	  	&	 4.53(2) 	&	 11.5 	&	 m3a - (1i$_{11/2}$)$^2$ \\
2683.3(2) 	&	  	&	 2$^+$ 	&	 	&	 4.945(14) 	&	 0.76 	&	 (2g$_{7/2}$)$^2$ \\
2695.9(1) 	&	  	&	 2$^+$ 	&	  	&	 7.75(3) 	&	 6.6 	&	 (2g$_{9/2}$)$^2$ \\
2709.4(2) 	&	  	&	 (2$^+$) 	&	  	&	 1.548(11) 	&	 6.6 	&	 m1b - (1j$_{15/2}$)$^2$ \\
2721.3(7) 	&	  	&	 (4$^+$) 	&	  	&	 1.01(2) 	&	 0.04 	&	 m2a - (3d$_{5/2}$)$^2$ \\
2733.4(3) 	&	  	&	 0$^+$ 	&	  	&	 2.62(2) 	&	 0.007 	&	 (3d$_{5/2}$)$^2$  \\
2746.3(1) 	&	  	&	 2$^+$ 	&	  	&	 5.06(3) 	&	 0.05 	&	 (3d$_{5/2}$)$^2$ \\
2755.2(2) 	&	  	&	 (6$^+$) 	&	  	&	 1.733(14) 	&	 0.061 	&	 m1b - (2g$_{7/2}$)$^2$ + (2g$_{9/2}$)$^2$ \\
2769.0(2) 	&	  	&	 2$^+$ 	&	  	&	 3.83(3) 	&	 1.05 	&	 m2a - (2g$_{9/2}$)$^2$ \\
2793.2(2) 	&	  	&	 0$^+$ 	&	  	&	 2.93(4) 	&	 0.006 	&	 (3d$_{5/2}$)$^2$  \\
2803.8(1) 	&	  	&	 2$^+$ 	&	  	&	 3.425(14) 	&	 \multicolumn{1}{c}{30} 	&	 (1i$_{11/2}$)$^2$ \\
2816.5(7) 	&	  	&	 (4$^+$) 	&	 	&	 1.70(3) 	&	 0.08 	&	 m2a - (3d$_{5/2}$)$^2$ \\
2823.4(3) 	&	  	&	 (4$^+$) 	&	  	&	 3.33(4) 	&	 0.067 	&	 m2a - (3d$_{5/2}$)$^2$ \\
2835.5(4) 	&	  	&	 ($3^-$,4$^+$) 	&	  	&	 1.89(2) 	&	 \multicolumn{1}{c}{0.2} 	&	 m1a* - (1i$_{13/2}$)(1j$_{15/2}$) \\
 	&	  	&	  	&	  	&	  	&	 \multicolumn{1}{c}{22} 	&	 m3a - (1i$_{11/2}$)$^2$ \\
2842.5(2) 	&	  	&	 2$^+$ 	&	  	&	 2.94(2) 	&	 \multicolumn{1}{c}{25} 	&	 (1i$_{11/2}$)$^2$ \\
2847.2(4) 	&	  	&	 (4$^+$) 	&	  	&	 1.97(3) 	&	 0.14 	&	 m2a - (3d$_{5/2}$)$^2$ \\
2852.6(4) 	&	  	&	 0$^+$ 	&	  	&	 3.73(2) 	&	 0.01 	&	 (3d$_{5/2}$)$^2$  \\
2885.2(8) 	&	  	&	  	&	  	&	 1.71(4) 	&	  	&	  \\
2888.5(4) 	&	  	&	 2$^+$ 	&	  	&	 2.40(4) 	&	 \multicolumn{1}{c}{27} 	&	 (1i$_{11/2}$)$^2$ \\
2975.7(2) 	&	  	&	 2$^+$ 	&	  	&	 1.66(2) 	&	 0.09 	&	 (2g$_{9/2}$)$^2$ \\
2990.8(3) 	&	  	&	 (0$^+$) 	&	  	&	 3.21(2) 	&	 0.006 	&	 (3d$_{5/2}$)$^2$ \\
3000.9(2) 	&	  	&	 0$^+$ 	&	  	&	 3.88(7) 	&	 0.057 	&	 (3d$_{5/2}$)$^2$
\label{table_energy}
%\end{tabular}
%\end{ruledtabular}
\end{longtable*}

The $(p,t)$ cross sections are given in Table\,\ref{longtable} and presented in Fig.\,\ref{fig:crosssections}. As observed in all well-deformed actinide nuclei\,\cite{Wirt04, Levon09, Levon13, Levon15}, the $0^+$ ground state is strongly excited in the $(p,t)$ reaction. The $(p,t)$ ground-state cross section is remarkably stable and only relatively little variations are observed, i.e. $\sigma_{\mathrm{total}} \sim 170$\,$\mu$b. In fact, this can be attributed to a little variation in the $R_{4/2}$ ratio  between target and residual nucleus\,\cite{Clark09}, i.e. $\delta R_{4/2} \lesssim 0.2$, and could indicate the stability of quadrupole deformation in the heavy actinides. We will give more details in Sec.\,\ref{sec:disc}.
\\

The ground-state rotational band members are excited up to $J^{\pi} = 8^+$ in our experiment. Nevertheless, angular distributions could only be measured up to $J^{\pi} = 6^+$. As one can see from Fig.\,\ref{fig:crosssections}, the cross section is successively decreasing with spin. This fact will later on be used to assign specific excited states to a rotational band. Another interesting observation is the clear gap between the firmly assigned $0^+_4$ state at 1115.7\,keV and the next firmly assigned $0^+$ state at 1887.3\,keV. The neutron-pairing gap $\Delta_n$ is located at about 545\,keV in $^{240}$Pu\,\cite{Audi03}. Therefore, non-collective 2QP excitations are expected above an energy of $2 \Delta_n \approx 1090$\,keV,  see also Ref.\,\cite{Spiek13a}. For $J \geq 2$, no gap is observed. Still, it is interesting that the next fairly strongly excited $2^+$ state is found at an energy of 1934.2\,keV ($\sigma_{\mathrm{total}} = 12.65(2)$\,$\mu$b).
\\

In general, states with unnatural parity are only weakly excited in the $(p,t)$ reaction. Only a few $3^+$ and $5^+$ states could be identified as their bandheads were strongly excited, see Fig.\,\ref{fig:others_angdist}. The only unnatural negative-parity state observed is the $K^{\pi} = 1^-_1 , J^{\pi} = 2^-_1$ band member. One should additionally note that aside from $J^{\pi} = 3^-$ states, which are comparably strongly excited via $L = 3$ transfers, negative-parity states are also weakly excited via the $(p,t)$ reaction. Nevertheless, all previously known $K$ projections of the one-octupole-phonon excitation were observed and a new but very tentative candidate for the $K^{\pi} = 3^-$ projection at an energy of 1550.3\,keV is proposed. Furthermore, two $1^-$ states and rotational band members newly proposed might have been observed at energies of 1540.1\,keV and 1807.4\,keV, respectively. We will also discuss these states in Secs.\,\ref{sec:negstat}, \ref{sec:moi} and \ref{sec:disc}.
\\

\subsection{$J^{\pi} = 0^+$ states}
\label{sec:0+}

\begin{figure*}[!t]
\centering
\includegraphics[width=0.85\linewidth]{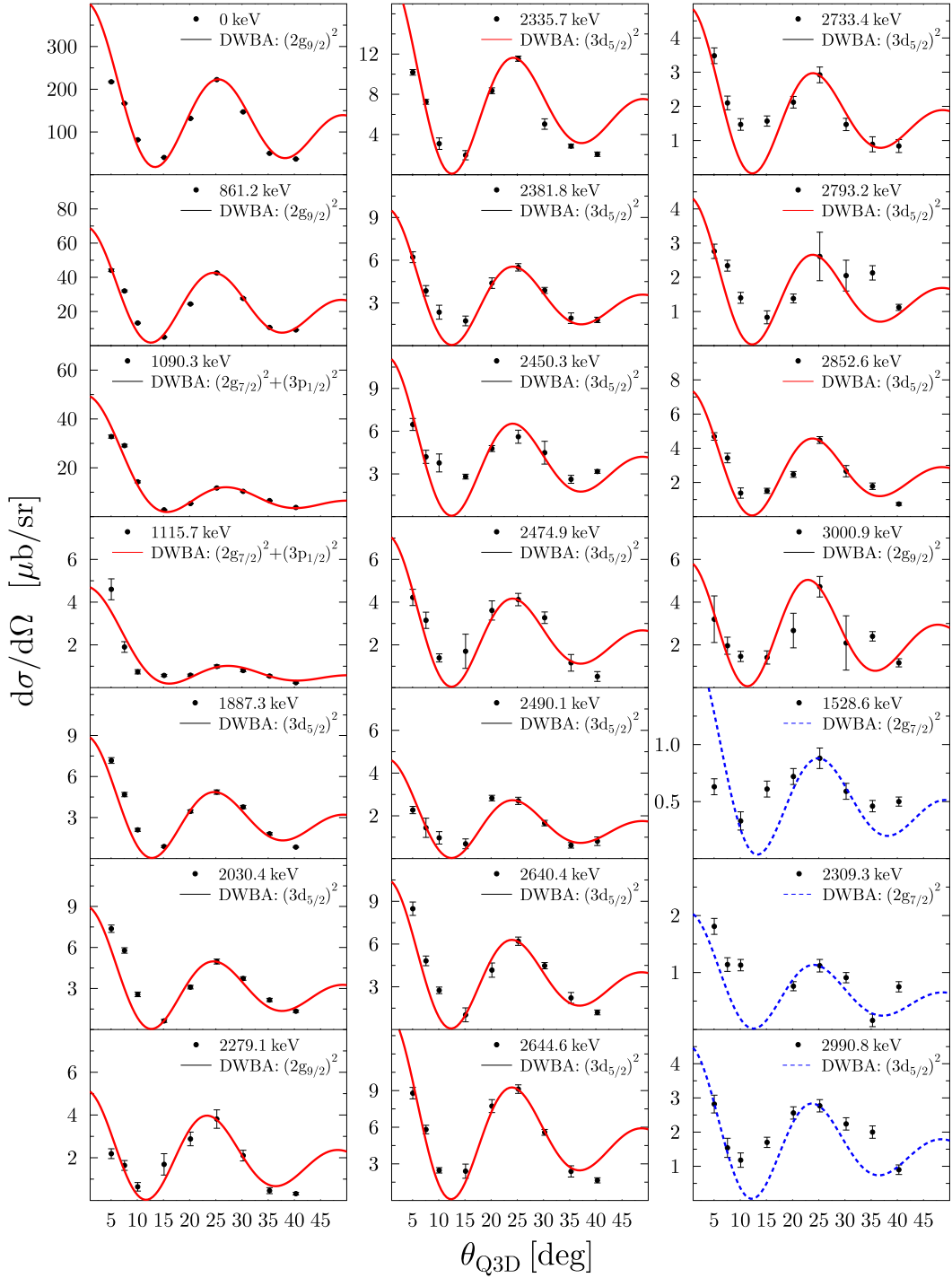}
\caption{\label{fig:0+_angdist}(Color online) Angular distributions of the $J^{\pi}$=~0$^+$ states in ${}^{240}$Pu. The angular distributions for the corresponding $L$~=~0 transfer calculated with the CHUCK3 code \cite{CHUCK} are shown with lines. Red lines correspond to firm assignments and blue-dashed lines to tentative assignments, respectively. Two-neutron transfer configurations of orbitals close to the Fermi surface were chosen.}
\end{figure*}

$0^+$ states are strongly excited in the $(p,t)$ reaction. In total 17 excited and firmly assigned $0^+$ states were observed up to an excitation energy of 3\,MeV. In addition, three states are tentatively assigned $0^+$ states. Their angular distributions are shown in Fig.\,\ref{fig:0+_angdist}. All experimental distributions could be described by assuming a direct excitation in the $(p,t)$ reaction. The (2g$_{9/2}$)$^2$ and (3d$_{5/2}$)$^2$ transfer configurations provided the best description of the angular distributions. The summed relative strength of all $0^+$ states adds up to 68.45(8)\,$\%$ of the ground-state transfer cross section, which is comparable to the cases of $^{228,230}$Th\,\cite{Levon09, Levon13} but slightly less strength than observed in the $^{234}$U$(p,t){}^{232}$U reaction\,\cite{Levon15}. 
\\
\\
Note, that prior to this experiment only two excited $0^+$ states were known. These included the proposed double-octupole phonon $J^{\pi} = 0^+_2$ state at $E_{\mathrm{x}} = 861.2$\,keV and the $J^{\pi} = 0^+_3$ state at $E_{\mathrm{x}} = 1090.3$\,keV. Structure implications for both states have already been discussed in our previous publication\,\cite{Spiek13a} and will be further discussed in Sec.\,\ref{sec:disc}. We want to comment on two specific excited states which were previously discussed to be possible $0^+$ states.
\\

{\it 1407.5\,keV:}  In Refs.\,\cite{Parek82, Hseuh81, Schmo70, Thomp75} a state at an energy of 1410.75(11)\,keV was controversially discussed. Ref.\,\cite{Schmo70} interpreted it as being the bandhead of a two-phonon octupole vibrational band with an energy of $E$(two-phonon) $\thickapprox$ $2 \cdot E$(one-phonon) and, consequently, assigned $K^{\pi}= 0^+$. Additional evidence came from a $2^+$ state at 1438.5\,keV on top of it and, thus, a rotational band with a MoI close to the one-phonon octupole vibrational band. In addition, Schmorak {\it et al.} observed only {\it E1} transitions which depopulated these two states to the one-phonon octupole vibrational band\,\cite{Schmo70}. Furthermore, these states were not populated in single-neutron-transfer reactions. In Ref.\,\cite{Hseuh81}, the assignment was rejected due to new neutron capture data and a $K^{\pi}= 0^-$ assigment was proposed for the state at 1410.8\,keV, as well as a $J^{\pi} = 2^-$ assignment for the state at 1438.5\,keV. Microscopically, these two states were interpreted as proton-quasiparticle excitations built mainly out of the [642$\tfrac{5}{2}^+$]$_\mathrm{p}$ and [523$\tfrac{5}{2}^-$]$_\mathrm{p}$ configurations. This interpretation would be one reason for the non-observation of these states in single-neutron-transfer experiments. In the $(d,d')$ reaction, Thompson {\it et al.}\,\cite{Thomp75} excited a state at 1407(3)\,keV which the authors attributed to the two-phonon octupole band reported in Ref.\,\cite{Schmo70}. A state at 1407.5(6)\,keV was also excited in our $(p,t)$ experiment with a relative strength of $\sigma/\sigma_{\mathrm{0^+_1}}\sim$~0.4, see Table\,\ref{longtable}. Compared to the other 0$^+$ angular distributions of Fig.\,\ref{fig:0+_angdist}, a different shape was observed. Therefore, a K$^{\pi}$= 0$^-$ assigment might be favored. Despite this, it is questionable if this excited state is indeed the proposed proton-quasiparticle excitation of Ref.\,\cite{Hseuh81}, which should in first order not be excited in a two-neutron transfer. Instead, based on our data a spin-parity assignment of J$^{\pi}$= 5$^-$ is proposed and the state is recognized as a possible bandmember of the $K= 2$ one-octupole phonon projection. It is very likely that this state corresponds to the state which was also excited in the $(d,d^{\prime})$ experiment of Ref.\,\cite{Thomp75}.
\\
A strong argument for the $K^{\pi} = 0^+$ assignment of Schmorak {\it et al.}\,\cite{Schmo70} was the observation of a $2^+$ state above this proposed $0^+$ state as mentioned earlier. In our $(p,t)$ experiment, an excited state at an energy of 1441.4 keV with an integrated cross section of 2.489(9) $\mu$b was observed. It is described well by a single-step transfer to a $2^+$ state, see Table\,\ref{longtable} and Fig.\,\ref{fig:others_angdist}. Therefore, a $J^{\pi}= 2^+$ assignment is strongly favored. However, its larger cross section is in conflict with the expectation of a decreasing cross section within a rotational band. Consequently, the band assignment of this $2^+$ state to the state at 1407.5\,keV is questionable and should be dropped.
\\

{\it 1528.6\,keV:} The previously tentatively assigned $0^+$ state was adopted at an energy of 1525.86(8)\,keV\,\cite{Sing08}. The population and the decay of this state has been observed in the $\beta^-$ decay of the 7.22 min $J^{\pi}$=~(1$^+$) isomer of ${}^{240}$Np\,\cite{Hseuh81} and in the ${}^{239}$Pu(n,$\gamma$) capture reaction with neutrons of 2 keV\,\cite{Chrie85}. In Ref.\,\cite{Sing08}, a tentatively assigned $J^{\pi} = 2^+$ state at an energy of 1558.85(5)\,keV is listed as its rotational bandmember. This $2^+$ assignment is in conflict with the new $(p,t)$ data which favors a $J^{\pi}= 6^+$ assignment for a state at an energy of 1559.0(1) keV. This $6^+$ state will be discussed later on. Nonetheless, it cannot be excluded by the $(p,t)$ data that the weakly excited state at 1528.6(6)\,keV is indeed a $0^+$ state. A fair agreement with the experimental angular distribution has been achieved with a two-neutron transfer configuration of (2g$_{7/2}$)$^2$ in a single-step transfer. However, the agreement is still poor compared to other $0^+$ states, see Fig.\,\ref{fig:0+_angdist}. It should be mentioned that the deviation between the observed and listed level energy is quite large. The observation of such a large deviation is rare in the present $(p,t)$ study. Therefore, it might also be possible that the state's weak excitation hints at a spin-parity assignment different from $0^+$. The present data favor a $J^{\pi} = 5^+$ assignment, see Fig.\,\ref{fig:others_angdist}. In addition, the state might belong to a $K^{\pi} = 4^+$ rotational band at an energy of 1456.5\,keV, see Table\,\ref{tab:posbands}.

\subsection{$J^{\pi} = 2^+$ states}

\begin{figure*}[!t]
\centering
\includegraphics[width=0.95\linewidth]{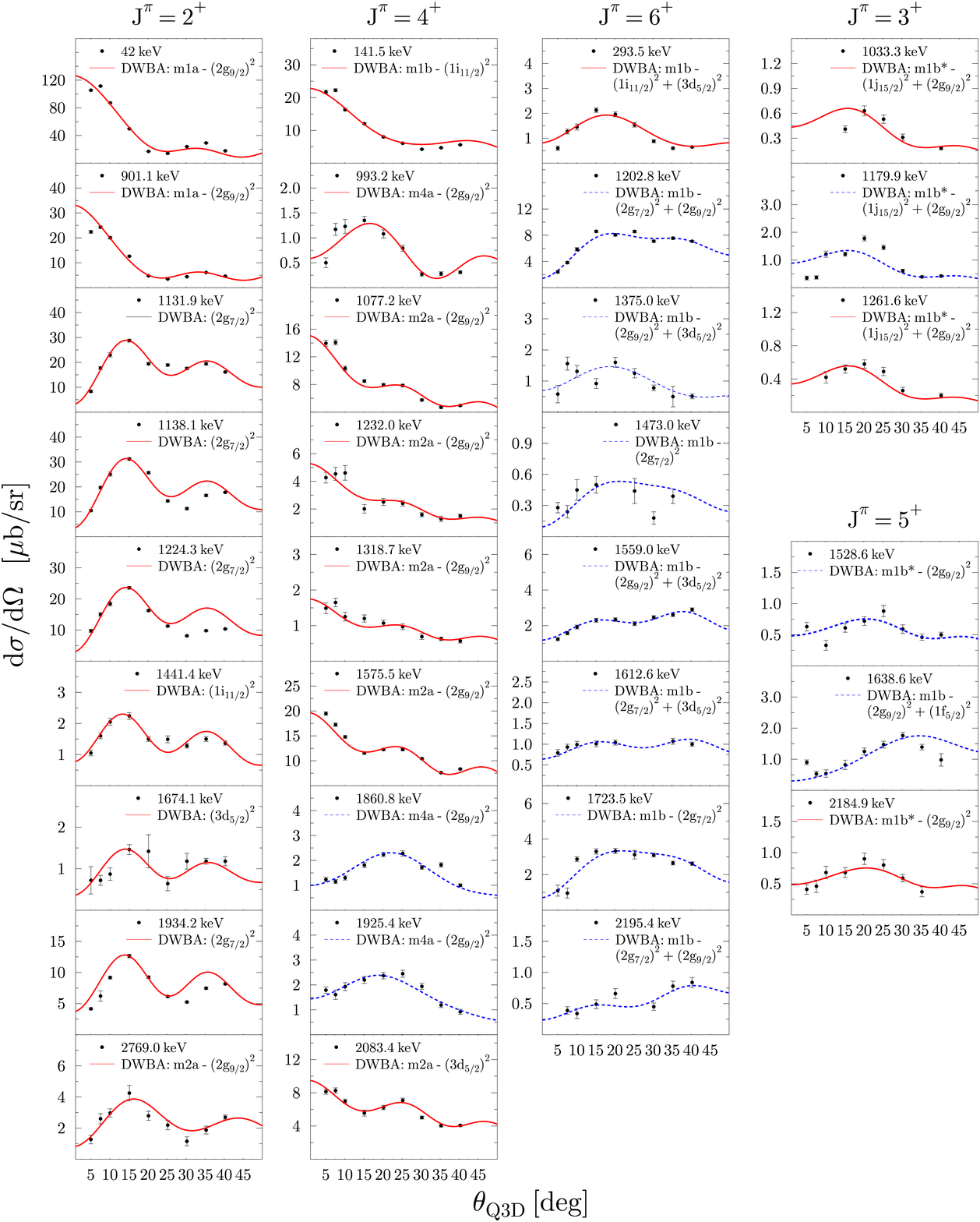}
\caption{\label{fig:others_angdist}(Color online) Same as Fig.\,\ref{fig:0+_angdist} but for $J^{\pi} = 2^+,3^+,4^+,5^+$, and $6^+$ states in $^{240}$Pu.}
\end{figure*}

In total, 28 excited $2^+$ states were firmly identified. In addition, there are nine tentatively assigned $2^+$ states up to an energy of 3\,MeV. Previously, most known states in ${}^{240}$Pu were either firmly assigned $2^+$ states or a possible spin-parity assignment of $(1,2^+)$ was listed\,\cite{Sing08}. The states were mostly observed in neutron capture reactions\,\cite{Sing08, Chrie85} and the $\beta$ decays leading to ${}^{240}$Pu\,\cite{Parek82, Hseuh81}. In our study, the $2^+$ assignments were confirmed for almost all states listed in the Nuclear Data Sheets\,\cite{Sing08}. Some angular distributions and corresponding DWBA calculations are shown in Fig.\,\ref{fig:others_angdist}. In contrast to the 0$^+$ states, multistep processes had to be included for several states. These have been highlighted in Table\,\ref{longtable} and Fig.\,\ref{fig:others_angdist} with their excitation scheme and the transfer configurations. Several two-neutron transfer configurations were used allowing for a good agreement with the experimental data.
\\

{\it 1138.1\,keV:} The first $2^+$ state, which is not a rotational bandmember of a $K^{\pi} = 0^+$ band, has been identified at an energy of 1138.1(1)\,keV. Due to the excellent energy resolution of the experimental setup, it has been possible to separate both $2^+$ states at 1131.9(1)\,keV and 1138.1(1)\,keV unambiguously. According to the classical picture of quadrupole vibrations in deformed nuclei, one might expect that this state is the $\gamma$-vibrational state. Indeed, in Ref.\,\cite{Parek82} it has been proposed as such. Unfortunately, no $B(E2)$ value has been measured up to now and, thus, a definite statement is not possible. Both $2^+$ states at approx. 1.1\,MeV are strongly excited in the $(p,t)$ reaction, see Table\,\ref{longtable}. Thus, no classification in terms of the $(p,t)$ cross section is possible. In Ref.\,\cite{Sing08} two rotational bandmembers at energies of 1177.63\,keV ($J^{\pi}= 3^+$), and 1232.46\,keV ($J^{\pi}= 4^+$) are listed, which both have tentative spin-parity assignments. While for the latter the spin-parity assignment is confirmed, the assumption of a doublet\,\cite{Sing08} at an energy of roughly 1177 keV might be confirmed. Within the scope of this work, assuming only $J^{\pi}= 3^+$ did not yield a satisfactory agreement with the data, see Fig.\,\ref{fig:others_angdist}. Implications coming from the moment of inertia of $2.88(5) \times  10^6$\,MeV\,fm$^2$/c will be discussed later.
\\

{\it 1224.3\,keV:} Nearby, another strongly excited $2^+$ state at an energy of 1224.3(2)\,keV has now been firmly assigned. In addition, the previously known and tentatively assigned $3^+$ state at 1261.6(6)\,keV and newly assigned $4^+$ as well as $6^+$ states at 1318.7(1)\,keV and 1473.0(5)\,keV, respectively, are proposed as its bandmembers. If these assigments are correct, a MoI of $2.87(6) \times 10^6$\,MeV\,fm$^2$/c is determined. This MoI is located between the moments of inertia of the $K^{\pi}=0^+_3$ and the $K^{\pi} = 0^+_2$ rotational bands. 

\subsection{$J^{\pi} = 4^+$ states}

The strength distribution of $4^+$ states, which has been observed in the $(p,t)$ reaction, is completely different to what was observed in the cases of the $0^+$ and $2^+$ states. While in the latter two cases, the respective ground-state bandmembers were the most strongly excited states, a $4^+$ state at an energy of 1575.5(1)\,keV has the largest $(p,t)$ cross section. In addition, three strongly excited states are found at excitation energies of 1077.2(1)\,keV, 1774.8(1)\,keV, and 2083.4(1)\,keV, respectively. Multistep processes had to be included in the DWBA calculations for almost all excited $4^+$ states, see Table\,\ref{longtable} and Fig.\,\ref{fig:others_angdist}. This is a fact, which was in its extent unexpected due to the experience with previous $(p,t)$ studies in $^{228,230}$Th\,\cite{Levon09, Levon13} but, however, multistep excitations were also observed for the case of $^{232}$U\,\cite{Levon15}. Nevertheless, it has been possible to assign 30 $4^+$ states out of which twelve are firmly and 18 tentatively assigned, respectively. Two of these states have been proposed as $K^{\pi}= 4^+$ rotational bandheads, see Tab.\,\ref{tab:posbands}. Many other are bandmembers of $K^{\pi}= 0^+$ or $2^+$ rotational bands, respectively.
\\

{\it 1077.2\,keV:} The state at 1077.2(1)\,keV is recognized as rotational bandmember of a $K^{\pi} = 3^+$ neutron QP-band\,\cite{Parek82} whose bandhead is found at an energy of 1033.3(5)\,keV in our experiment, see Fig.\,\ref{fig:others_angdist}. While the $5^+$ state of this band is not observed, the tentatively assigned and also strongly excited $6^+$ member is observed at an energy of 1202.8(2)\,keV.
\\

{\it 1575.5\,keV:} The total $(p,t)$ cross section of the $4^+$ state at 1575.5(1)\,keV is remarkable ($\sigma_{\mathrm{total}} = 17.48(2)$\,$\mu$b). It is recognized as the seventh strongest excited state. In the $(p,t)$ studies of Maher {\it et al.} it has been observed for the first time\,\cite{Maher72}. The $(d,d^{\prime})$ experiment of Thompson {\it et al.} excited this state as well and a rather strong excitation was observed\,\cite{Thomp75}. However, none of them was able to assign a spin. Even though, multistep processes had to be considered, the experimental as well as the DWBA angular distribution clearly reflect the shape of a positive-parity $L= 4$ transfer. Furthermore, a rotational band built upon this $4^+$ state is observed. Its newly assigned $5^+$ and $6^+$ bandmembers are located at 1638.6(6)\,keV and 1723.5(1)\,keV, respectively. The $6^+$ state is the second strongest excited 6$^+$ state, see Table\,\ref{longtable} and Fig.\,\ref{fig:others_angdist}. The MoI of this $K^{\pi} = 4^+$ rotational band has a value of $2.884(2) \times 10^6$\,MeV\,fm$^2$/c.
\\

{\it 1774.8\,keV:} The excited state at 1774.8(1)\,keV does not belong to any rotational band nor any rotational bandmembers have been found. Previously, a state at an energy of 1775.27(15)\,keV was experimentally observed. Due to its observation in $\beta$ decay\,\cite{Parek82} and neutron-capture reactions\,\cite{Chrie85}, it has been assigned $J^{\pi}= (1^-)$. It is unlikely that the current $(p,t)$ study populated the same state since a strong population of a $4^+$ state in neutron capture has not been observed in ${}^{240}$Pu\,\cite{Chrie85}. Therefore, it is assumed that this firmly assigned and strongly excited $4^+$ state ($\sigma_{\mathrm{total}} = 12.28(2)$\,$\mu$b) has been seen for the first time.
\\

{\it 2083.4\,keV:} This state with a cross section of $\sigma_{\mathrm{total}} = 8.82(2)$\,$\mu$b is a member of a strongly excited $K^{\pi} = 2^+$ rotational band at an energy of 1934.2(1)\,keV. The $5^+$ member is observed at an energy of 2184.9(4)\,keV, see Table\,\ref{longtable} and Fig.\,\ref{fig:others_angdist} leading to a MoI of $1.821(2) \cdot 10^6$\,MeV\,fm$^2$\,c$^2$. It is worth mentioning that close in energy a $K^{\pi} = 0^+$ rotational band has also been found with a rather small MoI, i.e. $I \leq 2 \cdot 10^6$\,MeV\,fm$^2$\,c$^2$, see Table\,\ref{tab:posbands}.

\begin{figure*}[!t]
\centering
\includegraphics[width=1\linewidth]{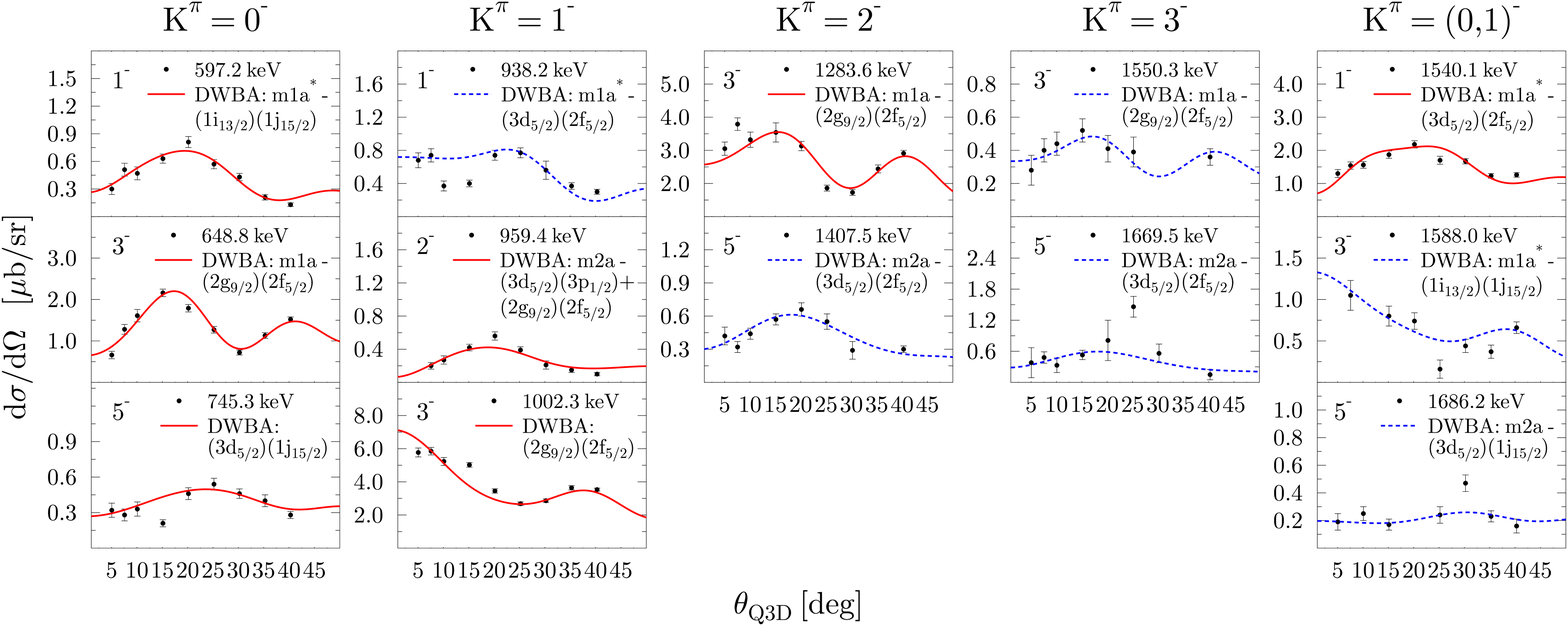}
\caption{\label{fig:octu_angdist}(Color online) Angular distributions of one-phonon octupole states and of the tentatively assigned $K^{\pi} = (0)^-$ bandmembers in $^{240}$Pu. See Sec.\,\ref{sec:dwba} and Fig.\,\ref{fig:0+_angdist} for information about the DWBA calculations.}
\end{figure*}

\subsection{$J^{\pi} = 6^+$ states}

Besides the $J^{\pi} = 6^+$ state of the ground-state rotational band and the tentatively assigned $J^{\pi} = 6^+$ state of the $K^{\pi} = 0^+_2$ rotational band, no $6^+$ state is listed in the Nuclear Data Sheets\,\cite{Sing08}. While the latter is not observed in the current $(p,t)$ study due to the proximity to the strongly excited $J^{\pi} = 2^+$ states of the $K^{\pi}= 0^+_3$ and $K^{\pi}= 2^+_1$ rotational bands, respectively, ten previously unknown states are tentatively assigned $J^{\pi} =6^+$, see Fig.\,\ref{fig:others_angdist} for some examples.
\\

For all excited $6^+$ states it was necessary to add multistep excitation to the DWBA calculations. Here, different two-neutron transfer configurations for the direct population (single-step process) and the indirect populations (multistep processes) had to be considered. In Fig.\,\ref{fig:others_angdist}, this is highlighted by two transfer configurations, where ($j_1)^2$ is the configuration of the direct and ($j_2)^2$ of the indirect population of the respective state. This procedure had to be used to reproduce the experimental angular distribution of the well-known $6^+$ ground-state bandmember at 293.5(6)\,keV as well as for the $3^+$ and $5^+$ states shown in Fig.\,\ref{fig:others_angdist}.
\\

{\it 1202.8\,keV:} The most strongly excited $6^+$ state ($\sigma_{\mathrm{total}}= 12.23(2)$\,$\mu$b) is observed at an energy of 1202.8(2)\,keV and has been newly assigned to the strongly excited $K^{\pi}= 3^+$ neutron-quasiparticle band at 1033.3(5)\,keV \cite{Parek82}, see Table\,\ref{tab:posbands}. Previously this state had been populated in a $(d,d^{\prime})$ experiment\,\cite{Thomp75} but no spin and parity could be assigned.
\\

{\it 1375.0\,keV:} The $J^{\pi} = 6^+$ assignment to the state at 1375.0(6)\,keV proposed in Ref.\,\cite{Hoog96} by means of the $K^{\pi}={0^+_3}, J^{\pi}= 6^+ \rightarrow K^{\pi} = 0^+_1, J^{\pi} = 6^+$ decay observed in electron conversion, is confirmed by our $(p,t)$ study. Furthermore, it is also recognized as a rotational bandmember of the $K^{\pi} = 0^+_3$ band at 1090.3(1)\,keV, see Table\,\ref{tab:posbands}.
\\

{\it 1559.0\,keV:} The state at an energy of 1559.0(1)\,keV could not be assigned to any rotational band. Nonetheless, it has a comparably strong cross section of 4.033(11)\,$\mu$b and is, despite the general problem for all $6^+$ states, perfectly fitted by a positive-parity $L = 6$ transfer. As already discussed in Sec.\,\ref{sec:0+}, a state at an energy of 1558.87(5)\,keV with a spin-parity assignment of $J^{\pi}= (2^+)$ was observed, which was also an assigned bandmember of a tentative $K^{\pi}= 0^+$ band at 1525.86(8)\,keV\,\cite{Sing08}. In Sec.\,\ref{sec:0+} it has already been pointed out that this band assignment is very likely to be wrong. The main reasons are found to be the very different excitation cross sections as well as the contradicting spin-parity assignments, see Figs.\,\ref{fig:0+_angdist} and \ref{fig:others_angdist}. Nonetheless, due to its observation in the ${}^{240}$Np $\beta^-$ decay of the 7.22\,min $J^{\pi}= (1^+)$ isomer and its population in a neutron capture reaction with neutron energies of 2\,keV\,\cite{Sing08}, there is certain doubt that a $6^+$ state has been populated in this former studies. By now, it has to be assumed that two different levels have been observed.

\subsection{Negative-parity states}
\label{sec:negstat}

The different $K$ projections of the one-octupole phonon excitation have also been observed. The respective bandmembers are presented in Table\,\ref{tab:negbands} and Fig.\,\ref{fig:octu_angdist}. The new but tentative assignment of the $K^{\pi} = 3^-$ projection is supported by its derived MoI which is comparable to the other $K$ projections. Note that it is only based on two states observed for this band. Furthermore, despite the $K^{\pi} = 0^-$ projection the MoI are rather close to the one of the proposed $K^{\pi} = 0^+_2$ double-octupole phonon band\,\cite{Wang09, Jolo13, Spiek13a}.
\\

{\it $K^{\pi} = 2^-_1$:}  An angular distribution for the proposed $2^-$ bandhead could not be measured due to its unnatural parity, which resulted in a very small $(p,t)$ cross section. However, at a laboratory angle of 20$^{\circ}$, which is the expected peak of its angular distribution, a differential cross section of 0.3(1) $\mu$b/sr was measured. The energy of 1241.8(6)\,keV is very close to the adopted energy of 1240.8(3)\,keV\,\cite{Sing08}. If the $K^{\pi} = 2^-_1$ bandhead is assumed to be correct, then a $J^{\pi} = 5^-$ state is found at an energy of 1407.5(6)\,keV which fits into the rotational band, see Fig.\,\ref{fig:rot_bands} and Table\,\ref{tab:negbands}. It is, thus, proposed to recognize this rotational sequence as the $K^{\pi} = 2^-$ one-octupole phonon projection.
\\

{\it $K^{\pi} = (3^-)$:} A candidate for the $K^{\pi} = 3^-$ one-octupole phonon projection is newly proposed with its bandhead at an energy of 1550.3(6)\,keV. On top of it, a $J^{\pi} = 5^-$ state is observed at an energy of 1669.5(9)\,keV. As already the bandhead is weakly populated, the total cross section of this state is even smaller which might explain the differential cross section at an angle of 25$^{\circ}$ as this is completely off. Neglecting the differential cross section at this angle would result in a total cross section of roughly 0.5\,$\mu$b. Besides this deviation, the negative-parity $L = 5$ transfer matches the experimental angular distribution. The proposed $K= 3$ bandhead of Ref.\,\cite{Parek82} at 1675\,keV was also observed in $(d,d^{\prime})$ but no spin assignment had been possible\,\cite{Thomp75}. In the our $(p,t)$ experiment a state at an energy of 1674.1(4)\,keV with a total cross section of 1.86(2)\,$\mu$b is observed. A single-step positive-parity $L = 2$ transfer with a two-neutron transfer configuration of (3d$_{5/2}$)$^2$ matches the experimental distribution well, see Fig.\,\ref{fig:others_angdist}. Therefore, a $J^{\pi} = 3^-$ assignment should be dropped. Instead, a spin-parity assignment of $J^{\pi} = 2^+$ is favored. 
\\
We do already note that the most strongly excited $3^-$ state above the $K^{\pi} = 2^-, J^{\pi} = 3^-$ state is experimentally observed at 1973.5\,keV, see Fig.\,\ref{fig:crosssections}. The IBM calculations which will be discussed in Sec.\,\ref{sec:disc} expect the $K^{\pi} = 3^-$ bandhead at an excitation energy of 2013.5\,keV. Further experiments will be needed to identify the $K^{\pi} = 3^-$ one-octupole phonon band unambiguously.
\\

{\it Additional negative-parity states:} An additional $K^{\pi} = 0^-$ negative-parity rotational band is observed at 1540.1(1)\,keV with $R(E1)_{2^+_1/0^+_1} = 1.82(6)$\,\cite{Sing08}. In Ref.\,\cite{Parek82} its bandhead was proposed to be a quasiparticle excitation. Newly assigned are now tentative $J^{\pi}= 3^-$ (1588.0(6)\,keV) and $5^-$ (1686.2(4)\,keV) bandmembers, see Fig.\,\ref{fig:octu_angdist}. If the $K^{\pi} = 0^-$ band assignment is correct, a MoI of $3.80(6) \times ~ 10^6$\,MeV\,fm$^2$/c could be calculated, which is comparable to the MoI of the $K^{\pi} = 0^-_1$ rotational band. Above this rotational band, additional negative-parity states are excited in the $(p,t)$ reaction, see Table\,\ref{longtable}. A second $1^-$ state is found at an energy of 1807.4(2) keV which has also been recognized in earlier studies and is listed with a spin-parity assignment of ($1^-$, $2^+$). The current study favors a spin-parity assignment of $1^-$. A tentative $J^{\pi} = 5^-$ state is found at an energy of 1967.2(13)\,keV. If they are assumed to be the members of a $K= 1$ band, a MoI of 3.4 $\times ~ 10^6$ MeV~fm$^2$/c could be derived. Its $J^{\pi} = 3^-$ state is presumably not observed due to strongly excited $0^+$ and $4^+$ states at its expected energy. Above this very tentative $K= 1$ band, three additional states with a possible $3^-$ assignment are observed at energies of 1919.5(6)\,keV, 1973.5(1)\,keV and 2835.5(4)\,keV. Compared to the one-octupole phonon $1^-$ states, the states at 1540.1(1)\,keV and 1807.4(2)\,keV are rather strongly excited with total cross sections of 2.584(9)\,$\mu$b and 1.298(8)\,$\mu$b, respectively. Their $(p,t)$ strength is inverted with respect to the one-octupole phonon projections.
\\

\subsection{The identification of rotational bands}
\label{sec:moi}

\begin{figure}[th]
\centering
\includegraphics[width=0.95\linewidth]{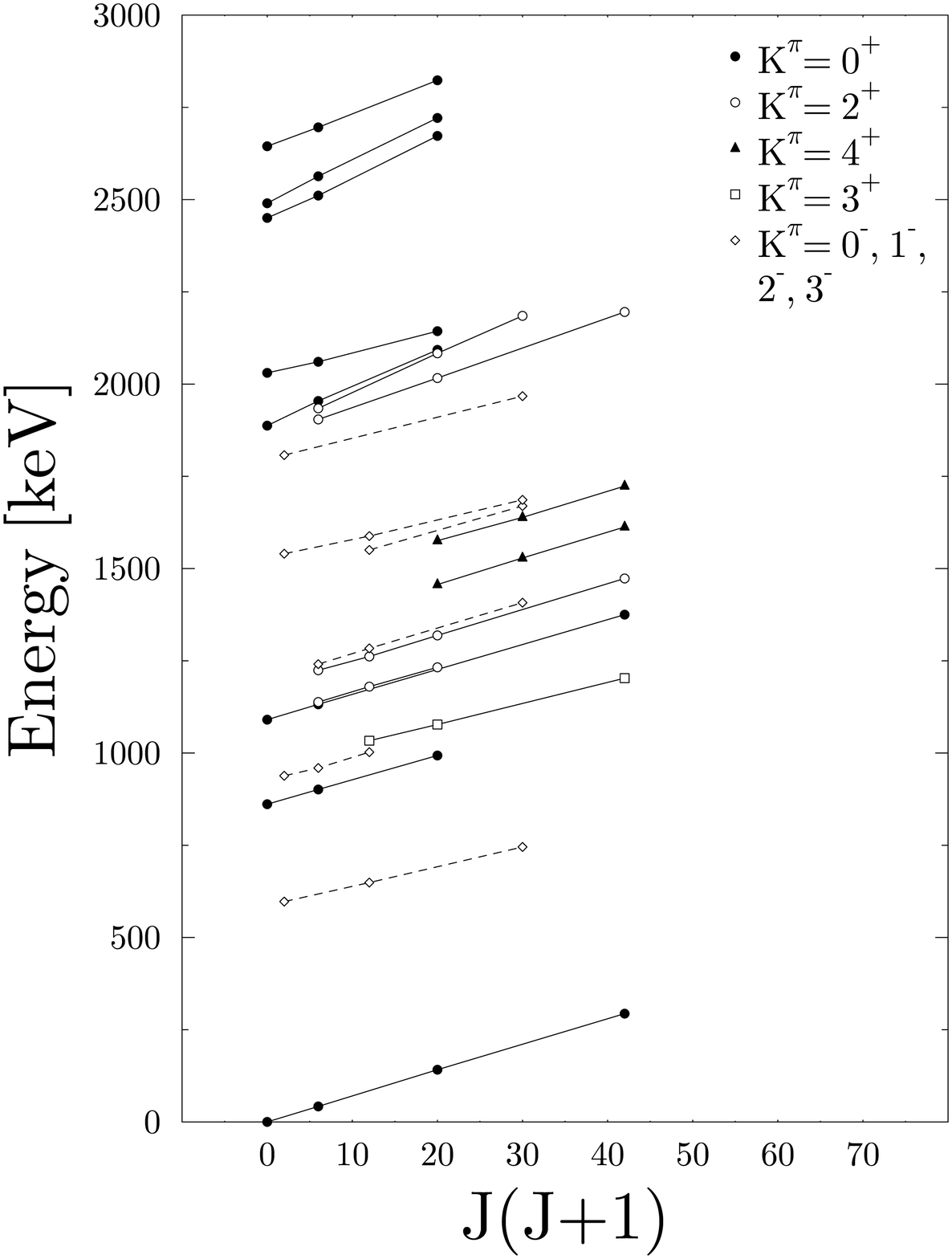}
\caption{\label{fig:rot_bands}Sequences of excited states, which might belong to a common rotational band. Positive-parity $K^{\pi} = 0^+$ (black circles), $K^{\pi} = 2^+$ (open circles), $K^{\pi} = 4^+$ (black triangles), and $K^{\pi} = 3^+$ (open squares) rotational bands are shown with solid lines. Negative-parity rotational bands (open diamonds) are shown with dashed lines.}
\end{figure}

\renewcommand{\thefootnote}{\fnsymbol{myfnsymbols}}
%\renewcommand*{\arraystretch}{1.8} 
%\setfnsymbol{myfnsymbols}
\begin{table*}[!t]
\caption{The positive-parity rotational bands observed in ${}^{240}$Pu by means of the $(p,t)$ reaction. Given are the $K$ projection and the energies of the bandmembers in keV. The last column presents the moment of inertia (MoI) derived for the respective band. ``$-$'' indicates that these states have not been observed in the present experiment.}
%\begin{adjustbox}{width=\linewidth}
\begin{ruledtabular}
\begin{tabularx}{\textwidth}{ccddddddd}
%ccddddddd
$\#$ & $K$ & \multicolumn{1}{c}{$0^+$} & \multicolumn{1}{c}{$2^+$} & \multicolumn{1}{c}{$3^+$} & \multicolumn{1}{c}{$4^+$} & \multicolumn{1}{c}{$5^+$} & \multicolumn{1}{c}{$6^+$} & \multicolumn{1}{c}{MoI} \\
 &  & & & & & & & \multicolumn{1}{c}{[10$^6$\,MeV\,fm$^2$/c$^2$]} \\
\hline
1 & $0^+$ & 0.0(1) & 42.0(1) &  & 141.5(1) &  & 293.5(6) & 2.77(2)\\
2 & $0^+$ & 861.2(1) & 901.1(1) &  & 993.2(4) &  & \multicolumn{1}{c}{-} & 2.93(2)\\
3 & $3^+$ &  &  & 1033.3(5) & 1077.2(1) & \multicolumn{1}{c}{-} & 1202.8(2)\footnotemark[1] &  3.47(4)\\
4 & $0^+$ & 1090.3(1) & 1131.9(1)\footnotemark[1] &  & \multicolumn{1}{c}{-} &  & 1375.0(6) & 2.836(12)\\
5 & $2^+$ &  & 1138.1(1) & 1179.9(4) & 1232.0(5) & \multicolumn{1}{c}{-} & \multicolumn{1}{c}{-} &  2.88(5)\\
6 & $2^+$ &  & 1224.3(2) & 1261.6(6) & 1318.7(1) & \multicolumn{1}{c}{-} & 1473.0(5) &  2.87(6)\\
7 & $4^+$ &  &  &  & 1456.5(1) & 1528.6(6) & 1612.6(2) &  2.735(4)\\
8 & $4^+$ &  &  &  & 1575.5(1) & 1638.6(6) & 1723.5(1) &  2.884(3)\\
9 & $0^+$ & 1887.3(1) & 1954.2(3) &  & 2092.7(1) &  & \multicolumn{1}{c}{-} & 1.851(9)\\
10 & $2^+$ &  & 1904.1(1) & \multicolumn{1}{c}{-} & 2016.3(2)\footnotemark[1] & \multicolumn{1}{c}{-} & 2195.4(6) &  2.402(10)\\
11 & $2^+$ &  & 1934.2(1) & \multicolumn{1}{c}{-} & 2083.4(1) & 2184.9(4) & \multicolumn{1}{c}{-} &  1.821(2)\\
12 & $0^+$ & 2030.4(1) & 2060.4(1) &  & 2143.4(3) &  & \multicolumn{1}{c}{-} &  3.58(2)\\
13 & $0^+$ & 2450.3(7) & 2510.9(1)\footnotemark[1] &  & 2672.6(2) &  & \multicolumn{1}{c}{-} &  1.80(2)\\
14 & $0^+$ & 2490.1(7) & 2563.2(4) &  & 2721.3(7) &  & \multicolumn{1}{c}{-} &  1.66(2)\\
15 & $0^+$ & 2644.6(2) & 2695.9(1)\footnotemark[1] &  & 2823.4(3) &  & \multicolumn{1}{c}{-} &  2.200(11) \\
\end{tabularx}
\end{ruledtabular}
\footnotetext[1]{Tentative placing because of larger cross section.}
\label{tab:posbands}
%\end{adjustbox}
\end{table*}

Having an $R_{4/2}$ ratio of 3.31 and a very collective $B(E2;2^+_1 \rightarrow 0^+_1)$ value of 287(11)\,W.u.\,\cite{ENSDF}, $^{240}$Pu is expected to feature rotational bands. In addition to the ground-state rotational band, several of such sequences of excited states are observed in $^{240}$Pu, see Fig.\,\ref{fig:rot_bands}. These can be described by the simple rotational formula:

\begin{equation}
E_{\mathrm{rot}} = \frac{{\hbar}^2}{2I} \left[ J(J+1) - K(K+1) \right] + E_K
\label{eq:roteq}
\end{equation}

Here, $I$ denotes the moment of inertia (MoI), $K$ the projection of the bandhead's total angular momentum onto the symmetry axis, $J$ the spin of the respective bandmember, and $E_K$ corresponds to the excitation energy of the bandhead. Thereby, a rotational band is unambiguously identified by the energy $E_K$ of its bandhead and its respective $K$ quantum number. Within a rotational band its members share the same MoI and only ``smooth'' variations of its value with increasing spin are observed, emphasized by the straight lines in Fig.\,\ref{fig:rot_bands}. A sequence of states has been accepted as a rotational band, if the DWBA yielded the spin-parity assignment in order to accept a given state as a rotational bandmember, if at least three bandmembers were identified, and if a decrease of the total cross section with increasing spin was observed, see also Fig.\,\ref{fig:crosssections} for the $K^{\pi} = 0^+_1$ rotational bandmembers. In the latter case, small deviations were excepted as multistep processes could alter the total cross section. In Table~\ref{tab:posbands}, these states are marked with ``a''. However, we want to stress that the criterium of a decreasing cross section with increasing spin is already violated by the uniformly strong population of the respective negative-parity $J^{\pi} = 3^-$ rotational bandmembers. Furthermore, the small $(p,t)$ cross sections of the unnatural-parity states point out that the reaction might be configuration and $L$ transfer sensitive.
\\ 
In general, mixing effects, centrifugal stretching as well as band crossing at higher spins can alter the moments of inertia\,\cite{Cast00}. However, at low spins and in the absence of mixing it might be expected that the MoI hints at the intrinsic structure of a rotational band since it is directly linked to the excitation's intrinsic shape. The moments of inertia derived for positive- and negative-parity rotational bands are given in Tables\,\ref{tab:posbands} and \ref{tab:negbands}, respectively.
\\

The largest MoI previously known is observed for the $K^{\pi} = 0^-$ projection of the one-octupole phonon excitation, i.e. $3.69(5) \cdot 10^6$\,MeV\,fm$^2$/c$^2$. The rotational band has already been studied up to highest spins\,\cite{Wied99, Wang09}. Below the 2QP energy, the $K^{\pi} = 0^+_2$ rotational band has the second largest MoI, i.e. $2.93(2) \cdot 10^6$\,MeV\,fm$^2$/c$^2$, see also Ref.\,\cite{Wang09}. As already mentioned, this rotational band was discussed to be of double-octupole phonon nature, see Refs.\,\cite{Wang09, Jolo13, Spiek13a}. In fact, it was speculated before whether these large moments of inertia could be attributed to double-octupole phonon excitations, see, {\it e.g.}, Refs.\,\cite{Levon09, Levon13, Levon15}. Certainly, the $K^{\pi} = 0^+_2$ MoI is larger than the the corresponding ground-state band as well as the $K^{\pi} = 0^+_3$ MoI, see Table\,\ref{tab:posbands}.
\\

Above an energy of 1.6\,MeV moments of inertia are observed which are well below the one of the ground-state band. Note that the excitation energies of these rotational bands are above the ``excitation'' gap which was observed for the $0^+$ states, see Fig.\,\ref{fig:crosssections}.

\begin{table*}[ht]
\begin{ruledtabular}
\caption{\label{tab:negbands}The negative-parity rotational bands observed in ${}^{240}$Pu by means of the $(p,t)$ reaction. The $K$ projection and the energies of the bandmembers in keV as well as the moments of inertia derived are given. ``$-$'' indicates that these states have not been observed in the present experiment. The $K = 3$ band is only tentatively assigned.}
\begin{tabularx}{\textwidth}{ccdddddd}
$\#$ & $K^{\pi}$ & 1^- & 2^- & 3^- & 4^- & 5^- & \multicolumn{1}{c}{MoI} \\
 &  & & & & & & \multicolumn{1}{c}{[10$^6$\,MeV\,fm$^2$/c$^2$]} \\
\hline
\multicolumn{8}{c}{One-octupole phonon rotational bands}\\
\hline
1 & $0^-$ & 597.2(4) &  & 648.8(4) &  & 745.3(8) & 3.69(5)\\
2 & $1^-$ & 938.2(3) & 959.4(5) & 1002.3(3) & - & - & 3.19(11)\\
3 & $2^-$ &  & - & 1283.6(2) & - & 1407.5(2) & 2.77(2)^{\mathrm{a}} \\
4 & $3^-$ &  &  & 1550.3(6) & - & 1669.5(9) & 2.93(3)^{\mathrm{b}}\\
\hline
\multicolumn{8}{c}{Negative-parity rotational bands}\\
\hline
5 & $(0^-,1^-)$ & 1540.1(1) & - & 1588.0(6) & - & 1686.2(4) & 3.80(6)
\end{tabularx}
\end{ruledtabular}
\footnotetext[1]{1240.8(3) keV assumed to be $K^{\pi} = 2^-$ bandhead.}
\footnotetext[2]{MoI calculated on the basis of the two possible band members observed.}
\end{table*}

\section{Discussion}
\label{sec:disc}

\subsection{The origin of excited $0^+$ states}

As already mentioned in the introduction, the structure of excited $0^+$ states in rare-earth nuclei and in the actinides has been controversially discussed for decades. A few specific structures were discussed in depth, which were pairing-isomeric states, quadrupole-type excitations, double-octupole phonon excitations, $\alpha$-cluster structures, as well as the 2QP pairing vibrational state and non-collective 2QP excitations. The 2QP excitations are expected above an energy, which is twice the energy of the neutron-pairing gap $\Delta_n$, i.e. $2 \Delta_n \approx 1090$\,keV in $^{240}$Pu. States below this excitation energy might, thus, be of rather pure collective nature. 
\\

\begin{figure}[!h]
\centering
\includegraphics[width=0.9\linewidth]{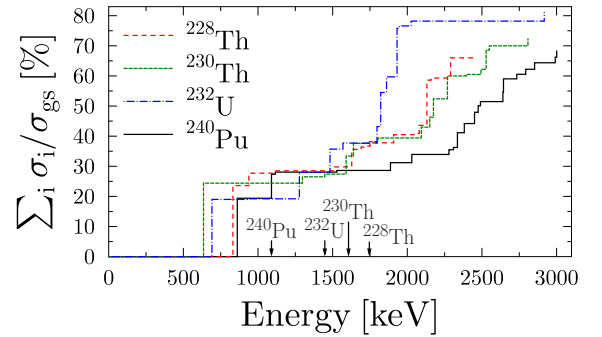}
\caption{\label{fig:reltrans_2QP}(Color online) Running relative transfer strength for $0^+$ states as a function of energy up to 3\,MeV. The 2QP energies are shown with solid arrows, respectively. The neutron-pairing energy $\Delta_{\mathrm{n}}$ was calculated from the odd-even mass differences~\cite{Audi03}. Except for ${}^{240}$Pu, the data are from Refs.~\cite{Wirt04, Levon09, Levon13, Levon15}.}
\end{figure}

In our previous study\,\cite{Spiek13a}, we have already identified the $0^+_2$ state at 861.2\,keV as the double-octupole phonon $0^+$ state in $^{240}$Pu, i.e. $N_{pf} = 2$ in the IBM. The identification was based on a stringent comparison of its known $\gamma$-decay properties, i.e. $E1/E2$ ratio and $E0$ transition to the ground state, as well as its $(p,t)$ cross section to the corresponding quantities predicted by the $spdf$ IBM. We have also shown that the $0^+_3$ state both experimentally and theoretically does exhibit very different $\gamma$-decay properties and that it corresponds to a quadrupole-type excitation, i.e. $sd$ state in the IBM but not the conventional $\beta$ vibration as defined by its decay properties. Furthermore, we emphasized that in all actinides rather strongly excited $0^+$ states ($\sigma_i/\sigma_{\mathrm{0^+_1}} \approx 5-10$\,$\%$) were observed at excitation energies of $E_x \approx 2\Delta_n$, see Fig.\,\ref{fig:reltrans_2QP}. In fact, these states might correspond to the 2QP pairing vibrational states. Above this energy and without any further experimentally measured observables, one might only speculate about the nature of the $0^+$ states and, therefore, a comparison to theory is needed.
\\

First, we will compare our experimental data to the predictions of the $spdf$ IBM. The following Hamiltonian was used,

\begin{align}
\label{eq:hamiltonian}
\hat{H}_{spdf}~=~&\epsilon_{d}\hat{n}_{d} + \epsilon_{p}\hat{n}_{p} + \epsilon_{f}\hat{n}_{f} - \kappa \hat{Q}_{spdf} \cdot \hat{Q}_{spdf} \nonumber
\\
&+ a_3 \left\lbrack \left(\hat{d}^{\dagger}\tilde{d}\right)^{(3)} \cdot \left(\hat{d}^{\dagger}\tilde{d}\right)^{(3)} \right\rbrack^{(0)},
\end{align}

and its parameters were determined to describe the low-spin members of the ground-state as well as $K^{\pi} = 0^-_1$ and $K^{\pi} = 0^+_3$  rotational bands. The boson energies are $\epsilon_d = 0.31$, $\epsilon_p = 2.1$ and $\epsilon_f = 0.68$\,MeV, while the quadrupole-coupling strength $\kappa$ and the strength $a_3$ of the $l=3$ term of the O(5) casimir operator are set to 0.015 and 0.014\,MeV, respectively. In the quadrupole operator $\hat{Q}_{spdf}$, $\chi_{sd}$ equals -$\sqrt{7}/2$ and $\chi_{pf}$ has been set to -1. As in previous IBM studies of the Pu isotopes\,\cite{Zamf03} and for consistency, the boson number $N_B = 15$ was counted with respect to the proposed neutron-subshell closure at 164.
\\

\begin{figure*}[t]
\centering
\includegraphics[width=0.75\linewidth]{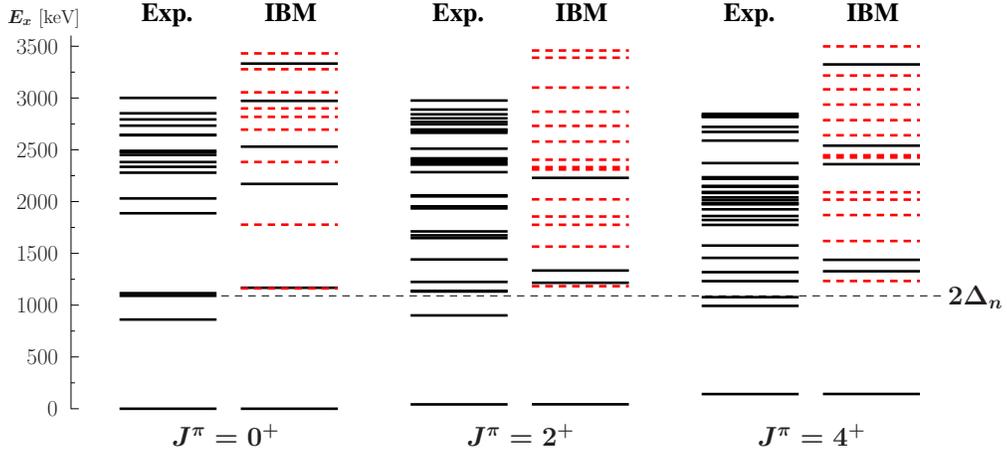}
\caption{\label{fig:exp_ibm_01}(Color online) Firmly assigned $0^+$, $2^+$, and $4^+$ states (Exp. with solid lines) as well as excited $0^+$, $2^+$, and $4^+$ states as predicted by the $spdf$ IBM. $sd$ states (solid lines) and $N_{pf} = 2$, i.e. double-dipole/octupole states (red dashed lines) are shown. In addition, the 2QP energy is depicted (black dashed line), i.e. $2\Delta_n$.}
\end{figure*}

A comparison of all experimentally firmly assigned $0^+$ states with the predicted excited states of the $spdf$ IBM and the corresponding collective structure is presented in Fig.\,\ref{fig:exp_ibm_01}. We should note immediately that one cannot expect to reproduce the complete experimental spectrum since certain states will be outside of the model space of the $spdf$ IBM. However, as already stressed in our previous publication\,\cite{Spiek13a}, two $0^+$ states are predicted close to the 2QP energy. One is the double-octupole phonon state (red dashed line), the other a quadrupole-type excitation (solid black line). The pronounced energy gap between these two $0^+$ states and the next excited $0^+$ state is observed experimentally and theoretically.
\\

In previous publications, it has been speculated whether one could use the moments of inertia derived for the rotational bands to discriminate between different underlying structures\,\cite{Levon09, Levon13, Levon15}. An inspection of Fig.\,\ref{fig:exp_ibm_01} reveals that the energy spacing between states of the $K^{\pi} = 0^+_2$ and $K^{\pi} = 0^+_3$ rotational bands increases with increasing spin. Indeed, this is due to different moments of inertia. The IBM predicts $5.3 \times 10^6$\,MeV\,fm$^2$/c for the $K^{\pi} = 0^+_2$ band and $2.4 \times 10^6$\,MeV\,fm$^2$/c for $K^{\pi} = 0^+_3$ band, respectively. Even though the MoI do not match the experimental values exactly, the MoI of the double-octupole phonon band is larger which is also observed in experiment, see Table\,\ref{tab:posbands}. The same holds for the $K = 2$ projections, i.e. $4.0 \times 10^6$\,MeV\,fm$^2$/c for the theoretical $N_{pf} = 2$ band at 1564\,keV and $2.6 \times 10^6$\,MeV\,fm$^2$/c for the predicted $sd$ band at 1334\,keV, respectively. The MoI of the $K^{\pi}=0^-_1$ one-octupole phonon band is $4.2 \times 10^6$\,MeV\,fm$^2$/c in the model. This value corresponds to an agreement in terms of excitation energies for the $1^-$ and $3^-$ which is as good as 1\,$\%$, and a deviation for the $5^-$ state which is less than 5\,$\%$. Note that within the IBM, bandmembers are identified in terms of $E2$ transitions between them and by an increase of $\langle \hat{n}_d \rangle$ with angular momentum\,\cite{Cast00}.
\\

To study the uniqueness of the MoI-based identification, we had a look at the rotational bands built upon the $0^+_3$, $0^+_4$, and $0^+_5$ IBM state, respectively. Their structure can be inferred from Fig.\,\ref{fig:exp_ibm_01}. The moments of inertia derived are: $1.6 \times 10^6$\,MeV\,fm$^2$/c for the $K^{\pi} = 0^+_3$, $2.0 \times 10^6$\,MeV\,fm$^2$/c for the $K^{\pi} = 0^+_4$, and $6.0 \times 10^6$\,MeV\,fm$^2$/c for the $K^{\pi} = 0^+_5$ band, respectively. Since the $0^+_4$ state has a double-octupole phonon structure and its MoI is smaller than the one of the $0^+_5$ state having an $sd$ structure, no unique identification in terms of MoI seems possible. However, it is very interesting that the first $K^{\pi} = 0^+$ band above the energy gap ($E_{\mathrm{x}} = 1887.3$\,keV) has a MoI of $1.851(9) \times 10^6$\,MeV\,fm$^2$/c. Therefore, it might indeed correspond to the  $K^{\pi} = 0^+_4$ rotational band predicted by the IBM. Still, without further information from experiments with complementary probes and without knowledge of the $\gamma$-decay behavior, the situation for the higher-lying states remains elusive.
\\

\begin{figure*}[!t]
\centering
\includegraphics[width=0.98\linewidth]{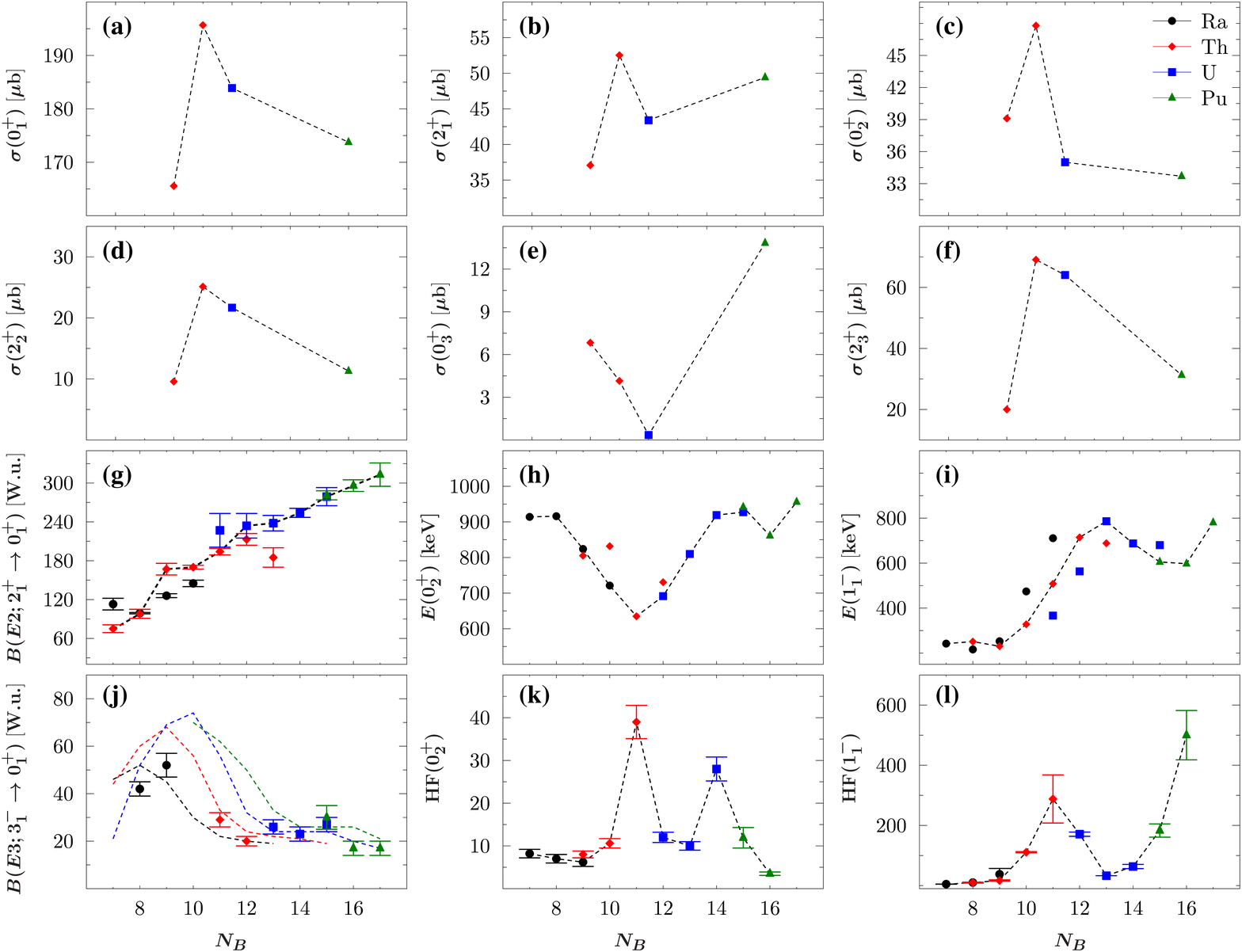}
\caption{\label{fig:ptobs_comp}(color online){\bf (a)-(f)} $(p,t)$ cross sections for the $0^+_1$, $2^+_1$, $0^+_2$, $2^+_2$, $0^+_3$, and $2^+_3$ states as a function of boson number $N_B$ for $^{228,230}$Th ($N_B = 10,11$)\,\cite{Levon09, Levon13}, $^{232}$U ($N_B = 12$)\,\cite{Levon15}, and $^{240}$Pu ($N_B = 16$). {\bf (g)-(l)} the $B(E2;2^+_1 \rightarrow 0^+_1)$ values, excitation energies $E(0^+_2)$ and $E(1^-_1)$, experimental (symbols) and theoretical $B(E3;3^-_1 \rightarrow 0^+_1)$ values (lines) from Refs.\,\cite{McG74a, Woll93a, Gaff13, Robl11a}, $\alpha$-decay hindrance factors HF$(0^+_2)$ and HF$(1^-_1)$. If not stated otherwise, the quantities given correspond to the adopted values of Ref.\,\cite{ENSDF}.}
\end{figure*}

The number of $0^+$ states and the $(p,t)$ cross sections have been discussed to be sensitive measures of general nuclear structure evolution, see, {\it e.g.}, Refs.\,\cite{Clark09, Cej10a, Zhang17a}. In fact, the strength pattern for the first three $0^+$ states in $^{240}$Pu is as expected from Ref.\,\cite{Clark09} since $\delta R_{4/2} < 0.1$. Predictions for the evolution of the corresponding matrix elements with boson number around the phase-transitional point towards stable quadrupole deformation were given in Ref.\,\cite{Zhang17a}. The experimental cross sections $\sigma(p,t)$ for the actinides studied are presented in Fig.\,\ref{fig:ptobs_comp}\,{\bf (a)}-{\bf (f)}. Discontinuities of the observables are expected at the critical point\,\cite{Zhang17a}. These are observed for all $(p,t)$ observables at $N_B = 11$, i.e. $^{230}$Th, despite $\sigma(0^+_3)$ where it might be observed at $N_B = 12$. One might, thus, claim that the phase transition from $U(5)$ to $SU(3)$, i.e. to stable quadrupole deformation following Ref.\,\cite{Zhang17a} takes place between $N_B = 10$ and $N_B = 12$ in the actinides. That is between nuclei with $R_{4/2}$ ratios of 3.23 and 3.29, which seems odd. We note that both the quadrupole and octupole phase transition have been studied theoretically in the lighter actinides, i.e. Ra and Th isotopes using the deformation-constrained EDF-IBM mapping approach\,\cite{Nomu13a,Nomu14a}. Here, the theoretically predicted quadrupole phase transition seems to occur at $N_B = 8$, i.e. $^{224}$Th ($R_{4/2} = 2.9$), compare Fig.\,1 of Ref.\,\cite{Nomu14a}. The same critical point is theoretically observed in the Ra isotopes.  These are also the actinide nuclei which are frequently discussed in terms of stable octupole deformation, see, {\it e.g.}, Ref.\,\cite{Butl96, Gaff13, Nomu14a}. Past the expected phase transition at $N_B = 8$, the evolution of the $B(E2;2^+_1 \rightarrow 0^+_1)$ strength might also change its slope, see Fig.\,\ref{fig:ptobs_comp}\,{\bf (g)}, as expected from Ref.\,\cite{Zhang17a}. Here, a kink in the evolution of the quadrupole equilibrium deformation $\beta_e$ is predicted near the $U(5) \rightarrow SU(3)$ transition. 
\\

\begin{table}[t] %add [H] placement to break table across pages
\centering
\caption{\label{tab:pt_ratios}The $(p,t)$ cross-section ratios $R\left( 5^{\circ} / 25^{\circ} \right)$ normalized to the corresponding ground-state ratio are given for the $0^+_2$ and $0^+_3$ states. The data for $^{228,230}$Th and $^{232}$U have been taken from Refs.\,\cite{Levon09,Levon13,Levon15}. Uncertainties are less than 10\,$\%$.}
\begin{ruledtabular}
\begin{tabular}{cccc}
Nucleus & $n$ & $E_x$\,[keV] & $R_{0^+_n/0^+_1} \left( 5^{\circ} / 25^{\circ} \right)$ \\
\hline
$^{228}$Th & 2 & 831.9 & 1.5 \\
 & 3 & 938.7 & 2.5 \\
$^{230}$Th & 2 & 635.1 & 2.1 \\
 & 3 & 1297.1 & 1.2 \\
$^{232}$U & 2 & 691.4 & 2.1 \\
 & 3 & 927.2 & 1.7 \\
$^{240}$Pu & 2 & 861.2 & 1.1 \\
& 3 & 1090.3 & 2.9
%Lines of table here ending with \\
\end{tabular}
\end{ruledtabular}
\end{table}

Interestingly, a clear minimum of $E(0^+_2)$ is observed at $N_B = 11$, see Fig.\,\ref{fig:ptobs_comp}\,{\bf (h)}. Usually, such a smooth evolution of the energy of an excited $0^+$ state has been interpreted as a signature of shape coexistence, see, {\it e.g.}, Ref.\,\cite{Heyd11a} for a recent review. We also pointed out in our previous publication\,\cite{Spiek13a} that close-lying excited $0^+$ states are observed in some actinides and could hint at the existence of double-octupole phonon states. To show that the structure of the $0^+_2$ states seems to be changing in the actinides, we have compiled the $(p,t)$ cross-section ratios $R\left( 5^{\circ} / 25^{\circ} \right) = \sigma(5^{\circ})/\sigma(25^{\circ})$ which are normalized to the corresponding ground-state ratio in Tab.\,\ref{tab:pt_ratios}. The double-octupole phonon candidates proposed in $^{228}$Th\,\cite{Levon13}, $^{232}$U\,\cite{Ard94a, Levon15} and $^{240}$Pu\,\cite{Wang09, Spiek13a} have ratios with $R < 2$. They show distinctly different $(p,t)$ angular distributions than the corresponding third- or second-excited $0^+$ state, compare, {\it e.g.}, Fig.\,\ref{fig:0+_angdist}.
\\
To shed some more light on the structure of the $0^+_2$ states and a possible connection to the negative-parity states, we compiled the adopted $\alpha$-decay hindrance factors HF$(0^+_2)$ and HF$(1^-_1)$, see Fig.\,\ref{fig:ptobs_comp}\,{\bf (k)} and {\bf (l)}. In fact, these are very sensitive measures of nuclear-structure changes between the mother and daugther nuclei as shown in, {\it e.g.}, Refs.\,\cite{vanDu00a, Bucu12a, Bucu13a}. As we can see, very prominent and localized maxima are observed at $N_B = 11$ and $N_B = 14$ for HF$(0^+_2)$. This indicates that the ground-state structures of $^{234}$U and $^{240}$Pu are very different from the structure of the $0^+_2$ states in $^{230}$Th and $^{236}$U, respectively. The same observation holds for HF$(1^-_1)$. Again, a pronounced maximum is observed at $N_B =11$ even though it is broader. The second maximum is observed for $N_B = 16$, i.e. $^{240}$Pu indicating that the $1^-$ of $^{240}$Pu and the ground state of $^{244}$Cm might have very different structures as the $\alpha$-decay is strongly hindered. We note that a favored transition in even-even nuclei would have an HF value of smaller than unity given the definition of the hindrance factor, compare Ref.\,\cite{vanDu00a}. 
\\

\begin{figure}[!t]
\centering
\includegraphics[width=0.75\linewidth]{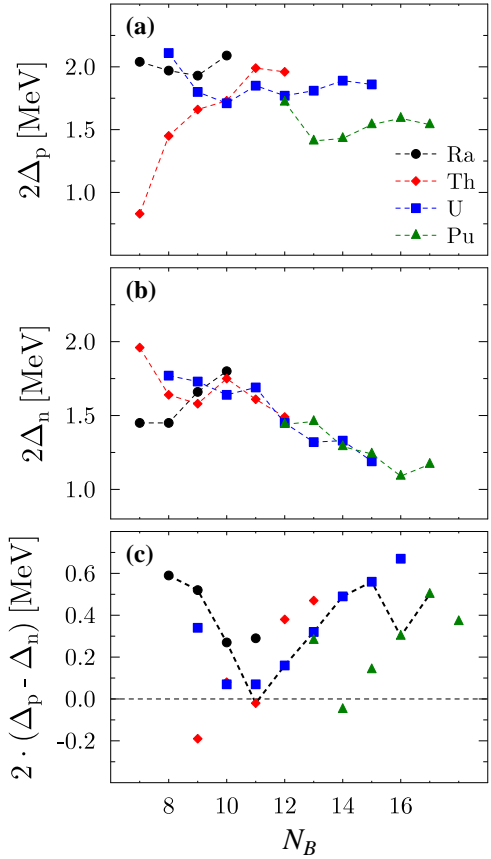}
\caption{\label{fig:pairing}(color online) The pairing gaps as a function of the total boson number $N_B$ in the even-even Ra, Th, U, and Pu isotopes as calculated from the odd-even mass differences\,\cite{Audi03}. {\bf (a)} Twice the energy of the proton-pairing gap $\Delta_p$. {\bf (b)} The same for the neutron-pairing gap $\Delta_n$. {\bf (c)} The energy difference between the two gaps. Note that $N_B$ has been shifted by one in {\bf (c)}. These values seem to follow the evolution of $E(0^+_2)$ as indicated by the dashed line, compare Fig.\,\ref{fig:ptobs_comp}\,{\bf (h)}.}
\end{figure}

P.\,van~Duppen and M.\,Huyse introduced a schematic two-level mixing in both the mother and daughter nucleus to explain a possible origin of large $\alpha$-decay hindrance factors in the neutron-deficient Po and Rn isotopes\,\cite{vanDu00a}. Here, they considered the normal proton configurations, i.e. $\pi$(2p-0h) and $\pi$(4p-0h), and additionally the corresponding proton two-particle-two-hole (2p-2h) intruder configurations, i.e. $\pi$(4p-2h) and $\pi$(6p-2h). For a certain mixing of the normal and intruder states in either both nuclei or just one of them, large HF values would be expected. Fig.\,\ref{fig:pairing} indicates that the same idea might apply to the neutron-rich side of the actinides. As mentioned earlier, 2p-2h excitations of a certain species are expected above twice the energy of the pairing gap, i.e. $2\Delta_{p,n}$. These energies calculated from the odd-even mass differences are shown in Fig.\,\ref{fig:pairing}\,{\bf (a)} and {\bf (b)}. The difference of both quantities shifted by one in $N_B$ is given in panel {\bf (c)} supporting the idea that resulting configurations will be observed in the next, i.e. $N_{\nu} + 1$ nucleus. The first time this difference is zero, coincides with the minimal energy of the $0^+_2$ state in $^{230}$Th, i.e. $N_B = 11$, see Fig.\,\ref{fig:ptobs_comp}\,{\bf (h)}. Strong mixing of the corresponding $\pi$(2p-2h) and $\nu$(2p-2h) states is, thus, expected in this nucleus and would result in a lowering of one of the resulting final states while the other would be pushed up in energy. Interestingly, a comparably large energy-separation of the $0^+_2$ and $0^+_3$ states was observed in $^{230}$Th supporting this scenario, see, {\it e.g.}, Refs.\,\cite{Spiek13a, Levon09}. We finally note that P.\,van~Duppen and M.\,Huyse showed that an admixture as small as 20\,$\%$ of this intruder configuration to the ground state of $^{196}$Po could explain the HF of 85 in the $\alpha$-decay of $^{200}$Rn to the excited $0^+$ state at 558\,keV in $^{196}$Po\,\cite{vanDu00a}.

A similar scenario could, therefore, also apply to the hindered $\alpha$-decay to the $1^-_1$ state. Recent calculations using covariant density functional theory (CDFT) suggest that $^{230}$Th is still octupole-deformed while $^{234}$U is octupole-soft in its ground state\,\cite{Agbe16a}. The 2p-2h excitations of $\Delta l = \Delta j = 3$ character, i.e. octupole excitations are more strongly admixed to the ground state of $^{230}$Th than to $^{234}$U. The explanation for the strong HF$(1^-_1)$ at $N_B = 11$ might, thus, also be found in terms of octupole-type admixtures to the ground states.

While for $^{230}$Th both HF$(0^+_2)$ and HF$(1^-_1)$ spike, the situation for $^{240}$Pu appears more complex. The HF$(0^+_2)$ spikes for the $^{240}$Pu to $^{236}$U $\alpha$-decay while HF$(1^-_1)$ spikes for the $^{244}$Cm to $^{240}$Pu $\alpha$-decay, respectively. It is tempting to interpret the observation in terms of the local minimum of $\Delta_n$ for $^{240}$Pu, see Fig.\,\ref{fig:pairing}\,{\bf (b)}. However, we consider a scenario founded on the octupole degree of freedom and on the ground-state structure of $^{240}$Pu to explain the experimental observations. $^{240}$Pu is located in the second octupole minimum, see also Fig.\,\ref{fig:ptobs_comp}\,{\bf (i)}. Even though $^{240}$Pu is considered octupole-soft in its ground state, the recent CDFT calculations indicate a gain in binding energy due to octupole deformation in contrast to $^{236}$U\,\cite{Agbe16a}. It might be this additional admixture to the ground state of $^{240}$Pu which hinders the $\alpha$-decay to the $0^+_2$ state of $^{236}$U. Indeed, if there is an octupole admixture to the ground state of $^{240}$Pu one might expect that the $\alpha$-decay to the $1^-_1$ state of $^{236}$U is less hindered. At the same time, the second octupole minimum is rather localized at $N = 146$\,\cite{Agbe16a}. Therefore, the ground state of $^{244}$Cm is not expected to show enhanced octupole correlations. We note that no functional considered in Ref.\,\cite{Agbe16a} predicts a gain in binding energy due to octupole deformation beyond $N = 146$. Consequently, it might be these missing correlations in the ground state of $^{244}$Cm which could explain the HF$(1^-_1)$ observed in the $\alpha$-decay to $^{240}$Pu and the evolution of the octupole correlations in general which might cause the observed hindrance factors. We note that this interpretation is in line with Ref.\,\cite{She91a} where HF's were discussed to be sensitive measures of reflection asymmetry in the Ra and Th isotopes. Unfortunately, no experimental data is yet available to calculate the polarization effect due to the odd particle close to $^{240}$Pu mentioned in Ref.\,\cite{She91a}. Still, the $\alpha$-decay of the odd-even nuclei might shed some additional light. 

The ground-state spins $J^{\pi}_{\mathrm{gs}}$ of the nuclei which we will consider in the vicinity of $^{240}$Pu are as follows: $7/2^-$ for $^{237}$Pu, $1/2^+$ for $^{239}$Pu, $5/2^+$ for $^{241}$Pu, $1/2^+$ for $^{241}$Cm, $5/2^+$ for $^{243}$Cm, and $7/2^+$ for $^{245}$Cm\,\cite{ENSDF}. The octupole-driving single-particle orbitals above $N = 126$ are 2g$_{9/2}$ and 1j$_{15/2}$\,\cite{Butl96}. However, due to the strongly upsloping $5/2^- \lbrack 503 \rbrack$ Nilsson orbital, the 2f$_{5/2}$ and 1i$_{11/2}$ octupole interaction will also contribute in $^{240}$Pu. First we consider the $\alpha$-decay of $^{241}$Cm to $^{237}$Pu. It is quite likely that the ground-state configuration of $^{237}$Pu is $7/2^- \lbrack 743 \rbrack$ while $^{241}$Cm might have $1/2^+ \lbrack 631 \rbrack$\,\cite{Ahm75a}. The first originates from the spherical 1j$_{15/2}$ orbital and the latter from the 3d$_{5/2}$ orbital. The ground state to ground state transition is hindered with an HF of 34. On the contrary, the decay to the $1/2^+$ state at 145.5\,keV is the most favored transition with an HF of about 2.6\,\cite{Ahm75a}. Here, we clearly see the influence of the configurations involved on the $\alpha$-decay hindrance factors. In the scenario of a parity-mixed state, i.e. $\Omega^+ \lbrack N n_z \Lambda \rbrack \otimes \Omega^- \lbrack N' n_z' \Lambda' \rbrack$ discussed in Ref.\,\cite{She91a}, and in the absence of pronounced reflection asymmetry the $\alpha$-decay between the states of the same parity will be favored, which is exactly what was observed above. We note that for the $\alpha$-decay of $^{237}$Pu to $^{233}$U the transitions to the low-lying $7/2^-$ at 320.8\,keV and 503.5\,keV are the most favored\,\cite{ENSDF}. 

We now want to combine the information on the possible Nilsson orbitals to arrive at a consistent picture for $^{240}$Pu. It seems rather clear that the ground-state configuration of $^{239}$Pu is $1/2^+ \lbrack 6 3 1 \rbrack$, i.e. 3d$_{5/2}$ while it is $5/2^+ \lbrack 6 2 2 \rbrack$, i.e. 1i$_{11/2}$ for $^{241}$Pu. Taking the newly adopted $\beta_2 = 0.29$ value for $^{240}$Pu\,\cite{Prity16a}, these assumptions appear legitimate when considering the odd-even Cm isotopes as well. Here the ground-state configurations would be the same as for $^{239,241}$Pu in $^{241,243}$Cm and, additionally, $7/2^+ \lbrack 6 2 4 \rbrack$, i.e. 2g$_{9/2}$ in $^{245}$Cm. We, thus, conclude that the dominant ground-state Nilsson configuration of $^{244}$Cm is $\{ 5/2^+ \lbrack 6 2 2 \rbrack \}^2$. Now let us assume that the $1^-$ state in $^{240}$Pu has the following Nilsson configuration $\{ 5/2^- \lbrack 5 0 3 \rbrack \}^{-1} \{ 5/2^+ \lbrack 6 2 2 \rbrack \}^1$, i.e. an octupole excitation from the 2f$_{5/2}$ orbital to the 1i$_{11/2}$ orbital which would give a $K^{\pi} = 0^-$ rotational band. For the reasons stated above, the $\alpha$-decay from $^{244}$Cm to this structure would be hindered since a neutron in the $\Omega^-$ orbital is involved. This $\alpha$-decay is experimentally hindered, see Fig.\,\ref{fig:ptobs_comp}\,{\bf (l)}. In contrast to the $0^+_2$ state of $^{230}$Th, the same state in $^{240}$Pu exhibits fast $E1$ transitions to the $K^{\pi} = 0^-_1$ rotational band, see Ref.\,\cite{Spiek13a}. Negative-parity orbitals must be involved. If the structure of the $0^+_2$ state contains the following octupole-type 2p-2h admixture $\{ 5/2^- \lbrack 5 0 3 \rbrack \}^{-2} \{ 5/2^+ \lbrack 6 2 2 \rbrack \}^2$, a less hindered decay compared to the decay to the $1^-$ state would be expected since the decay proceeds between neutrons which are in the same $\Omega^+$ orbital. A less hindered $\alpha$-decay to the $0^+_2$ state of $^{240}$Pu is observed in Fig.\,\ref{fig:ptobs_comp}\,{\bf (k)}. The $\alpha$-decay observables might, thus, further support the double-octupole interpretation of the $0^+_2$ in $^{240}$Pu. 
\\

% maybe not included?
%To support this hypothesis, we will have a closer look at the double-octupole $0^+$ state proposed in $^{232}$U ($0^+_3$ at 927\,keV), see, {\it e.g.}, Refs.\,\cite{Ard94a, Levon15}. In $^{232}$U both the $0^+_2$ at 691\,keV and the $0^+_3$ have comparably weakly HF's of 12 and 15, respectively\,\cite{Ard94a}. Depopulating $E1$ transitions are only observed for the $K^{\pi} = 0^+_3$ band. The ground state of $^{235}$Pu has $J^{\pi} = 5/2^+$ which is likely due to the $5/2^+ \lbrack 6 3 3 \rbrack$ Nilsson configuration. We, thus, assume the $\{ 5/2^+ \lbrack 6 3 3 \rbrack \}^2$ ground-state configuration in $^{236}$Pu, i.e. 2g$_{9/2}$. The ground state of $^{232}$U will have a large $\{ 5/2^- \lbrack 7 5 2 \rbrack \}^2$ admixture, i.e. 1j$_{15/2}$. We note that the $\{ 5/2^+ \lbrack 6 3 3 \rbrack \}^2$ configuration might be very close to the aforementioned in $^{232}$U\,\cite{Bohr75, ENSDF} and, thus, also strongly admixed to the ground state. Analogously to the case of $^{240}$Pu, the $\{ 5/2^- \lbrack 5 7 2 \rbrack \}^{-2} \{ 5/2^+ \lbrack 6 3 3 \rbrack \}^2$ configuration could, thus, explain the observations for the $0^+_3$ state. 

As mentioned earlier, another interpretation of the $K^{\pi} = 0^+_2$ band's structure based on $\alpha$-clustering was recently published and the enhanced $E1$ decay rates could be reproduced nicely\,\cite{Shneid15a}. In this work, the $0^+_2$ state corresponds to the lowest excitation in the mass-asymmetry coordinate $\xi$. We note that, possibly, this interpretation could also provide a qualitative understanding of the $\alpha$-decay hindrance factors in $^{240}$Pu. Before $\alpha$-decay of $^{244}$Cm to $^{240}$Pu will take place, the dinuclear system of $\alpha$-particle and $^{240}$Pu exists. The heavier fragment can be in different rotational states of its ground-state band. If reflection asymmetry is already present in the ground-state band, $\alpha$-decay to negative-parity states will be less hindered. In fact, this is observed in Fig.\,\ref{fig:ptobs_comp}\,{\bf (l)} for $\alpha$-decays leading to daughter nuclei with signs of reflection asymmetry in their spectra, {\it e.g.}, nuclei with $N_B \leq 9$. At the same time, this interpretation would indicate that reflection asymmetry in the ground state of $^{230}$Th and $^{240}$Pu is not pronounced since larger HF($1^-_1$) values are experimentally observed. Clearly, the minimum at $N_B = 13$ and 14 should be explained as well if this interpretation is correct. A hindered $\alpha$-decay to the $0^+_2$ state would be expected since it does not belong to the ground-state band. The fine structure observed does, however, indicate that the microscopic structure of the states as outlined above needs to be considered.

\begin{figure*}[!t]
\centering
\includegraphics[width=0.95\linewidth]{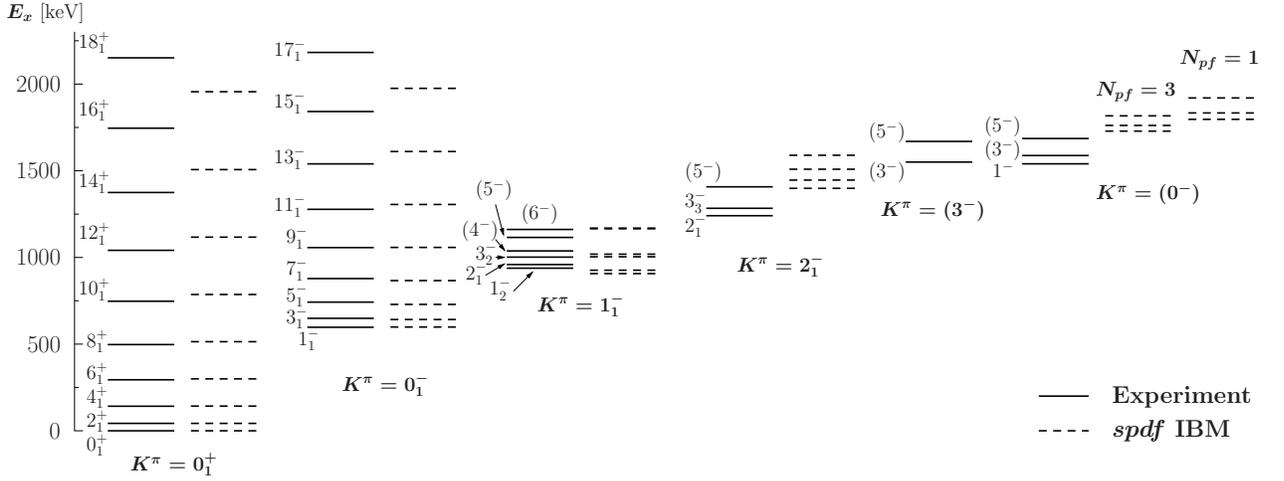}
\caption{\label{fig:exp_ibm_02}Comparison of the experimentally well-established ground-state and $K^{\pi} = 0^-_1$ rotational bands\,\cite{ENSDF} (solid lines) with the predictions of the $spdf$ IBM (dashed lines). In addition, possible negative-parity bands are compared to the corresponding bands of the IBM. Note, that the $5^-$ and $6^-$ state of the $K^{\pi} = 1^-_{\mathrm{IBM}}$ band are nearly degenerate. The $4^-$ state of the $K^{\pi} = 2^-$ band has not been observed experimentally. Unlike all other negative-parity bands in the figure, one of the $K^{\pi}=0^-_{\mathrm{IBM}}$ bands shown here has a $N_{pf} = 3$ structure (right part of the figure). The other negative-parity bands correspond to one-octupole phonon excitations. The one-octupole phonon $K = 3$ projection is predicted at an energy of 2013.5\,keV. The present $K^{\pi} = 3^-$ assignment to the state at 1550\,keV might, therefore, not be unambiguous.}
\end{figure*}

\subsection{Octupole excitations or $\alpha$-clustering?}

Already in the previous section and introduction, we have stressed that reflection asymmetry is the origin of low-lying negative-parity states. However, this reflection asymmetry could either be caused due to the octupole degree of freedom or due to mass-asymmetry, i.e. $\alpha$-clustering. Both interpretations can describe the low-lying excitation spectra and $\gamma$-decay observables observed for the states of the $K^{\pi} = 0^-_1$ and $0^+_2$ rotational bands, see, {\it e.g.}, Refs.\,\cite{Zamf03, Spiek13a, Shneid15a} and Fig.\,\ref{fig:exp_ibm_02}. Therefore, further observables are needed to distinguish between the two origins of reflection asymmetry in the actinides.
\\
\\

As mentioned earlier, a tentatively assigned $K^{\pi} = 0^-$ rotational band has been observed at 1540\,keV with a MoI of $3.80(6) \cdot 10^6$\,MeV$\mathrm{fm^2/c^2}$, see Table\,\ref{tab:negbands} and Fig.\,\ref{fig:exp_ibm_02}. The IBM predicts two additional $K^{\pi} = 0^-$ bands at 1729\,keV ($N_{pf} = 3$) and 1796\,keV ($N_{pf} = 1$) with MoI's of $6.3$ and $4.2 \cdot 10^6$\,MeV$\mathrm{fm^2/c^2}$, respectively. Based on its decay properties, the state at 1796\,keV corresponds to a one-octupole phonon excitation built upon the $K^{\pi} = 0^+_3$ rotational band of the $spdf$ IBM-1, i.e. 

\begin{align*}
&\frac{B(E1;1^-_4 \rightarrow 0^+_3)}{B(E1;1^-_1 \rightarrow 0^+_1)} = 0.78, \\ 
&\frac{B(E3;3^-_5 \rightarrow 0^+_3)}{B(E3;3^-_1 \rightarrow 0^+_1)} = 1.2 \\    
\end{align*}

In $^{152}$Sm, such $\gamma$-decays were indeed observed and interpreted as new signatures of shape coexistence\,\cite{Garrett09}. However, these $\gamma$-decays have not been observed so far for the $1^-$ states seen in our $(p,t)$ experiment at 1540\,keV and 1807\,keV\,\cite{ENSDF}. Interestingly, for a tentatively assigned $1^-$ at 1608\,keV, which was not observed in our experiment, the $\gamma$-decay to the $0^+_3$ state has been detected\,\cite{ENSDF}. As already indicated, the $1^-_3$ IBM state has a three-octupole phonon structure and, consequently, strong decays to the double-octupole phonon $K^{\pi} = 0^+_2$ rotational band are expected. The present $spdf$ IBM-1 calculcations predict:

\begin{align*}
&\frac{B(E1;1^-_3 \rightarrow 0^+_2)}{B(E1;1^-_1 \rightarrow 0^+_1)} = 2.7, \\ 
&\frac{B(E3;3^-_4 \rightarrow 0^+_2)}{B(E3;3^-_1 \rightarrow 0^+_1)} = 1.8 \\    
\end{align*} 

Similar to the $K^{\pi} = 0^+_2$ rotational band (Band D), the $K^{\pi} = 0^-_2$ rotational band (Band E) at 1302\,keV is built on the lowest excited state in the mass-asymmetry coordinate $\xi$ in the $\alpha$-cluster model of Ref.\,\cite{Shneid15a}. A MoI of  $3.6 \cdot 10^6$\,MeV$\mathrm{fm^2/c^2}$ is predicted, i.e. very close to the experimentally observed MoI of the $K^{\pi} = 0^-$ band at 1540\,keV. Compared to the $K^{\pi} = 0^-_1$ states, the states of this band predicted by the model of Ref.\,\cite{Shneid15a} are expected to decay as follows:

\begin{align*}
&\frac{B(E1;1^-_3 \rightarrow 0^+_2)}{B(E1;1^-_1 \rightarrow 0^+_1)} = 3.2, \\ 
&\frac{B(E3;3^-_3 \rightarrow 0^+_2)}{B(E3;3^-_1 \rightarrow 0^+_1)} = 1.9 \\    
\end{align*}
 
The $E1$ decay rates are the most promising signatures to distinguish between the different structures. Unfortunately, no experimental level lifetimes are available to quantify the reduced transition strengths of the $\gamma$-decay branches mentioned. However, to obtain a clearer picture we had a closer look at the partly known $\gamma$-decay behavior of these states which is shown in Table\,\ref{tab:be1be2}. Here, we compiled the $E1$ decays to the members of the ground-state rotational band and the $E2$ decays to the members of the $K^{\pi} = 0^-_1$ rotational band to calculate the $B(E1)/B(E2)$ ratios:

\begin{align*}
\frac{B(E1)}{B(E2)} = 0.767 \cdot \frac{E^5_{\gamma,E2} \cdot I_{\gamma,E1}}{E^3_{\gamma,E1} \cdot I_{\gamma,E2}} \ \mathrm{[10^{-6} \ fm^{-2}]}
\end{align*}

\begin{table}[t] %add [H] placement to break table across pages
\centering
\caption{\label{tab:be1be2}The experimental $B(E1)/B(E2)$ ratios (R$_{\mathrm{E1/E2}}$)\,\cite{ENSDF} in comparison to the predicted quantities of the $spdf$ IBM and the $\alpha$-cluster model of Ref.\,\cite{Shneid15a}.}
\begin{ruledtabular}
\begin{tabular}{ccccc}
$E_x$ & $J^{\pi}_{\mathrm{i}}$ & $J^{\pi}_{\mathrm{f},E2}$ & $J^{\pi}_{\mathrm{f},E1}$ & R$_{\mathrm{E1/E2}}$  \\
$\mathrm{[keV]}$ & & & & [10$^{-6}$~fm$^{-2}$]\\
\hline
\multicolumn{5}{c}{Experiment} \\
\hline
1540 & $1^-$ & $1^-_1$ & $0^+_1$ & 1.37(14) \\
& & $1^-_1$ & $2^+_1$ & 2.4(2) \\
& & $3^-_1$ & $0^+_1$ & 5.7(9) \\
& & $3^-_1$ & $2^+_1$ & 10(2) \\
1807 & $1^-$ & $1^-_1$ & $0^+_1$ & 0.04(3) \\
& & $1^-_1$ & $2^+_1$ & 0.17(6) \\
& & $3^-_1$ & $0^+_1$ & 0.09(6) \\
& & $3^-_1$ & $2^+_1$ & 0.34(12) \\
1608\footnote{not observed in the present $(p,t)$ experiment} & $1^-$ & $3^-_1$ & $0^+_1$ & 0.06(2) \\
& & $3^-_1$ & $0^+_3$ & 0.04(2) \\
\hline
\multicolumn{5}{c}{$spdf$ IBM-1} \\
\hline
1729 & $1^-$ & $1^-_1$ & $0^+_1$ & 0 \\
($N_{pf} = 3$)& & $1^-_1$ & $2^+_1$ & 0 \\
($n_p/n_f = 0.2$)& & $3^-_1$ & $0^+_1$ & 0 \\
& & $3^-_1$ & $2^+_1$ & 0 \\
& & $3^-_1$& $0^+_2$ & 13657 \\
& & $3^-_1$& $0^+_3$ & 5.57 \\
1796 & $1^-$ & $1^-_1$ & $0^+_1$ & 0.0009 \\
($N_{pf} = 1$)& & $1^-_1$ & $2^+_1$ & 0.004 \\
($n_p/n_f = 0.2$)& & $3^-_1$ & $0^+_1$ & 0.0005 \\
& & $3^-_1$ & $2^+_1$ & 0.002 \\
& & $3^-_1$& $0^+_2$ & 0.04 \\
& & $3^-_1$& $0^+_3$ & 0.14 \\
2238 & $1^-$ & $1^-_1$ & $0^+_1$ & 0.7 \\
($N_{pf} = 1$)& & $1^-_1$ & $2^+_1$ & 1.9 \\
($n_p/n_f = 4.3$)& & $3^-_1$ & $0^+_1$ & 0.3 \\
& & $3^-_1$ & $2^+_1$ & 0.9 \\
\hline
\multicolumn{5}{c}{$\alpha$-cluster model of Ref.\,\cite{Shneid15a} (Band E)} \\
\hline
1302 & $1^-$ & $1^-_1$ & $0^+_1$ & 0.005 \\
& & $1^-_1$ & $2^+_1$ & 0.02 \\
& & $3^-_1$ & $0^+_1$ & 0.002 \\
& & $3^-_1$ & $2^+_1$ & 0.007 \\
%Lines of table here ending with \\
\end{tabular}
\end{ruledtabular}
\end{table}

\noindent Up to now the $1^-$ state at 1540\,keV seems to be the only excited state which exhibits fast $E1$ decays to the ground-state band. None of the theoretically predicted states below 2\,MeV does show a similar $\gamma$-decay behavior. Therefore, this state seems to be out of the scope of the present calculations and its possible non-collective nature might also be the reason for its comparably large $(p,t)$ cross-section. Furthermore, the nature of the also strongly excited state at 1807\,keV remains unclear. An observation of the $\gamma$-decays to the $0^+_2$, $0^+_3$ state or other states might provide further clues about its structure. In contrast to the aforementioned enhanced $E1$ decay rates of the 1540\,keV $1^-$ state, the state at 1608\,keV does show more hindered $E1$-decay rates. Besides that the specific value differs by about three orders of magnitude, small R$_{E1/E2}$ ratios, i.e. hindered $E1$ decays to the ground-state band are also predicted by the IBM. The scenario mentioned above might, thus, be possible. We note that both models, i.e. the $\alpha$-cluster model of Ref.\,\cite{Shneid15a} and the $spdf$ IBM predict $B(E1;1^-_i \rightarrow 0^+_1)$ values of smaller than $0.03 \cdot 10^{-3} \ \mathrm{e^2 fm^2}$ for the $1^-_i$ states with $i > 1$ below 2\,MeV. Fast $E1$ transitions are observed above 2\,MeV for the present IBM calculations, see Table\,\ref{tab:be1be2} for one example of a $J^{\pi} = 1^-$ state at 2238\,keV. 
\\

It has to be mentioned that the intraband $B(E2)$ values of Band E in the $\alpha$-cluster model might be overestimated which could also explain the small predicted $B(E1)/B(E2)$ ratios. Presently, the intrinsic structure of the core, i.e. the single-particle structure is the same for all excitations in the mass-asymmetry coordinate. This is a good approximation for states built on the same excitation in this coordinate. However, it is expected that the single-particle structure of states built on the lowest and excited states of the mass-asymmetry coordinate will be different. A different single-particle structure could lead to a reduction of the $B(E2)$ strengths while the $E1$ transition strengths would be almost unaffected. 

\begin{figure}[!t]
\centering
\includegraphics[width=0.75\linewidth]{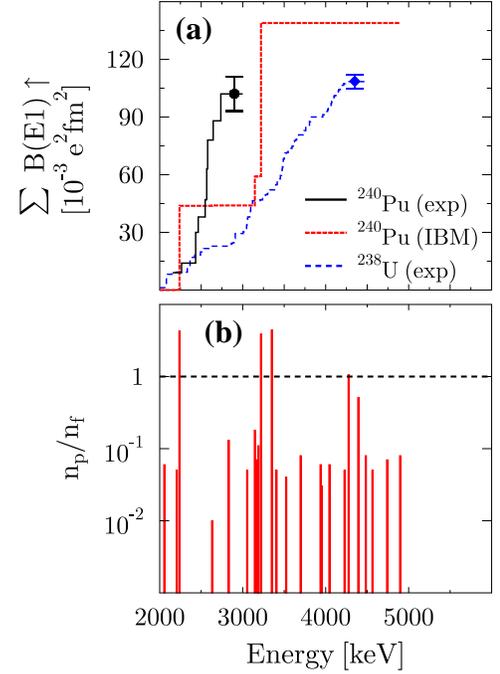}
\caption{\label{fig:e1strength}(color online) {\bf (a)} Possible running sum of experimental $B(E1;0^+_1 \rightarrow 1^-_i)$ strength in $^{240}$Pu (black solid line) between 2\,MeV and 4.5\,MeV\,\cite{Quit12a}. No parities were determined in Ref.\,\cite{Quit12a}. Thus, some of the $J = 1$ states will likely have a positive-parity assignment. For comparison, the experimentally determined running sum of $B(E1;0^+_1 \rightarrow 1^-_i)$ strength of firmly assigned $J^{\pi} = 1^-$ states in $^{238}$U up to 4.5\,MeV (blue dashed line) is also presented\,\cite{Hamm12a}. No clearly resolved strength is observed above this energy since the level density is too high. Missing strength of about $60 \cdot 10^{-3}$\,$\mathrm{e^2fm^2}$ up to the neutron-separation threshold was estimated. The $E1$ strength predicted by the IBM is shown as well (fine-dashed red line). {\bf (b)} the $n_p/n_f$ ratios as predicted by the $spdf$ IBM. States with $n_p/n_f > 1$ correspond to dominant $\alpha$-cluster $1^-$ states\,\cite{Spiek15a}.}
\end{figure}

$B(E1)$ strength above 2\,MeV has been measured by means of the nuclear resonance fluorescence (NRF) technique\,\cite{Quit12a}. Unfortunately, only strength between 2\,MeV and 3\,MeV has been reported which does not allow for a very stringent comparison to the predictions of the IBM. To get an idea of the missing strength, the experimental data on $^{238}$U was added to Fig.\,\ref{fig:e1strength} which shows a comparison of the experimental data to the IBM strength. As in our previous studies\,\cite{Levon13, Spiek13a, Spiek15a, Levon15}, the one-body $E1$ operator was used:

\begin{align}
\label{Eq:E1}
\hat{T}(E1) = &e_1 \lbrack \chi_{sp} (s^{\dagger} \tilde{p} + p^{\dagger} \tilde{s})^{(1)} + (p^{\dagger} \tilde{d} + d^{\dagger} \tilde{p})^{(1)} \nonumber \\
&+ \chi_{df} (d^{\dagger} \tilde{f} + f^{\dagger} \tilde{d})^{(1)} \rbrack
\end{align}

Its parameters were set to $e_1 = 0.018$\,$\mathrm{eb^{1/2}}$, $\chi_{sp} = 0.11$ and $\chi_{df} = -0.22$. These parameters simultaneously provide a good description of the $E1$ $\gamma$-decay ratios observed for the low-spin members of the $K^{\pi} = 0^-_1$ rotational band and a reasonable agreement in terms of the summed $B(E1;0^+_1 \rightarrow 1^-_i)$ strength observed for $^{238}$U and $^{240}$Pu. Besides the theoretical strength which is generated by an $E1$ excitation to the state at 3146\,keV with $n_p/n_f = 0.18$, all signficant $E1$ strength above 2\,MeV is caused by two states with $n_p/n_f > 1$ at 2238\,keV and 3221\,keV, respectively. In two of our previous publications, we have shown that states with $n_p/n_f > 1$ might be connected to $\alpha$-clustering in rare-earth\,\cite{Spiek15a} and $A < 100$ nuclei\,\cite{Spiek17a}. The $E1$ strength built upon the ground state in the actinides might as well be generated by the $p$-boson, i.e. $\alpha$-clustering. Of course, the strength predicted by the IBM is far less fragmented than the experimental strength, see Fig.\,\ref{fig:e1strength}\,{\bf (a)}. Still, the IBM does predict several $1^-$ states, compare Fig.\,\ref{fig:e1strength}\,{\bf (b)}. Obviously, the different configurations are far more mixed than anticipated by the present calculations.
\\

\begin{figure}[t]
\centering
\includegraphics[width=0.75\linewidth]{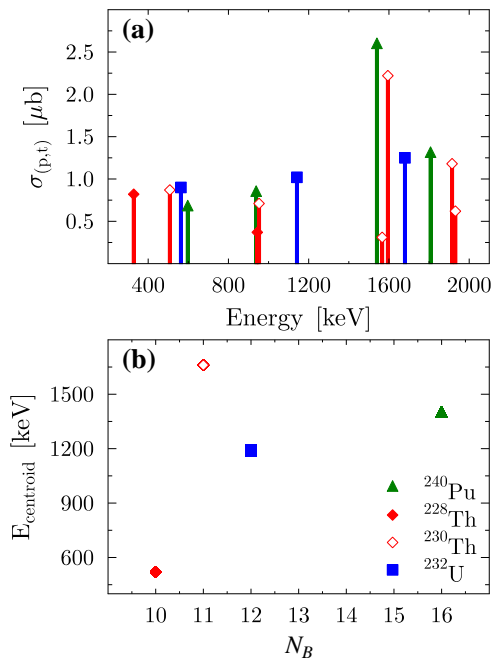}
\caption{\label{fig:1-_pt}(color online) {\bf (a)} $(p,t)$ cross sections $\sigma_{(p,t)}$ depicted as a function of excitation energy for the $J^{\pi} = 1^-$ states observed in $^{228,230}$Th\,\cite{Levon09, Levon13}, $^{232}$U\,\cite{Levon15} and $^{240}$Pu. {\bf (b)} centroid energies calculated from the excitation energies and $\sigma_{(p,t)}$ of the $J^{\pi} = 1^-$ states as a function of the boson number $N_B$.}
\end{figure}

We may conclude that enhanced $E1$ transitions between the ground state and any $1^-$ state are triggered by the $(s^{\dagger} \tilde{p} + p^{\dagger} \tilde{s})^{(1)}$ part of the one-body $E1$ operator. The possible shortcomings of the $spdf$ IBM to describe $R_{E1/E2}$ below 2\,MeV might, thus, have two reasons. One reason could be that the $p$-boson admixture, i.e. $\alpha$-cluster admixture to the low-lying $1^-$ states is currently underestimated. The dominant $p$-boson state at 2238\,keV did indeed exhibit enhanced $E1$ decay rates, see Table\,\ref{tab:be1be2}. Interestingly, the $R_{E1/E2}$ ratios predicted by the $\alpha$-cluster model of Ref.\,\cite{Shneid15a} might also appear too low. Furthermore, it might be necessary to consider higher-order terms for the $E1$ operator. By definition, no enhanced $E1$ transitions to the ground state are presently expected for states with little $p$-boson admixture or pronounced multiphonon structure, {\it e.g.}, a three-octupole phonon state. We note that the dipole term mentioned in Refs.\,\cite{Zamf03, Levon13, Spiek13a, Levon15} which introduces an admixture of negative-parity bosons to the ground-state band does not alter this statement.
\\

As already stressed above the $(p,t)$ cross section $\sigma_{(p,t)}$ of the 1540\,keV state is larger by a factor of about five compared to the $K^{\pi} = 0^-_1, J^{\pi} = 1^-_1$ state. The $(p,t)$ cross sections of all $1^-$ states observed in the actinides are shown in Fig.\,\ref{fig:1-_pt}\,{\bf (a)}. In addition, the centroid energies have been calculated, see Fig.\,\ref{fig:1-_pt}\,{\bf (b)}. At least four observations are interesting regarding the previous discussion. First of all, $\sigma_{(p,t)}$ for the $J^{\pi} = 1^-_1$ states is almost constant from Th to Pu. Second, $\sigma(1^-_2)_{(p,t)}$ might be larger than or equal to $\sigma(1^-_1)_{(p,t)}$ in $^{232}$U and $^{240}$Pu while the strength pattern is inverted in $^{228,230}$Th. Third, significant cross sections to higher-lying excited $1^-$ states at about 1.6\,MeV to 1.8\,MeV are observed in $^{230}$Th\,\cite{Levon09}, $^{232}$U\,\cite{Levon15} and $^{240}$Pu but not in $^{228}$Th\,\cite{Levon13}. And the fourth point, a discontinuity is once again observed at $N_B = 11$, see Fig.\,\ref{fig:1-_pt}\,{\bf (b)}. Unfortunately, no $\gamma$ transitions of the 1594\,keV state in $^{230}$Th have been measured up to now\,\cite{ENSDF}. The present data and the similarity to $^{240}$Pu do, however, suggest that this $1^-$ state might also show enhanced $E1$-decay rates.
\\
\\ 

\section{Conclusion}

A high-resolution $(p,t)$ experiment using the Q3D spectrograph was performed to study low-spin states in $^{240}$Pu up to an excitation energy of 3\,MeV. In total 209 excited states were identified and many of these were seen for the first time. To assign spin and parity to the states, angular distributions were measured and compared to the predictions of coupled-channels DWBA calculations. Several rotational bands built upon the low-lying bandheads excited in our experiment were also identified and their moments of inertia could be calculated.
\\
\\

In this publication we have discussed the origin of $J^{\pi} = 0^+$ and negative-parity states in detail using presently available experimental data on these states in $^{240}$Pu. As in our previous work\,\cite{Spiek13a}, we have pointed out that considering negative-parity single-particle states being admixed to the $K^{\pi} = 0^+_2$ band is crucial to understand the experimental observables. To clarify whether the octupole degree of freedom or $\alpha$-clustering are causing the enhanced $E1$ decays, we also took a closer look at the $\alpha$-decay hindrance factors measured for the $0^+_2$ and $1^-_1$ states in the actinides. However, both mechanisms provide reasonable explanations of the HF's. These observables also emphasize the importance to further understand the evolution of negative-parity single-particle states in the actinides. We attempted to connect the $J^{\pi} = 1^-$ states predicted by the $spdf$ IBM-1 and the $\alpha$-cluster model of Ref.\,\cite{Shneid15a} to experimentally observed states. Besides a possible one-octupole phonon excitation built on the $0^+_3$ state, no clear structure could be identified. Still, the $1^-$ state at 1540\,keV sticks out since it is the only state above 1\,MeV and below 2\,MeV which decays via enhanced $E1$ transitions to the ground-state band and which is comparably strongly excited in the present $(p,t)$ experiment. The latter suggests an enhanced pairing character compared to the other $1^-$ states which needs to be explained. We note that a small $\log ft$ value of 6.0 was reported for this $1^-$ state in the $\beta^-$-decay of 7.22\,min $^{240}$Np$^m$ parent state\,\cite{Hseuh81}. A value of 6.3 was observed for the transition to the $K^{\pi} = 0^-_1, J^{\pi} = 1^-$ state and the authors argued that this might hint either at a $7/2^+ \lbrack 6 2 4 \rbrack$ or a $7/2^- \lbrack 7 4 3 \rbrack$ admixture to the parent state, i.e. an admixture of the relevant $\Delta j, \Delta l = 3$ orbitals since the comparably small $\log ft$ value suggests a one-particle transition\,\cite{Hseuh81}.  
\\
\\

\begin{table}[t] %add [H] placement to break table across pages
\centering
\caption{\label{tab:be1be2_final}Excitation energies and experimental $B(E1)/B(E2)$ ratios (R$_{\mathrm{E1/E2}}$)\,\cite{ENSDF} of the possible double-octupole phonon or $\alpha$-cluster $K^{\pi} = 0^+, J^{\pi} = 0^+$ states in the actinides. The $0^+$ state given corresponds to the $n^{\mathrm{th}}$ $0^+$ state in the nucleus, respectively.}
\begin{ruledtabular}
\begin{tabular}{cccccc}
Nucleus & $n$ & $E_x$ & $J^{\pi}_{f,E1}$ & $J^{\pi}_{f,E2}$& $R_{E1/E2}$\\
  & & $\mathrm{[keV]}$ & & & [$10^{-6} \mathrm{fm^{-2}}$] \\
\hline
$^{224}$Ra  & 2 & 916.4 & $1^-_1$& $2^+_1$ & $\approx 0.2$ \\
$^{226}$Ra & 2 & 824.6 & $1^-_1$& - & -\footnote{no $E2$ transition observed.} \\
$^{228}$Ra & 2 & 721.2 & $1^-_1$ & $2^+_1$ & 2.7(4) \\
$^{226}$Th & 2 & 805.2 &  $1^-_1$& - & -$^{\mathrm{a}}$ \\
$^{228}$Th & 2 & 831.9& $1^-_1$ & $2^+_1$ & 5.1(4) \\
$^{230}$Th & (3) & 1297.1 & $1^-_1$ & $2^+_1$ & 0.71(4) \\
$^{232}$Th & (3) & 1078.6 & $1^-_1$ & $2^+_1$ & -\footnote{no $\gamma$-intensities measured.}\\
$^{232}$U & 3 & 927.3& $1^-_1$ & $2^+_1$ & 44(7) \\
$^{234}$U & 3 & 1044.5 &$1^-_1$ & $2^+_1$ & 3.9(3)\\
$^{238}$U & 2 & 927.2& - & $2^+_1$ & -\footnote{assignment based on $R_{E1/E2}$ of $J^{\pi} = 2^+$ bandmember.}\\
$^{238}$Pu & 2 & 941.5& $1^-_1$ & $2^+_1$ & $\leq 0.5$\\
$^{240}$Pu & 2 & 861.2& $1^-_1$ & $2^+_1$ & 13.7(3)
%Lines of table here ending with \\
\end{tabular}
\end{ruledtabular}
\end{table}

We have also shown that the $0^+_2$ states of $^{230}$Th and $^{240}$Pu exhibit different and distinct structures, i.e. at least two configurations exist which mix with each other in the actinides. The study of the pairing gaps suggests that both proton- and neutron-pairing states need to be considered. We propose that the $0^+_2$ state in $^{230}$Th is caused solely by the mixing of these two pairing states. In contrast to $^{240}$Pu, no enhanced $E1$ transitions are observed from this state. Therefore, we once again emphasize that the identification of enhanced $E1$ transitions from excited states is an important observable to distinguish between different underlying structures. Without these observables from $\gamma$-ray spectroscopy experiments, the structure of higher-lying $0^+$ states remains elusive. We have shown that identifying certain structures by means of the moments of inertia might be misleading. Still, based on a comparison to the present IBM calculations the $0^+_4$ state at 1887\,keV might as well have a double-octupole phonon structure. Finally, we have compiled all $0^+$ states in the actinides which will either have a double-octupole phonon or $\alpha$-cluster structure in Table\,\ref{tab:be1be2_final}. The fact that the $0^+_3$ states are the candidates in the $N = 140$ and $N = 142$ Th and U isotones supports the idea that another configuration drops drastically in energy at these neutron numbers. The deformed subshell closure proposed at $N \sim 142$ might be important to understand this ``intruder'' configuration\,\cite{Bucu13a}. In addition, we proposed that for the chosen kinematics of the present $(p,t)$ experiments the cross-section ratio $R\left( 5^{\circ} / 25^{\circ} \right)$ could be sensitive to the underlying structure of the low-lying $0^+$ states, see Tab.\,\ref{tab:pt_ratios}. To test this observable, $(p,t)$ experiments should be performed to study $0^+$ states in the nuclei listed in Tab.\,\ref{tab:be1be2_final}, {\it e.g.}, $^{226}$Ra$(p,t){}^{224}$Ra, $^{238}$U$(p,t){}^{236}$U, $^{240}$Pu$(p,t){}^{238}$Pu and $^{244}$Pu$(p,t){}^{242}$Pu.
\\
\\

We hope that our studies will trigger further investigations of $K^{\pi} = 0^+$ bands and negative-parity states in the actinides. Major goals should be to measure the $B(E1; 0^+_1 \rightarrow 1^-_i)$ strength below 2\,MeV in NRF experiments, to determine the $\gamma$-decay behavior of the low-spin states in, {\it e.g.}, $(n,\gamma)$ reactions, and, since the proton-pairing configuration might also be important, to perform two-proton-transfer experiments in the actinides, {\it e.g.}, $^{230}$Th$(^{16}\mathrm{O},{}^{14}\mathrm{C}){}^{228}$Ra or $^{234}$U$(^{16}\mathrm{O},{}^{14}\mathrm{C}){}^{232}$Th. In addition, the $E3$-transition rates from the low-lying $K^{\pi} = 0^+$ to the $K^{\pi} = 0^-_1$ rotational band should be determined as it has been done for the case of $^{148}$Nd\,\cite{Ibbot97a} and which might ultimately support the double-octupole phonon interpretation. 

\appendix*
\section{Exclusion of significant target contaminants}

\begin{figure*}[!t]
\centering
\includegraphics[width=0.9\linewidth]{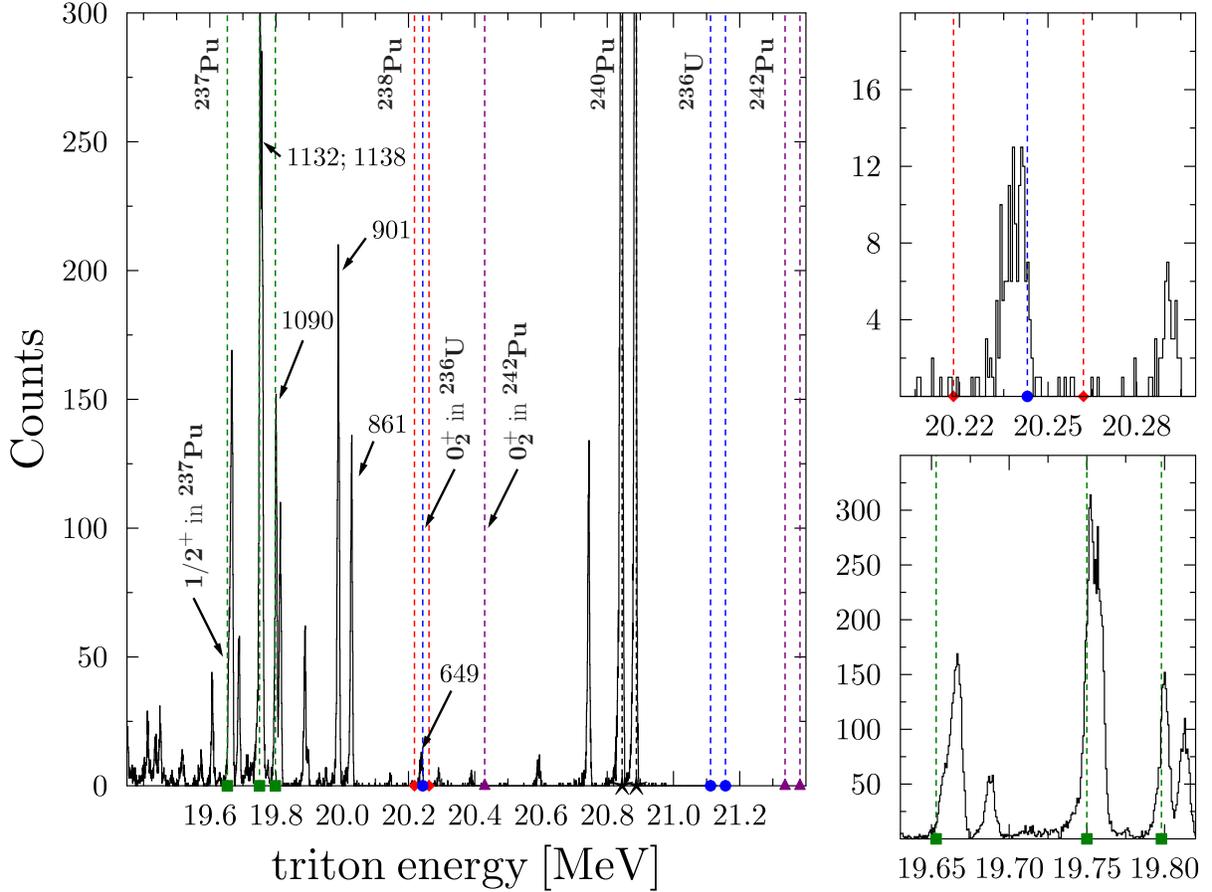}
\caption{\label{fig:contaminants}$^{242}$Pu$(p,t){}^{240}$Pu spectrum detected at 10$^{\circ}$ as a function of the residual triton energy. Some prominent excited states of $^{240}$Pu are marked with their excitation energy given in keV. In addition, the expected triton energies for the ground as well as first excited state of $^{237}$Pu (green), $^{238}$Pu (red) , $^{242}$Pu (purple), and $^{236}$U (blue) are marked with dashed lines and symbols, respectively. The first $J^{\pi}=1/2^+$ state of $^{237}$Pu has been added as well since it would be excited via a $l=0$ transfer in the $^{239}$Pu$(p,t){}^{237}$Pu reaction. In addition, the expected triton energies for the $J^{\pi}=0_2^+$ state in $^{236}$U and $^{242}$Pu are highlighted. The spectra in the right panels are zoomed into the regions of interest to identify or exclude possible contaminants.}
\end{figure*}

According to the information given by the Oak Ridge National Laboratory our $^{242}$Pu target was enriched to 99.93$\%$, i.e. leaving 0.07$\%$ for possible contaminations from other Pu isotopes. $^{242}$Pu decays with a half-live of $3.75(33) \cdot 10^5$ years to $^{238}$U\,\cite{ENSDF}. Our target was newly produced in 2009 and the experiment was conducted in November 2011, i.e. 99.999$\%$ of the initial target nuclei were still present.

As seen in Fig.\,\ref{fig:contaminants}, no significant contamination from tritons originating from the $^{238}$U$(p,t){}^{236}$U reaction was observed in the focal plane. In fact, neither the ground state nor the first $2^+$ state could be detected in the focal plane with the chosen magnetic settings. Note that also no signs of the first excited $0^+$ state have been seen, marked with ``$0_2^+$ in $^{236}$U'' in Fig.\,\ref{fig:contaminants}. The angular distribution of the $3_1^-$ state at 649\,keV could be nicely described, see Fig.\,\ref{fig:octu_angdist}. No additional contribution was needed.
\\
\\

Similar arguments hold for possible tritons resulting from the $^{244}$Pu$(p,t){}^{242}$Pu and $^{240}$Pu$(p,t){}^{238}$Pu reactions. No events were detected at the expected energy of the first excited $0^+$ state of $^{242}$Pu, marked with ``$0_2^+$ in $^{242}$Pu'' in Fig.\,\ref{fig:contaminants}, and almost no events were detected at the expected energies of the ground and $2_1^+$ state of $^{238}$Pu.
\\
\\

The situation for possible contaminations stemming from the $(p,t)$ reaction on $^{239}$Pu is different. The ground state of $^{239}$Pu is $J^{\pi}= 1/2^+$ and $J^{\pi} = 7/2^-$ in $^{237}$Pu. The first excited state of $^{237}$Pu is a $9/2^-$ state. The expected $1/2^+$ ($^{239}$Pu)$\ \rightarrow \ 7/2^-$ ($^{237}$Pu) and $1/2^+$ ($^{239}$Pu)$\ \rightarrow \ 9/2^-$ ($^{237}$Pu) triton energies coincide with the tails of the strongly excited $J^{\pi} = 0^+$ and $2^+$ states at 1090\,keV and 1138\,keV in $^{240}$Pu, respectively. As we have shown in this publication, the measured angular distributions clearly resemble $l=0$ and $l=2$ transfers for these states, compare Figs.\,\ref{fig:0+_angdist} and \ref{fig:others_angdist}. A ground state to ground state transfer ($^{239}$Pu$\ \rightarrow \ ^{237}$Pu) would correspond to a $l=3$ angular distribution while the other transfer would correspond to a $l=5$ angular distribution (parity changes). Both are not detected leaving little evidence for a significant $^{239}$Pu contamination in our target. To strengthen this point we also had a closer look at a possible $l=0$ transfer leading to the first excited $1/2^+$ state at 146\,keV in $^{237}$Pu. Also this state would be located in the tail of a rather strongly excited $J^{\pi}=4^+$ state at 1232\,keV for which we observed a $l=4$ rather than a $l=0$ angular distribution, compare Fig.\,\ref{fig:others_angdist}.
\\
\\

In conclusion, we have no reason to believe that there are significant contaminants present in our $^{242}$Pu target which could corrupt our results.

\begin{acknowledgments}
We gratefully acknowledge the help of the accelerator staff at MLL in Munich. Furthermore, we thank A.~Levon for discussing his experimental results. We also want to thank J.~Jolie as well as R.V.~Jolos for helpful discussions. T.M.S. acknowledges the  support from the  Russian Government
Subsidy Program of the Competitive Growth of Kazan Federal University. This work was supported in part by the Deutsche Forschungsgemeinschaft under contracts ZI 510/7-1, 436 RUM 17/1/07 and KR-2326/4-1 as well as by the Romanian UEFISCDI Project No. PN-II-ID-PCE-2011-3-0140.
\end{acknowledgments}

\bibliography{240Pu_prc}

\providecommand{\noopsort}[1]{}\providecommand{\singleletter}[1]{#1}%
\begin{thebibliography}{74}%
\makeatletter
\providecommand \@ifxundefined [1]{%
 \@ifx{#1\undefined}
}%
\providecommand \@ifnum [1]{%
 \ifnum #1\expandafter \@firstoftwo
 \else \expandafter \@secondoftwo
 \fi
}%
\providecommand \@ifx [1]{%
 \ifx #1\expandafter \@firstoftwo
 \else \expandafter \@secondoftwo
 \fi
}%
\providecommand \natexlab [1]{#1}%
\providecommand \enquote  [1]{``#1''}%
\providecommand \bibnamefont  [1]{#1}%
\providecommand \bibfnamefont [1]{#1}%
\providecommand \citenamefont [1]{#1}%
\providecommand \href@noop [0]{\@secondoftwo}%
\providecommand \href [0]{\begingroup \@sanitize@url \@href}%
\providecommand \@href[1]{\@@startlink{#1}\@@href}%
\providecommand \@@href[1]{\endgroup#1\@@endlink}%
\providecommand \@sanitize@url [0]{\catcode `\\12\catcode `\$12\catcode
  `\&12\catcode `\#12\catcode `\^12\catcode `\_12\catcode `\%12\relax}%
\providecommand \@@startlink[1]{}%
\providecommand \@@endlink[0]{}%
\providecommand \url  [0]{\begingroup\@sanitize@url \@url }%
\providecommand \@url [1]{\endgroup\@href {#1}{\urlprefix }}%
\providecommand \urlprefix  [0]{URL }%
\providecommand \Eprint [0]{\href }%
\providecommand \doibase [0]{http://dx.doi.org/}%
\providecommand \selectlanguage [0]{\@gobble}%
\providecommand \bibinfo  [0]{\@secondoftwo}%
\providecommand \bibfield  [0]{\@secondoftwo}%
\providecommand \translation [1]{[#1]}%
\providecommand \BibitemOpen [0]{}%
\providecommand \bibitemStop [0]{}%
\providecommand \bibitemNoStop [0]{.\EOS\space}%
\providecommand \EOS [0]{\spacefactor3000\relax}%
\providecommand \BibitemShut  [1]{\csname bibitem#1\endcsname}%
\let\auto@bib@innerbib\@empty
%</preamble>
\bibitem [{\citenamefont {Robledo}\ and\ \citenamefont
  {Bertsch}(2011)}]{Robl11a}%
  \BibitemOpen
  \bibfield  {author} {\bibinfo {author} {\bibfnamefont {L.~M.}\ \bibnamefont
  {Robledo}}\ and\ \bibinfo {author} {\bibfnamefont {G.~F.}\ \bibnamefont
  {Bertsch}},\ }\href {\doibase 10.1103/PhysRevC.84.054302} {\bibfield
  {journal} {\bibinfo  {journal} {Phys. Rev. C}\ }\textbf {\bibinfo {volume}
  {84}},\ \bibinfo {pages} {054302} (\bibinfo {year} {2011})}\BibitemShut
  {NoStop}%
\bibitem [{\citenamefont {Gaffney}\ \emph {et~al.}(2013)\citenamefont
  {Gaffney}, \citenamefont {Butler}, \citenamefont {Scheck}, \citenamefont
  {Hayes}, \citenamefont {Wenander} \emph {et~al.}}]{Gaff13}%
  \BibitemOpen
  \bibfield  {author} {\bibinfo {author} {\bibfnamefont {L.~P.}\ \bibnamefont
  {Gaffney}}, \bibinfo {author} {\bibfnamefont {P.}~\bibnamefont {Butler}},
  \bibinfo {author} {\bibfnamefont {M.}~\bibnamefont {Scheck}}, \bibinfo
  {author} {\bibfnamefont {A.~B.}\ \bibnamefont {Hayes}}, \bibinfo {author}
  {\bibfnamefont {F.}~\bibnamefont {Wenander}},  \emph {et~al.},\ }\href@noop
  {} {\bibfield  {journal} {\bibinfo  {journal} {Nature}\ }\textbf {\bibinfo
  {volume} {497}},\ \bibinfo {pages} {199} (\bibinfo {year}
  {2013})}\BibitemShut {NoStop}%
\bibitem [{\citenamefont {Nomura}\ \emph {et~al.}(2014)\citenamefont {Nomura},
  \citenamefont {Vretenar}, \citenamefont {Nik\ifmmode \check{s}\else
  \v{s}\fi{}i\ifmmode~\acute{c}\else \'{c}\fi{}},\ and\ \citenamefont
  {Lu}}]{Nomu14a}%
  \BibitemOpen
  \bibfield  {author} {\bibinfo {author} {\bibfnamefont {K.}~\bibnamefont
  {Nomura}}, \bibinfo {author} {\bibfnamefont {D.}~\bibnamefont {Vretenar}},
  \bibinfo {author} {\bibfnamefont {T.}~\bibnamefont {Nik\ifmmode
  \check{s}\else \v{s}\fi{}i\ifmmode~\acute{c}\else \'{c}\fi{}}}, \ and\
  \bibinfo {author} {\bibfnamefont {B.-N.}\ \bibnamefont {Lu}},\ }\href
  {\doibase 10.1103/PhysRevC.89.024312} {\bibfield  {journal} {\bibinfo
  {journal} {Phys. Rev. C}\ }\textbf {\bibinfo {volume} {89}},\ \bibinfo
  {pages} {024312} (\bibinfo {year} {2014})}\BibitemShut {NoStop}%
\bibitem [{\citenamefont {Birkenbach}\ \emph {et~al.}(2015)\citenamefont
  {Birkenbach}, \citenamefont {Vogt}, \citenamefont {Geibel}, \citenamefont
  {Recchia}, \citenamefont {Reiter}, \citenamefont {Valiente-Dob\'on},
  \citenamefont {Bazzacco}, \citenamefont {Bowry}, \citenamefont {Bracco},
  \citenamefont {Bruyneel}, \citenamefont {Corradi}, \citenamefont {Crespi},
  \citenamefont {de~Angelis}, \citenamefont {D\'esesquelles}, \citenamefont
  {Eberth}, \citenamefont {Farnea}, \citenamefont {Fioretto}, \citenamefont
  {Gadea}, \citenamefont {Gengelbach}, \citenamefont {Giaz}, \citenamefont
  {G\"orgen}, \citenamefont {Gottardo}, \citenamefont {Grebosz}, \citenamefont
  {Hess}, \citenamefont {John}, \citenamefont {Jolie}, \citenamefont {Judson},
  \citenamefont {Jungclaus}, \citenamefont {Korten}, \citenamefont {Lenzi},
  \citenamefont {Leoni}, \citenamefont {Lunardi}, \citenamefont {Menegazzo},
  \citenamefont {Mengoni}, \citenamefont {Michelagnoli}, \citenamefont
  {Mijatovi\ifmmode~\acute{c}\else \'{c}\fi{}}, \citenamefont {Montagnoli},
  \citenamefont {Montanari}, \citenamefont {Napoli}, \citenamefont {Pellegri},
  \citenamefont {Pollarolo}, \citenamefont {Pullia}, \citenamefont {Quintana},
  \citenamefont {Radeck}, \citenamefont {Rosso}, \citenamefont
  {\ifmmode~\mbox{\c{S}}\else \c{S}\fi{}ahin}, \citenamefont {Salsac},
  \citenamefont {Scarlassara}, \citenamefont {S\"oderstr\"om}, \citenamefont
  {Stefanini}, \citenamefont {Steinbach}, \citenamefont {Stezowski},
  \citenamefont {Szilner}, \citenamefont {Szpak}, \citenamefont {Theisen},
  \citenamefont {Ur}, \citenamefont {Vandone},\ and\ \citenamefont
  {Wiens}}]{Birk15a}%
  \BibitemOpen
  \bibfield  {author} {\bibinfo {author} {\bibfnamefont {B.}~\bibnamefont
  {Birkenbach}}, \bibinfo {author} {\bibfnamefont {A.}~\bibnamefont {Vogt}},
  \bibinfo {author} {\bibfnamefont {K.}~\bibnamefont {Geibel}}, \bibinfo
  {author} {\bibfnamefont {F.}~\bibnamefont {Recchia}}, \bibinfo {author}
  {\bibfnamefont {P.}~\bibnamefont {Reiter}}, \bibinfo {author} {\bibfnamefont
  {J.~J.}\ \bibnamefont {Valiente-Dob\'on}}, \bibinfo {author} {\bibfnamefont
  {D.}~\bibnamefont {Bazzacco}}, \bibinfo {author} {\bibfnamefont
  {M.}~\bibnamefont {Bowry}}, \bibinfo {author} {\bibfnamefont
  {A.}~\bibnamefont {Bracco}}, \bibinfo {author} {\bibfnamefont
  {B.}~\bibnamefont {Bruyneel}}, \bibinfo {author} {\bibfnamefont
  {L.}~\bibnamefont {Corradi}}, \bibinfo {author} {\bibfnamefont {F.~C.~L.}\
  \bibnamefont {Crespi}}, \bibinfo {author} {\bibfnamefont {G.}~\bibnamefont
  {de~Angelis}}, \bibinfo {author} {\bibfnamefont {P.}~\bibnamefont
  {D\'esesquelles}}, \bibinfo {author} {\bibfnamefont {J.}~\bibnamefont
  {Eberth}}, \bibinfo {author} {\bibfnamefont {E.}~\bibnamefont {Farnea}},
  \bibinfo {author} {\bibfnamefont {E.}~\bibnamefont {Fioretto}}, \bibinfo
  {author} {\bibfnamefont {A.}~\bibnamefont {Gadea}}, \bibinfo {author}
  {\bibfnamefont {A.}~\bibnamefont {Gengelbach}}, \bibinfo {author}
  {\bibfnamefont {A.}~\bibnamefont {Giaz}}, \bibinfo {author} {\bibfnamefont
  {A.}~\bibnamefont {G\"orgen}}, \bibinfo {author} {\bibfnamefont
  {A.}~\bibnamefont {Gottardo}}, \bibinfo {author} {\bibfnamefont
  {J.}~\bibnamefont {Grebosz}}, \bibinfo {author} {\bibfnamefont
  {H.}~\bibnamefont {Hess}}, \bibinfo {author} {\bibfnamefont {P.~R.}\
  \bibnamefont {John}}, \bibinfo {author} {\bibfnamefont {J.}~\bibnamefont
  {Jolie}}, \bibinfo {author} {\bibfnamefont {D.~S.}\ \bibnamefont {Judson}},
  \bibinfo {author} {\bibfnamefont {A.}~\bibnamefont {Jungclaus}}, \bibinfo
  {author} {\bibfnamefont {W.}~\bibnamefont {Korten}}, \bibinfo {author}
  {\bibfnamefont {S.}~\bibnamefont {Lenzi}}, \bibinfo {author} {\bibfnamefont
  {S.}~\bibnamefont {Leoni}}, \bibinfo {author} {\bibfnamefont
  {S.}~\bibnamefont {Lunardi}}, \bibinfo {author} {\bibfnamefont
  {R.}~\bibnamefont {Menegazzo}}, \bibinfo {author} {\bibfnamefont
  {D.}~\bibnamefont {Mengoni}}, \bibinfo {author} {\bibfnamefont
  {C.}~\bibnamefont {Michelagnoli}}, \bibinfo {author} {\bibfnamefont
  {T.}~\bibnamefont {Mijatovi\ifmmode~\acute{c}\else \'{c}\fi{}}}, \bibinfo
  {author} {\bibfnamefont {G.}~\bibnamefont {Montagnoli}}, \bibinfo {author}
  {\bibfnamefont {D.}~\bibnamefont {Montanari}}, \bibinfo {author}
  {\bibfnamefont {D.}~\bibnamefont {Napoli}}, \bibinfo {author} {\bibfnamefont
  {L.}~\bibnamefont {Pellegri}}, \bibinfo {author} {\bibfnamefont
  {G.}~\bibnamefont {Pollarolo}}, \bibinfo {author} {\bibfnamefont
  {A.}~\bibnamefont {Pullia}}, \bibinfo {author} {\bibfnamefont
  {B.}~\bibnamefont {Quintana}}, \bibinfo {author} {\bibfnamefont
  {F.}~\bibnamefont {Radeck}}, \bibinfo {author} {\bibfnamefont
  {D.}~\bibnamefont {Rosso}}, \bibinfo {author} {\bibfnamefont
  {E.}~\bibnamefont {\ifmmode~\mbox{\c{S}}\else \c{S}\fi{}ahin}}, \bibinfo
  {author} {\bibfnamefont {M.~D.}\ \bibnamefont {Salsac}}, \bibinfo {author}
  {\bibfnamefont {F.}~\bibnamefont {Scarlassara}}, \bibinfo {author}
  {\bibfnamefont {P.-A.}\ \bibnamefont {S\"oderstr\"om}}, \bibinfo {author}
  {\bibfnamefont {A.~M.}\ \bibnamefont {Stefanini}}, \bibinfo {author}
  {\bibfnamefont {T.}~\bibnamefont {Steinbach}}, \bibinfo {author}
  {\bibfnamefont {O.}~\bibnamefont {Stezowski}}, \bibinfo {author}
  {\bibfnamefont {S.}~\bibnamefont {Szilner}}, \bibinfo {author} {\bibfnamefont
  {B.}~\bibnamefont {Szpak}}, \bibinfo {author} {\bibfnamefont
  {C.}~\bibnamefont {Theisen}}, \bibinfo {author} {\bibfnamefont
  {C.}~\bibnamefont {Ur}}, \bibinfo {author} {\bibfnamefont {V.}~\bibnamefont
  {Vandone}}, \ and\ \bibinfo {author} {\bibfnamefont {A.}~\bibnamefont
  {Wiens}},\ }\href {\doibase 10.1103/PhysRevC.92.044319} {\bibfield  {journal}
  {\bibinfo  {journal} {Phys. Rev. C}\ }\textbf {\bibinfo {volume} {92}},\
  \bibinfo {pages} {044319} (\bibinfo {year} {2015})}\BibitemShut {NoStop}%
\bibitem [{\citenamefont {Zimba}\ \emph {et~al.}(2016)\citenamefont {Zimba},
  \citenamefont {Sharpey-Schafer}, \citenamefont {Jones}, \citenamefont
  {Bvumbi}, \citenamefont {Masiteng}, \citenamefont {Majola}, \citenamefont
  {Dinoko}, \citenamefont {Lawrie}, \citenamefont {Lawrie}, \citenamefont
  {Negi}, \citenamefont {Papka}, \citenamefont {Roux}, \citenamefont
  {Shirinda}, \citenamefont {Easton},\ and\ \citenamefont
  {Khumalo}}]{Zimba16a}%
  \BibitemOpen
  \bibfield  {author} {\bibinfo {author} {\bibfnamefont {G.~L.}\ \bibnamefont
  {Zimba}}, \bibinfo {author} {\bibfnamefont {J.~F.}\ \bibnamefont
  {Sharpey-Schafer}}, \bibinfo {author} {\bibfnamefont {P.}~\bibnamefont
  {Jones}}, \bibinfo {author} {\bibfnamefont {S.~P.}\ \bibnamefont {Bvumbi}},
  \bibinfo {author} {\bibfnamefont {L.~P.}\ \bibnamefont {Masiteng}}, \bibinfo
  {author} {\bibfnamefont {S.~N.~T.}\ \bibnamefont {Majola}}, \bibinfo {author}
  {\bibfnamefont {T.~S.}\ \bibnamefont {Dinoko}}, \bibinfo {author}
  {\bibfnamefont {E.~A.}\ \bibnamefont {Lawrie}}, \bibinfo {author}
  {\bibfnamefont {J.~J.}\ \bibnamefont {Lawrie}}, \bibinfo {author}
  {\bibfnamefont {D.}~\bibnamefont {Negi}}, \bibinfo {author} {\bibfnamefont
  {P.}~\bibnamefont {Papka}}, \bibinfo {author} {\bibfnamefont
  {D.}~\bibnamefont {Roux}}, \bibinfo {author} {\bibfnamefont {O.}~\bibnamefont
  {Shirinda}}, \bibinfo {author} {\bibfnamefont {J.~E.}\ \bibnamefont
  {Easton}}, \ and\ \bibinfo {author} {\bibfnamefont {N.~A.}\ \bibnamefont
  {Khumalo}},\ }\href {\doibase 10.1103/PhysRevC.94.054303} {\bibfield
  {journal} {\bibinfo  {journal} {Phys. Rev. C}\ }\textbf {\bibinfo {volume}
  {94}},\ \bibinfo {pages} {054303} (\bibinfo {year} {2016})}\BibitemShut
  {NoStop}%
\bibitem [{\citenamefont {Butler}(2016)}]{Butl16a}%
  \BibitemOpen
  \bibfield  {author} {\bibinfo {author} {\bibfnamefont {P.~A.}\ \bibnamefont
  {Butler}},\ }\href {http://stacks.iop.org/0954-3899/43/i=7/a=073002}
  {\bibfield  {journal} {\bibinfo  {journal} {Journal of Physics G: Nuclear and
  Particle Physics}\ }\textbf {\bibinfo {volume} {43}},\ \bibinfo {pages}
  {073002} (\bibinfo {year} {2016})}\BibitemShut {NoStop}%
\bibitem [{\citenamefont {Agbemava}\ \emph {et~al.}(2016)\citenamefont
  {Agbemava}, \citenamefont {Afanasjev},\ and\ \citenamefont {Ring}}]{Agbe16a}%
  \BibitemOpen
  \bibfield  {author} {\bibinfo {author} {\bibfnamefont {S.~E.}\ \bibnamefont
  {Agbemava}}, \bibinfo {author} {\bibfnamefont {A.~V.}\ \bibnamefont
  {Afanasjev}}, \ and\ \bibinfo {author} {\bibfnamefont {P.}~\bibnamefont
  {Ring}},\ }\href {\doibase 10.1103/PhysRevC.93.044304} {\bibfield  {journal}
  {\bibinfo  {journal} {Phys. Rev. C}\ }\textbf {\bibinfo {volume} {93}},\
  \bibinfo {pages} {044304} (\bibinfo {year} {2016})}\BibitemShut {NoStop}%
\bibitem [{\citenamefont {Bucher}\ \emph {et~al.}(2016)\citenamefont {Bucher},
  \citenamefont {Zhu}, \citenamefont {Wu}, \citenamefont {Janssens},
  \citenamefont {Cline}, \citenamefont {Hayes}, \citenamefont {Albers},
  \citenamefont {Ayangeakaa}, \citenamefont {Butler}, \citenamefont {Campbell},
  \citenamefont {Carpenter}, \citenamefont {Chiara}, \citenamefont {Clark},
  \citenamefont {Crawford}, \citenamefont {Cromaz}, \citenamefont {David},
  \citenamefont {Dickerson}, \citenamefont {Gregor}, \citenamefont {Harker},
  \citenamefont {Hoffman}, \citenamefont {Kay}, \citenamefont {Kondev},
  \citenamefont {Korichi}, \citenamefont {Lauritsen}, \citenamefont
  {Macchiavelli}, \citenamefont {Pardo}, \citenamefont {Richard}, \citenamefont
  {Riley}, \citenamefont {Savard}, \citenamefont {Scheck}, \citenamefont
  {Seweryniak}, \citenamefont {Smith}, \citenamefont {Vondrasek},\ and\
  \citenamefont {Wiens}}]{Buch16a}%
  \BibitemOpen
  \bibfield  {author} {\bibinfo {author} {\bibfnamefont {B.}~\bibnamefont
  {Bucher}}, \bibinfo {author} {\bibfnamefont {S.}~\bibnamefont {Zhu}},
  \bibinfo {author} {\bibfnamefont {C.~Y.}\ \bibnamefont {Wu}}, \bibinfo
  {author} {\bibfnamefont {R.~V.~F.}\ \bibnamefont {Janssens}}, \bibinfo
  {author} {\bibfnamefont {D.}~\bibnamefont {Cline}}, \bibinfo {author}
  {\bibfnamefont {A.~B.}\ \bibnamefont {Hayes}}, \bibinfo {author}
  {\bibfnamefont {M.}~\bibnamefont {Albers}}, \bibinfo {author} {\bibfnamefont
  {A.~D.}\ \bibnamefont {Ayangeakaa}}, \bibinfo {author} {\bibfnamefont
  {P.~A.}\ \bibnamefont {Butler}}, \bibinfo {author} {\bibfnamefont {C.~M.}\
  \bibnamefont {Campbell}}, \bibinfo {author} {\bibfnamefont {M.~P.}\
  \bibnamefont {Carpenter}}, \bibinfo {author} {\bibfnamefont {C.~J.}\
  \bibnamefont {Chiara}}, \bibinfo {author} {\bibfnamefont {J.~A.}\
  \bibnamefont {Clark}}, \bibinfo {author} {\bibfnamefont {H.~L.}\ \bibnamefont
  {Crawford}}, \bibinfo {author} {\bibfnamefont {M.}~\bibnamefont {Cromaz}},
  \bibinfo {author} {\bibfnamefont {H.~M.}\ \bibnamefont {David}}, \bibinfo
  {author} {\bibfnamefont {C.}~\bibnamefont {Dickerson}}, \bibinfo {author}
  {\bibfnamefont {E.~T.}\ \bibnamefont {Gregor}}, \bibinfo {author}
  {\bibfnamefont {J.}~\bibnamefont {Harker}}, \bibinfo {author} {\bibfnamefont
  {C.~R.}\ \bibnamefont {Hoffman}}, \bibinfo {author} {\bibfnamefont {B.~P.}\
  \bibnamefont {Kay}}, \bibinfo {author} {\bibfnamefont {F.~G.}\ \bibnamefont
  {Kondev}}, \bibinfo {author} {\bibfnamefont {A.}~\bibnamefont {Korichi}},
  \bibinfo {author} {\bibfnamefont {T.}~\bibnamefont {Lauritsen}}, \bibinfo
  {author} {\bibfnamefont {A.~O.}\ \bibnamefont {Macchiavelli}}, \bibinfo
  {author} {\bibfnamefont {R.~C.}\ \bibnamefont {Pardo}}, \bibinfo {author}
  {\bibfnamefont {A.}~\bibnamefont {Richard}}, \bibinfo {author} {\bibfnamefont
  {M.~A.}\ \bibnamefont {Riley}}, \bibinfo {author} {\bibfnamefont
  {G.}~\bibnamefont {Savard}}, \bibinfo {author} {\bibfnamefont
  {M.}~\bibnamefont {Scheck}}, \bibinfo {author} {\bibfnamefont
  {D.}~\bibnamefont {Seweryniak}}, \bibinfo {author} {\bibfnamefont {M.~K.}\
  \bibnamefont {Smith}}, \bibinfo {author} {\bibfnamefont {R.}~\bibnamefont
  {Vondrasek}}, \ and\ \bibinfo {author} {\bibfnamefont {A.}~\bibnamefont
  {Wiens}},\ }\href {\doibase 10.1103/PhysRevLett.116.112503} {\bibfield
  {journal} {\bibinfo  {journal} {Phys. Rev. Lett.}\ }\textbf {\bibinfo
  {volume} {116}},\ \bibinfo {pages} {112503} (\bibinfo {year}
  {2016})}\BibitemShut {NoStop}%
\bibitem [{\citenamefont {Maquart}\ \emph {et~al.}(2017)\citenamefont
  {Maquart}, \citenamefont {Augey}, \citenamefont {Chaix}, \citenamefont
  {Companis}, \citenamefont {Ducoin}, \citenamefont {Dudouet}, \citenamefont
  {Guinet}, \citenamefont {Lehaut}, \citenamefont {Mancuso}, \citenamefont
  {Redon}, \citenamefont {St\'ezowski}, \citenamefont {Vancraeyenest},
  \citenamefont {Astier}, \citenamefont {Azaiez}, \citenamefont {Courtin},
  \citenamefont {Curien}, \citenamefont {Deloncle}, \citenamefont {Dorvaux},
  \citenamefont {Duch\^ene}, \citenamefont {Gall}, \citenamefont {Grahn},
  \citenamefont {Greenlees}, \citenamefont {Herzan}, \citenamefont {Hauschild},
  \citenamefont {Jakobsson}, \citenamefont {Jones}, \citenamefont {Julin},
  \citenamefont {Juutinen}, \citenamefont {Ketelhut}, \citenamefont {Leino},
  \citenamefont {Lopez-Martens}, \citenamefont {Nieminen}, \citenamefont
  {Petkov}, \citenamefont {Peura}, \citenamefont {Porquet}, \citenamefont
  {Rahkila}, \citenamefont {Rinta-Antila}, \citenamefont {Rousseau},
  \citenamefont {Ruotsalainen}, \citenamefont {Sandzelius}, \citenamefont
  {Sar\'en}, \citenamefont {Scholey}, \citenamefont {Sorri}, \citenamefont
  {Stolze},\ and\ \citenamefont {Uusitalo}}]{Maq17a}%
  \BibitemOpen
  \bibfield  {author} {\bibinfo {author} {\bibfnamefont {G.}~\bibnamefont
  {Maquart}}, \bibinfo {author} {\bibfnamefont {L.}~\bibnamefont {Augey}},
  \bibinfo {author} {\bibfnamefont {L.}~\bibnamefont {Chaix}}, \bibinfo
  {author} {\bibfnamefont {I.}~\bibnamefont {Companis}}, \bibinfo {author}
  {\bibfnamefont {C.}~\bibnamefont {Ducoin}}, \bibinfo {author} {\bibfnamefont
  {J.}~\bibnamefont {Dudouet}}, \bibinfo {author} {\bibfnamefont
  {D.}~\bibnamefont {Guinet}}, \bibinfo {author} {\bibfnamefont
  {G.}~\bibnamefont {Lehaut}}, \bibinfo {author} {\bibfnamefont
  {C.}~\bibnamefont {Mancuso}}, \bibinfo {author} {\bibfnamefont
  {N.}~\bibnamefont {Redon}}, \bibinfo {author} {\bibfnamefont
  {O.}~\bibnamefont {St\'ezowski}}, \bibinfo {author} {\bibfnamefont
  {A.}~\bibnamefont {Vancraeyenest}}, \bibinfo {author} {\bibfnamefont
  {A.}~\bibnamefont {Astier}}, \bibinfo {author} {\bibfnamefont
  {F.}~\bibnamefont {Azaiez}}, \bibinfo {author} {\bibfnamefont
  {S.}~\bibnamefont {Courtin}}, \bibinfo {author} {\bibfnamefont
  {D.}~\bibnamefont {Curien}}, \bibinfo {author} {\bibfnamefont
  {I.}~\bibnamefont {Deloncle}}, \bibinfo {author} {\bibfnamefont
  {O.}~\bibnamefont {Dorvaux}}, \bibinfo {author} {\bibfnamefont
  {G.}~\bibnamefont {Duch\^ene}}, \bibinfo {author} {\bibfnamefont
  {B.}~\bibnamefont {Gall}}, \bibinfo {author} {\bibfnamefont {T.}~\bibnamefont
  {Grahn}}, \bibinfo {author} {\bibfnamefont {P.}~\bibnamefont {Greenlees}},
  \bibinfo {author} {\bibfnamefont {A.}~\bibnamefont {Herzan}}, \bibinfo
  {author} {\bibfnamefont {K.}~\bibnamefont {Hauschild}}, \bibinfo {author}
  {\bibfnamefont {U.}~\bibnamefont {Jakobsson}}, \bibinfo {author}
  {\bibfnamefont {P.}~\bibnamefont {Jones}}, \bibinfo {author} {\bibfnamefont
  {R.}~\bibnamefont {Julin}}, \bibinfo {author} {\bibfnamefont
  {S.}~\bibnamefont {Juutinen}}, \bibinfo {author} {\bibfnamefont
  {S.}~\bibnamefont {Ketelhut}}, \bibinfo {author} {\bibfnamefont
  {M.}~\bibnamefont {Leino}}, \bibinfo {author} {\bibfnamefont
  {A.}~\bibnamefont {Lopez-Martens}}, \bibinfo {author} {\bibfnamefont
  {P.}~\bibnamefont {Nieminen}}, \bibinfo {author} {\bibfnamefont
  {P.}~\bibnamefont {Petkov}}, \bibinfo {author} {\bibfnamefont
  {P.}~\bibnamefont {Peura}}, \bibinfo {author} {\bibfnamefont {M.-G.}\
  \bibnamefont {Porquet}}, \bibinfo {author} {\bibfnamefont {P.}~\bibnamefont
  {Rahkila}}, \bibinfo {author} {\bibfnamefont {S.}~\bibnamefont
  {Rinta-Antila}}, \bibinfo {author} {\bibfnamefont {M.}~\bibnamefont
  {Rousseau}}, \bibinfo {author} {\bibfnamefont {P.}~\bibnamefont
  {Ruotsalainen}}, \bibinfo {author} {\bibfnamefont {M.}~\bibnamefont
  {Sandzelius}}, \bibinfo {author} {\bibfnamefont {J.}~\bibnamefont {Sar\'en}},
  \bibinfo {author} {\bibfnamefont {C.}~\bibnamefont {Scholey}}, \bibinfo
  {author} {\bibfnamefont {J.}~\bibnamefont {Sorri}}, \bibinfo {author}
  {\bibfnamefont {S.}~\bibnamefont {Stolze}}, \ and\ \bibinfo {author}
  {\bibfnamefont {J.}~\bibnamefont {Uusitalo}},\ }\href {\doibase
  10.1103/PhysRevC.95.034304} {\bibfield  {journal} {\bibinfo  {journal} {Phys.
  Rev. C}\ }\textbf {\bibinfo {volume} {95}},\ \bibinfo {pages} {034304}
  (\bibinfo {year} {2017})}\BibitemShut {NoStop}%
\bibitem [{\citenamefont {Bucher}\ \emph {et~al.}(2017)\citenamefont {Bucher},
  \citenamefont {Zhu}, \citenamefont {Wu}, \citenamefont {Janssens},
  \citenamefont {Bernard}, \citenamefont {Robledo}, \citenamefont
  {Rodr\'{\i}guez}, \citenamefont {Cline}, \citenamefont {Hayes}, \citenamefont
  {Ayangeakaa}, \citenamefont {Buckner}, \citenamefont {Campbell},
  \citenamefont {Carpenter}, \citenamefont {Clark}, \citenamefont {Crawford},
  \citenamefont {David}, \citenamefont {Dickerson}, \citenamefont {Harker},
  \citenamefont {Hoffman}, \citenamefont {Kay}, \citenamefont {Kondev},
  \citenamefont {Lauritsen}, \citenamefont {Macchiavelli}, \citenamefont
  {Pardo}, \citenamefont {Savard}, \citenamefont {Seweryniak},\ and\
  \citenamefont {Vondrasek}}]{Buch17a}%
  \BibitemOpen
  \bibfield  {author} {\bibinfo {author} {\bibfnamefont {B.}~\bibnamefont
  {Bucher}}, \bibinfo {author} {\bibfnamefont {S.}~\bibnamefont {Zhu}},
  \bibinfo {author} {\bibfnamefont {C.~Y.}\ \bibnamefont {Wu}}, \bibinfo
  {author} {\bibfnamefont {R.~V.~F.}\ \bibnamefont {Janssens}}, \bibinfo
  {author} {\bibfnamefont {R.~N.}\ \bibnamefont {Bernard}}, \bibinfo {author}
  {\bibfnamefont {L.~M.}\ \bibnamefont {Robledo}}, \bibinfo {author}
  {\bibfnamefont {T.~R.}\ \bibnamefont {Rodr\'{\i}guez}}, \bibinfo {author}
  {\bibfnamefont {D.}~\bibnamefont {Cline}}, \bibinfo {author} {\bibfnamefont
  {A.~B.}\ \bibnamefont {Hayes}}, \bibinfo {author} {\bibfnamefont {A.~D.}\
  \bibnamefont {Ayangeakaa}}, \bibinfo {author} {\bibfnamefont {M.~Q.}\
  \bibnamefont {Buckner}}, \bibinfo {author} {\bibfnamefont {C.~M.}\
  \bibnamefont {Campbell}}, \bibinfo {author} {\bibfnamefont {M.~P.}\
  \bibnamefont {Carpenter}}, \bibinfo {author} {\bibfnamefont {J.~A.}\
  \bibnamefont {Clark}}, \bibinfo {author} {\bibfnamefont {H.~L.}\ \bibnamefont
  {Crawford}}, \bibinfo {author} {\bibfnamefont {H.~M.}\ \bibnamefont {David}},
  \bibinfo {author} {\bibfnamefont {C.}~\bibnamefont {Dickerson}}, \bibinfo
  {author} {\bibfnamefont {J.}~\bibnamefont {Harker}}, \bibinfo {author}
  {\bibfnamefont {C.~R.}\ \bibnamefont {Hoffman}}, \bibinfo {author}
  {\bibfnamefont {B.~P.}\ \bibnamefont {Kay}}, \bibinfo {author} {\bibfnamefont
  {F.~G.}\ \bibnamefont {Kondev}}, \bibinfo {author} {\bibfnamefont
  {T.}~\bibnamefont {Lauritsen}}, \bibinfo {author} {\bibfnamefont {A.~O.}\
  \bibnamefont {Macchiavelli}}, \bibinfo {author} {\bibfnamefont {R.~C.}\
  \bibnamefont {Pardo}}, \bibinfo {author} {\bibfnamefont {G.}~\bibnamefont
  {Savard}}, \bibinfo {author} {\bibfnamefont {D.}~\bibnamefont {Seweryniak}},
  \ and\ \bibinfo {author} {\bibfnamefont {R.}~\bibnamefont {Vondrasek}},\
  }\href {\doibase 10.1103/PhysRevLett.118.152504} {\bibfield  {journal}
  {\bibinfo  {journal} {Phys. Rev. Lett.}\ }\textbf {\bibinfo {volume} {118}},\
  \bibinfo {pages} {152504} (\bibinfo {year} {2017})}\BibitemShut {NoStop}%
\bibitem [{\citenamefont {Butler}\ and\ \citenamefont
  {Nazarewicz}(1996)}]{Butl96}%
  \BibitemOpen
  \bibfield  {author} {\bibinfo {author} {\bibfnamefont {P.~A.}\ \bibnamefont
  {Butler}}\ and\ \bibinfo {author} {\bibfnamefont {W.}~\bibnamefont
  {Nazarewicz}},\ }\href@noop {} {\bibfield  {journal} {\bibinfo  {journal}
  {Rev. Mod. Phys.}\ }\textbf {\bibinfo {volume} {68}},\ \bibinfo {pages} {349}
  (\bibinfo {year} {1996})}\BibitemShut {NoStop}%
\bibitem [{\citenamefont {Wiedenh\"over}\ \emph {et~al.}(1999)\citenamefont
  {Wiedenh\"over} \emph {et~al.}}]{Wied99}%
  \BibitemOpen
  \bibfield  {author} {\bibinfo {author} {\bibfnamefont {I.}~\bibnamefont
  {Wiedenh\"over}} \emph {et~al.},\ }\href@noop {} {\bibfield  {journal}
  {\bibinfo  {journal} {Phys. Rev. Lett.}\ }\textbf {\bibinfo {volume} {83}},\
  \bibinfo {pages} {2143} (\bibinfo {year} {1999})}\BibitemShut {NoStop}%
\bibitem [{\citenamefont {Jolos}\ and\ \citenamefont {von
  Brentano}(2011)}]{Jolo11}%
  \BibitemOpen
  \bibfield  {author} {\bibinfo {author} {\bibfnamefont {R.~V.}\ \bibnamefont
  {Jolos}}\ and\ \bibinfo {author} {\bibfnamefont {P.}~\bibnamefont {von
  Brentano}},\ }\href@noop {} {\bibfield  {journal} {\bibinfo  {journal} {Phys.
  Rev. C}\ }\textbf {\bibinfo {volume} {84}},\ \bibinfo {pages} {024312}
  (\bibinfo {year} {2011})}\BibitemShut {NoStop}%
\bibitem [{\citenamefont {Jolos}\ \emph {et~al.}(2012)\citenamefont {Jolos},
  \citenamefont {von Brentano},\ and\ \citenamefont {Jolie}}]{Jolo12}%
  \BibitemOpen
  \bibfield  {author} {\bibinfo {author} {\bibfnamefont {R.~V.}\ \bibnamefont
  {Jolos}}, \bibinfo {author} {\bibfnamefont {P.}~\bibnamefont {von Brentano}},
  \ and\ \bibinfo {author} {\bibfnamefont {J.}~\bibnamefont {Jolie}},\
  }\href@noop {} {\bibfield  {journal} {\bibinfo  {journal} {Phys. Rev. C}\
  }\textbf {\bibinfo {volume} {86}},\ \bibinfo {pages} {024319} (\bibinfo
  {year} {2012})}\BibitemShut {NoStop}%
\bibitem [{\citenamefont {Wang}\ \emph {et~al.}(2009)\citenamefont {Wang} \emph
  {et~al.}}]{Wang09}%
  \BibitemOpen
  \bibfield  {author} {\bibinfo {author} {\bibfnamefont {X.}~\bibnamefont
  {Wang}} \emph {et~al.},\ }\href@noop {} {\bibfield  {journal} {\bibinfo
  {journal} {Phys. Rev. Lett.}\ }\textbf {\bibinfo {volume} {102}},\ \bibinfo
  {pages} {122501} (\bibinfo {year} {2009})}\BibitemShut {NoStop}%
\bibitem [{\citenamefont {Frauendorf}(2008)}]{Frauend08}%
  \BibitemOpen
  \bibfield  {author} {\bibinfo {author} {\bibfnamefont {S.}~\bibnamefont
  {Frauendorf}},\ }\href@noop {} {\bibfield  {journal} {\bibinfo  {journal}
  {Phys. Rev. C}\ }\textbf {\bibinfo {volume} {77}},\ \bibinfo {pages}
  {021304(R)} (\bibinfo {year} {2008})}\BibitemShut {NoStop}%
\bibitem [{\citenamefont {Jolos}\ \emph {et~al.}(2013)\citenamefont {Jolos},
  \citenamefont {von Brentano},\ and\ \citenamefont {Casten}}]{Jolo13}%
  \BibitemOpen
  \bibfield  {author} {\bibinfo {author} {\bibfnamefont {R.~V.}\ \bibnamefont
  {Jolos}}, \bibinfo {author} {\bibfnamefont {P.}~\bibnamefont {von Brentano}},
  \ and\ \bibinfo {author} {\bibfnamefont {R.~F.}\ \bibnamefont {Casten}},\
  }\href@noop {} {\bibfield  {journal} {\bibinfo  {journal} {Phys. Rev. C}\
  }\textbf {\bibinfo {volume} {88}},\ \bibinfo {pages} {034306} (\bibinfo
  {year} {2013})}\BibitemShut {NoStop}%
\bibitem [{\citenamefont {Spieker}\ \emph {et~al.}(2013)\citenamefont
  {Spieker}, \citenamefont {Bucurescu}, \citenamefont {Endres}, \citenamefont
  {Faestermann}, \citenamefont {Hertenberger}, \citenamefont {Pascu},
  \citenamefont {Skalacki}, \citenamefont {Weber}, \citenamefont {Wirth},
  \citenamefont {Zamfir},\ and\ \citenamefont {Zilges}}]{Spiek13a}%
  \BibitemOpen
  \bibfield  {author} {\bibinfo {author} {\bibfnamefont {M.}~\bibnamefont
  {Spieker}}, \bibinfo {author} {\bibfnamefont {D.}~\bibnamefont {Bucurescu}},
  \bibinfo {author} {\bibfnamefont {J.}~\bibnamefont {Endres}}, \bibinfo
  {author} {\bibfnamefont {T.}~\bibnamefont {Faestermann}}, \bibinfo {author}
  {\bibfnamefont {R.}~\bibnamefont {Hertenberger}}, \bibinfo {author}
  {\bibfnamefont {S.}~\bibnamefont {Pascu}}, \bibinfo {author} {\bibfnamefont
  {S.}~\bibnamefont {Skalacki}}, \bibinfo {author} {\bibfnamefont
  {S.}~\bibnamefont {Weber}}, \bibinfo {author} {\bibfnamefont {H.-F.}\
  \bibnamefont {Wirth}}, \bibinfo {author} {\bibfnamefont {N.-V.}\ \bibnamefont
  {Zamfir}}, \ and\ \bibinfo {author} {\bibfnamefont {A.}~\bibnamefont
  {Zilges}},\ }\href@noop {} {\bibfield  {journal} {\bibinfo  {journal} {Phys.
  Rev. C}\ }\textbf {\bibinfo {volume} {88}},\ \bibinfo {pages} {041303(R)}
  (\bibinfo {year} {2013})}\BibitemShut {NoStop}%
\bibitem [{\citenamefont {Levon}\ \emph {et~al.}(2013)\citenamefont {Levon},
  \citenamefont {Graw}, \citenamefont {Hertenberger}, \citenamefont {Pascu},
  \citenamefont {Thirolf}, \citenamefont {Wirth},\ and\ \citenamefont
  {Alexa}}]{Levon13}%
  \BibitemOpen
  \bibfield  {author} {\bibinfo {author} {\bibfnamefont {A.~I.}\ \bibnamefont
  {Levon}}, \bibinfo {author} {\bibfnamefont {G.}~\bibnamefont {Graw}},
  \bibinfo {author} {\bibfnamefont {R.}~\bibnamefont {Hertenberger}}, \bibinfo
  {author} {\bibfnamefont {S.}~\bibnamefont {Pascu}}, \bibinfo {author}
  {\bibfnamefont {P.~G.}\ \bibnamefont {Thirolf}}, \bibinfo {author}
  {\bibfnamefont {H.-F.}\ \bibnamefont {Wirth}}, \ and\ \bibinfo {author}
  {\bibfnamefont {P.}~\bibnamefont {Alexa}},\ }\href {\doibase
  10.1103/PhysRevC.88.014310} {\bibfield  {journal} {\bibinfo  {journal} {Phys.
  Rev. C}\ }\textbf {\bibinfo {volume} {88}},\ \bibinfo {pages} {014310}
  (\bibinfo {year} {2013})}\BibitemShut {NoStop}%
\bibitem [{\citenamefont {Levon}\ \emph {et~al.}(2015)\citenamefont {Levon},
  \citenamefont {Alexa}, \citenamefont {Graw}, \citenamefont {Hertenberger},
  \citenamefont {Pascu}, \citenamefont {Thirolf},\ and\ \citenamefont
  {Wirth}}]{Levon15}%
  \BibitemOpen
  \bibfield  {author} {\bibinfo {author} {\bibfnamefont {A.~I.}\ \bibnamefont
  {Levon}}, \bibinfo {author} {\bibfnamefont {P.}~\bibnamefont {Alexa}},
  \bibinfo {author} {\bibfnamefont {G.}~\bibnamefont {Graw}}, \bibinfo {author}
  {\bibfnamefont {R.}~\bibnamefont {Hertenberger}}, \bibinfo {author}
  {\bibfnamefont {S.}~\bibnamefont {Pascu}}, \bibinfo {author} {\bibfnamefont
  {P.~G.}\ \bibnamefont {Thirolf}}, \ and\ \bibinfo {author} {\bibfnamefont
  {H.-F.}\ \bibnamefont {Wirth}},\ }\href {\doibase 10.1103/PhysRevC.92.064319}
  {\bibfield  {journal} {\bibinfo  {journal} {Phys. Rev. C}\ }\textbf {\bibinfo
  {volume} {92}},\ \bibinfo {pages} {064319} (\bibinfo {year}
  {2015})}\BibitemShut {NoStop}%
\bibitem [{\citenamefont {Maher}\ \emph {et~al.}(1972)\citenamefont {Maher}
  \emph {et~al.}}]{Maher72}%
  \BibitemOpen
  \bibfield  {author} {\bibinfo {author} {\bibfnamefont {J.~V.}\ \bibnamefont
  {Maher}} \emph {et~al.},\ }\href@noop {} {\bibfield  {journal} {\bibinfo
  {journal} {Phys. Rev. C}\ }\textbf {\bibinfo {volume} {5}},\ \bibinfo {pages}
  {1380} (\bibinfo {year} {1972})}\BibitemShut {NoStop}%
\bibitem [{\citenamefont {Casten}\ \emph {et~al.}(1972)\citenamefont {Casten}
  \emph {et~al.}}]{Cast72}%
  \BibitemOpen
  \bibfield  {author} {\bibinfo {author} {\bibfnamefont {R.~F.}\ \bibnamefont
  {Casten}} \emph {et~al.},\ }\href@noop {} {\bibfield  {journal} {\bibinfo
  {journal} {Phys. Lett. B}\ }\textbf {\bibinfo {volume} {40}},\ \bibinfo
  {pages} {333} (\bibinfo {year} {1972})}\BibitemShut {NoStop}%
\bibitem [{\citenamefont {Rij}\ and\ \citenamefont {Kahana}(1972)}]{Rij72}%
  \BibitemOpen
  \bibfield  {author} {\bibinfo {author} {\bibfnamefont {W.~I.}\ \bibnamefont
  {Rij}}\ and\ \bibinfo {author} {\bibfnamefont {S.~H.}\ \bibnamefont
  {Kahana}},\ }\href@noop {} {\bibfield  {journal} {\bibinfo  {journal} {Phys.
  Rev. Lett.}\ }\textbf {\bibinfo {volume} {28}},\ \bibinfo {pages} {50}
  (\bibinfo {year} {1972})}\BibitemShut {NoStop}%
\bibitem [{\citenamefont {Friedman}\ and\ \citenamefont
  {Katori}(1973)}]{Fried73}%
  \BibitemOpen
  \bibfield  {author} {\bibinfo {author} {\bibfnamefont {A.}~\bibnamefont
  {Friedman}}\ and\ \bibinfo {author} {\bibfnamefont {K.}~\bibnamefont
  {Katori}},\ }\href@noop {} {\bibfield  {journal} {\bibinfo  {journal} {Phys.
  Rev. Lett.}\ }\textbf {\bibinfo {volume} {30}},\ \bibinfo {pages} {102}
  (\bibinfo {year} {1973})}\BibitemShut {NoStop}%
\bibitem [{\citenamefont {Friedman}\ \emph {et~al.}(1974)\citenamefont
  {Friedman} \emph {et~al.}}]{Fried74}%
  \BibitemOpen
  \bibfield  {author} {\bibinfo {author} {\bibfnamefont {A.}~\bibnamefont
  {Friedman}} \emph {et~al.},\ }\href@noop {} {\bibfield  {journal} {\bibinfo
  {journal} {Phys. Rev. C}\ }\textbf {\bibinfo {volume} {9}},\ \bibinfo {pages}
  {760} (\bibinfo {year} {1974})}\BibitemShut {NoStop}%
\bibitem [{\citenamefont {Ragnarsson}\ and\ \citenamefont
  {Broglia}(1976)}]{Ragn76}%
  \BibitemOpen
  \bibfield  {author} {\bibinfo {author} {\bibfnamefont {I.}~\bibnamefont
  {Ragnarsson}}\ and\ \bibinfo {author} {\bibfnamefont {R.~A.}\ \bibnamefont
  {Broglia}},\ }\href@noop {} {\bibfield  {journal} {\bibinfo  {journal} {Nucl.
  Phys. A}\ }\textbf {\bibinfo {volume} {263}},\ \bibinfo {pages} {315}
  (\bibinfo {year} {1976})}\BibitemShut {NoStop}%
\bibitem [{\citenamefont {Allmond}\ \emph {et~al.}(2017)\citenamefont
  {Allmond}, \citenamefont {Beausang}, \citenamefont {Ross}, \citenamefont
  {Humby}, \citenamefont {Basunia}, \citenamefont {Bernstein}, \citenamefont
  {Bleuel}, \citenamefont {Brooks}, \citenamefont {Brown}, \citenamefont
  {Burke}, \citenamefont {Darakchieva}, \citenamefont {Dudziak}, \citenamefont
  {Evans}, \citenamefont {Fallon}, \citenamefont {Jeppesen}, \citenamefont
  {LeBlanc}, \citenamefont {Lesher}, \citenamefont {McMahan}, \citenamefont
  {Meyer}, \citenamefont {Phair}, \citenamefont {Rasmussen}, \citenamefont
  {Scielzo}, \citenamefont {Stroberg},\ and\ \citenamefont
  {Wiedeking}}]{Allm17a}%
  \BibitemOpen
  \bibfield  {author} {\bibinfo {author} {\bibfnamefont {J.~M.}\ \bibnamefont
  {Allmond}}, \bibinfo {author} {\bibfnamefont {C.~W.}\ \bibnamefont
  {Beausang}}, \bibinfo {author} {\bibfnamefont {T.~J.}\ \bibnamefont {Ross}},
  \bibinfo {author} {\bibfnamefont {P.}~\bibnamefont {Humby}}, \bibinfo
  {author} {\bibfnamefont {M.~S.}\ \bibnamefont {Basunia}}, \bibinfo {author}
  {\bibfnamefont {L.~A.}\ \bibnamefont {Bernstein}}, \bibinfo {author}
  {\bibfnamefont {D.~L.}\ \bibnamefont {Bleuel}}, \bibinfo {author}
  {\bibfnamefont {W.}~\bibnamefont {Brooks}}, \bibinfo {author} {\bibfnamefont
  {N.}~\bibnamefont {Brown}}, \bibinfo {author} {\bibfnamefont {J.~T.}\
  \bibnamefont {Burke}}, \bibinfo {author} {\bibfnamefont {B.~K.}\ \bibnamefont
  {Darakchieva}}, \bibinfo {author} {\bibfnamefont {K.~R.}\ \bibnamefont
  {Dudziak}}, \bibinfo {author} {\bibfnamefont {K.~E.}\ \bibnamefont {Evans}},
  \bibinfo {author} {\bibfnamefont {P.}~\bibnamefont {Fallon}}, \bibinfo
  {author} {\bibfnamefont {H.~B.}\ \bibnamefont {Jeppesen}}, \bibinfo {author}
  {\bibfnamefont {J.~D.}\ \bibnamefont {LeBlanc}}, \bibinfo {author}
  {\bibfnamefont {S.~R.}\ \bibnamefont {Lesher}}, \bibinfo {author}
  {\bibfnamefont {M.~A.}\ \bibnamefont {McMahan}}, \bibinfo {author}
  {\bibfnamefont {D.~A.}\ \bibnamefont {Meyer}}, \bibinfo {author}
  {\bibfnamefont {L.}~\bibnamefont {Phair}}, \bibinfo {author} {\bibfnamefont
  {J.~O.}\ \bibnamefont {Rasmussen}}, \bibinfo {author} {\bibfnamefont {N.~D.}\
  \bibnamefont {Scielzo}}, \bibinfo {author} {\bibfnamefont {S.~R.}\
  \bibnamefont {Stroberg}}, \ and\ \bibinfo {author} {\bibfnamefont
  {M.}~\bibnamefont {Wiedeking}},\ }\href {\doibase 10.1140/epja/i2017-12253-2}
  {\bibfield  {journal} {\bibinfo  {journal} {Eur. Phys. Journal A}\ }\textbf
  {\bibinfo {volume} {53}},\ \bibinfo {pages} {62} (\bibinfo {year}
  {2017})}\BibitemShut {NoStop}%
\bibitem [{\citenamefont {Shneidman}\ \emph {et~al.}(2015)\citenamefont
  {Shneidman}, \citenamefont {Adamian}, \citenamefont {Antonenko},
  \citenamefont {Jolos},\ and\ \citenamefont {Zhou}}]{Shneid15a}%
  \BibitemOpen
  \bibfield  {author} {\bibinfo {author} {\bibfnamefont {T.~M.}\ \bibnamefont
  {Shneidman}}, \bibinfo {author} {\bibfnamefont {G.~G.}\ \bibnamefont
  {Adamian}}, \bibinfo {author} {\bibfnamefont {N.~V.}\ \bibnamefont
  {Antonenko}}, \bibinfo {author} {\bibfnamefont {R.~V.}\ \bibnamefont
  {Jolos}}, \ and\ \bibinfo {author} {\bibfnamefont {S.-G.}\ \bibnamefont
  {Zhou}},\ }\href {\doibase 10.1103/PhysRevC.92.034302} {\bibfield  {journal}
  {\bibinfo  {journal} {Phys. Rev. C}\ }\textbf {\bibinfo {volume} {92}},\
  \bibinfo {pages} {034302} (\bibinfo {year} {2015})}\BibitemShut {NoStop}%
\bibitem [{\citenamefont {L\"{o}ffler}\ \emph {et~al.}(1973)\citenamefont
  {L\"{o}ffler} \emph {et~al.}}]{loef73}%
  \BibitemOpen
  \bibfield  {author} {\bibinfo {author} {\bibfnamefont {M.}~\bibnamefont
  {L\"{o}ffler}} \emph {et~al.},\ }\href@noop {} {\bibfield  {journal}
  {\bibinfo  {journal} {Nucl. Instr. and Meth.}\ }\textbf {\bibinfo {volume}
  {111}},\ \bibinfo {pages} {1} (\bibinfo {year} {1973})}\BibitemShut {NoStop}%
\bibitem [{Sup()}]{Sup18a}%
  \BibitemOpen
  \href@noop {} {}\bibinfo {howpublished} {See Supplemental Material at {\bf
  add URL} for a triton spectrum used to exclude target
  contaminations.}\BibitemShut {Stop}%
\bibitem [{\citenamefont {Wirth}\ \emph {et~al.}(2000)\citenamefont {Wirth}
  \emph {et~al.}}]{Wirt00}%
  \BibitemOpen
  \bibfield  {author} {\bibinfo {author} {\bibfnamefont {H.-F.}\ \bibnamefont
  {Wirth}} \emph {et~al.},\ }\href@noop {} {\bibfield  {journal} {\bibinfo
  {journal} {Annual Report, Beschleunigerlaboratorium M\"unchen}\ ,\ \bibinfo
  {pages} {71}} (\bibinfo {year} {2000})}\BibitemShut {NoStop}%
\bibitem [{\citenamefont {Levon}\ \emph {et~al.}(2009)\citenamefont {Levon},
  \citenamefont {Graw}, \citenamefont {Eisermann}, \citenamefont
  {Hertenberger}, \citenamefont {Jolie}, \citenamefont {Shirikova},
  \citenamefont {Stuchbery}, \citenamefont {Sushkov}, \citenamefont {Thirolf},
  \citenamefont {Wirth},\ and\ \citenamefont {Zamfir}}]{Levon09}%
  \BibitemOpen
  \bibfield  {author} {\bibinfo {author} {\bibfnamefont {A.~I.}\ \bibnamefont
  {Levon}}, \bibinfo {author} {\bibfnamefont {G.}~\bibnamefont {Graw}},
  \bibinfo {author} {\bibfnamefont {Y.}~\bibnamefont {Eisermann}}, \bibinfo
  {author} {\bibfnamefont {R.}~\bibnamefont {Hertenberger}}, \bibinfo {author}
  {\bibfnamefont {J.}~\bibnamefont {Jolie}}, \bibinfo {author} {\bibfnamefont
  {N.~Y.}\ \bibnamefont {Shirikova}}, \bibinfo {author} {\bibfnamefont {A.~E.}\
  \bibnamefont {Stuchbery}}, \bibinfo {author} {\bibfnamefont {A.~V.}\
  \bibnamefont {Sushkov}}, \bibinfo {author} {\bibfnamefont {P.~G.}\
  \bibnamefont {Thirolf}}, \bibinfo {author} {\bibfnamefont {H.-F.}\
  \bibnamefont {Wirth}}, \ and\ \bibinfo {author} {\bibfnamefont {N.~V.}\
  \bibnamefont {Zamfir}},\ }\href {\doibase 10.1103/PhysRevC.79.014318}
  {\bibfield  {journal} {\bibinfo  {journal} {Phys. Rev. C}\ }\textbf {\bibinfo
  {volume} {79}},\ \bibinfo {pages} {014318} (\bibinfo {year}
  {2009})}\BibitemShut {NoStop}%
\bibitem [{\citenamefont {Kunz}()}]{CHUCK}%
  \BibitemOpen
  \bibfield  {author} {\bibinfo {author} {\bibfnamefont {P.~D.}\ \bibnamefont
  {Kunz}},\ }\href@noop {} {\bibinfo  {journal} {Program CHUCK}\ ,\ \bibinfo
  {pages} {University of Colorado, unpublished}}\BibitemShut {NoStop}%
\bibitem [{\citenamefont {Mortensen}\ \emph {et~al.}(1980)\citenamefont
  {Mortensen}, \citenamefont {Betts},\ and\ \citenamefont
  {Bockelmann}}]{Mort80}%
  \BibitemOpen
\bibfield  {journal} {  }\bibfield  {author} {\bibinfo {author} {\bibfnamefont
  {M.~H.}\ \bibnamefont {Mortensen}}, \bibinfo {author} {\bibfnamefont {R.~R.}\
  \bibnamefont {Betts}}, \ and\ \bibinfo {author} {\bibfnamefont {C.~K.}\
  \bibnamefont {Bockelmann}},\ }\href@noop {} {\bibfield  {journal} {\bibinfo
  {journal} {Phys. Rev. C}\ }\textbf {\bibinfo {volume} {21}},\ \bibinfo
  {pages} {2275} (\bibinfo {year} {1980})}\BibitemShut {NoStop}%
\bibitem [{\citenamefont {Catford}(2013)}]{catkin}%
  \BibitemOpen
  \bibfield  {author} {\bibinfo {author} {\bibfnamefont {W.~N.}\ \bibnamefont
  {Catford}},\ }\href@noop {} {\enquote {\bibinfo {title} {Catkin program},}\
  }\bibinfo {howpublished}
  {http://personal.ph.surrey.ac.uk/$\sim$phs1wc/kinematics/} (\bibinfo {year}
  {2013})\BibitemShut {NoStop}%
\bibitem [{\citenamefont {Audi}\ \emph {et~al.}(2003)\citenamefont {Audi} \emph
  {et~al.}}]{Audi03}%
  \BibitemOpen
  \bibfield  {author} {\bibinfo {author} {\bibfnamefont {G.}~\bibnamefont
  {Audi}} \emph {et~al.},\ }\href@noop {} {\bibfield  {journal} {\bibinfo
  {journal} {Nucl. Phys. A}\ }\textbf {\bibinfo {volume} {729}},\ \bibinfo
  {pages} {337} (\bibinfo {year} {2003})}\BibitemShut {NoStop}%
\bibitem [{\citenamefont {Wirth}\ \emph {et~al.}(2004)\citenamefont {Wirth}
  \emph {et~al.}}]{Wirt04}%
  \BibitemOpen
  \bibfield  {author} {\bibinfo {author} {\bibfnamefont {H.-F.}\ \bibnamefont
  {Wirth}} \emph {et~al.},\ }\href@noop {} {\bibfield  {journal} {\bibinfo
  {journal} {Phys. Rev. C}\ }\textbf {\bibinfo {volume} {69}},\ \bibinfo
  {pages} {044310} (\bibinfo {year} {2004})}\BibitemShut {NoStop}%
\bibitem [{\citenamefont {Baer}\ \emph {et~al.}(1973)\citenamefont {Baer},
  \citenamefont {Kraushaar}, \citenamefont {Moss}, \citenamefont {King},
  \citenamefont {Green}, \citenamefont {Kunz},\ and\ \citenamefont
  {Rost}}]{Baer73}%
  \BibitemOpen
  \bibfield  {author} {\bibinfo {author} {\bibfnamefont {H.~W.}\ \bibnamefont
  {Baer}}, \bibinfo {author} {\bibfnamefont {J.~J.}\ \bibnamefont {Kraushaar}},
  \bibinfo {author} {\bibfnamefont {C.~E.}\ \bibnamefont {Moss}}, \bibinfo
  {author} {\bibfnamefont {N.~S.~P.}\ \bibnamefont {King}}, \bibinfo {author}
  {\bibfnamefont {R.~E.~L.}\ \bibnamefont {Green}}, \bibinfo {author}
  {\bibfnamefont {P.~D.}\ \bibnamefont {Kunz}}, \ and\ \bibinfo {author}
  {\bibfnamefont {E.}~\bibnamefont {Rost}},\ }\href@noop {} {\bibfield
  {journal} {\bibinfo  {journal} {Ann. Phys. (N.Y.)}\ }\textbf {\bibinfo
  {volume} {76}},\ \bibinfo {pages} {437} (\bibinfo {year} {1973})}\BibitemShut
  {NoStop}%
\bibitem [{\citenamefont {Mahgoub}(2008)}]{Mahg08}%
  \BibitemOpen
  \bibfield  {author} {\bibinfo {author} {\bibfnamefont {M.}~\bibnamefont
  {Mahgoub}},\ }\emph {\bibinfo {title} {Neutron Transfer Reactions in the {\it
  fp}-shell Region}},\ \href@noop {} {Ph.D. thesis},\ \bibinfo  {school} {TU
  München} (\bibinfo {year} {2008})\BibitemShut {NoStop}%
\bibitem [{\citenamefont {Perey}\ and\ \citenamefont {Perey}(1976)}]{Pere76}%
  \BibitemOpen
  \bibfield  {author} {\bibinfo {author} {\bibfnamefont {C.~M.}\ \bibnamefont
  {Perey}}\ and\ \bibinfo {author} {\bibfnamefont {F.~G.}\ \bibnamefont
  {Perey}},\ }\href@noop {} {\bibfield  {journal} {\bibinfo  {journal} {At.
  Data Nucl. Data Tables}\ }\textbf {\bibinfo {volume} {17}},\ \bibinfo {pages}
  {1} (\bibinfo {year} {1976})}\BibitemShut {NoStop}%
\bibitem [{\citenamefont {Becchetti}\ and\ \citenamefont
  {Greenlees}(1969)}]{Becch69}%
  \BibitemOpen
  \bibfield  {author} {\bibinfo {author} {\bibfnamefont {F.~D.}\ \bibnamefont
  {Becchetti}}\ and\ \bibinfo {author} {\bibfnamefont {G.~W.}\ \bibnamefont
  {Greenlees}},\ }\href@noop {} {\bibfield  {journal} {\bibinfo  {journal}
  {Phys. Rev.}\ }\textbf {\bibinfo {volume} {182}},\ \bibinfo {pages} {1190}
  (\bibinfo {year} {1969})}\BibitemShut {NoStop}%
\bibitem [{\citenamefont {Flynn}\ \emph {et~al.}(1969)\citenamefont {Flynn},
  \citenamefont {Armstrong}, \citenamefont {Beery},\ and\ \citenamefont
  {Blair}}]{Flynn69}%
  \BibitemOpen
  \bibfield  {author} {\bibinfo {author} {\bibfnamefont {E.~R.}\ \bibnamefont
  {Flynn}}, \bibinfo {author} {\bibfnamefont {D.~D.}\ \bibnamefont
  {Armstrong}}, \bibinfo {author} {\bibfnamefont {J.~G.}\ \bibnamefont
  {Beery}}, \ and\ \bibinfo {author} {\bibfnamefont {A.~G.}\ \bibnamefont
  {Blair}},\ }\href@noop {} {\bibfield  {journal} {\bibinfo  {journal} {Phys.
  Rev.}\ }\textbf {\bibinfo {volume} {182}},\ \bibinfo {pages} {1113} (\bibinfo
  {year} {1969})}\BibitemShut {NoStop}%
\bibitem [{\citenamefont {M{\"o}ller}\ \emph {et~al.}(1995)\citenamefont
  {M{\"o}ller}, \citenamefont {Nix}, \citenamefont {Myers},\ and\ \citenamefont
  {Swiatecki}}]{Moel95}%
  \BibitemOpen
  \bibfield  {author} {\bibinfo {author} {\bibfnamefont {P.}~\bibnamefont
  {M{\"o}ller}}, \bibinfo {author} {\bibfnamefont {J.~R.}\ \bibnamefont {Nix}},
  \bibinfo {author} {\bibfnamefont {W.~D.}\ \bibnamefont {Myers}}, \ and\
  \bibinfo {author} {\bibfnamefont {W.~J.}\ \bibnamefont {Swiatecki}},\
  }\href@noop {} {\bibfield  {journal} {\bibinfo  {journal} {At. Data Nucl.
  Data Tables}\ }\textbf {\bibinfo {volume} {59}},\ \bibinfo {pages} {185}
  (\bibinfo {year} {1995})}\BibitemShut {NoStop}%
\bibitem [{\citenamefont {Bohr}\ and\ \citenamefont
  {Mottelson}(1998)}]{Bohr75}%
  \BibitemOpen
  \bibfield  {author} {\bibinfo {author} {\bibfnamefont {A.}~\bibnamefont
  {Bohr}}\ and\ \bibinfo {author} {\bibfnamefont {B.~R.}\ \bibnamefont
  {Mottelson}},\ }\href@noop {} {\emph {\bibinfo {title} {Nuclear Structure Bd.
  II}}}\ (\bibinfo  {publisher} {World Scientific, Singapore},\ \bibinfo {year}
  {1998})\BibitemShut {NoStop}%
\bibitem [{\citenamefont {Singh}\ and\ \citenamefont {Browne}(2008)}]{Sing08}%
  \BibitemOpen
  \bibfield  {author} {\bibinfo {author} {\bibfnamefont {B.}~\bibnamefont
  {Singh}}\ and\ \bibinfo {author} {\bibfnamefont {E.}~\bibnamefont {Browne}},\
  }\href@noop {} {\bibfield  {journal} {\bibinfo  {journal} {Nucl. Data
  Sheets}\ }\textbf {\bibinfo {volume} {109}},\ \bibinfo {pages} {2439}
  (\bibinfo {year} {2008})}\BibitemShut {NoStop}%
\bibitem [{\citenamefont {Clark}\ \emph {et~al.}(2009)\citenamefont {Clark}
  \emph {et~al.}}]{Clark09}%
  \BibitemOpen
  \bibfield  {author} {\bibinfo {author} {\bibfnamefont {R.~M.}\ \bibnamefont
  {Clark}} \emph {et~al.},\ }\href@noop {} {\bibfield  {journal} {\bibinfo
  {journal} {Phys. Rev. C}\ }\textbf {\bibinfo {volume} {80}},\ \bibinfo
  {pages} {011303(R)} (\bibinfo {year} {2009})}\BibitemShut {NoStop}%
\bibitem [{\citenamefont {Parekh}\ \emph {et~al.}(1982)\citenamefont {Parekh},
  \citenamefont {Peker}, \citenamefont {Katcoff},\ and\ \citenamefont
  {Franz}}]{Parek82}%
  \BibitemOpen
  \bibfield  {author} {\bibinfo {author} {\bibfnamefont {P.~P.}\ \bibnamefont
  {Parekh}}, \bibinfo {author} {\bibfnamefont {L.~K.}\ \bibnamefont {Peker}},
  \bibinfo {author} {\bibfnamefont {S.}~\bibnamefont {Katcoff}}, \ and\
  \bibinfo {author} {\bibfnamefont {E.-M.}\ \bibnamefont {Franz}},\ }\href@noop
  {} {\bibfield  {journal} {\bibinfo  {journal} {Phys. Rev. C}\ }\textbf
  {\bibinfo {volume} {26}},\ \bibinfo {pages} {2178} (\bibinfo {year}
  {1982})}\BibitemShut {NoStop}%
\bibitem [{\citenamefont {Hseuh}\ \emph {et~al.}(1981)\citenamefont {Hseuh},
  \citenamefont {Franz}, \citenamefont {Haustein}, \citenamefont {Katcoff},\
  and\ \citenamefont {Peker}}]{Hseuh81}%
  \BibitemOpen
  \bibfield  {author} {\bibinfo {author} {\bibfnamefont {H.-C.}\ \bibnamefont
  {Hseuh}}, \bibinfo {author} {\bibfnamefont {E.-M.}\ \bibnamefont {Franz}},
  \bibinfo {author} {\bibfnamefont {P.~E.}\ \bibnamefont {Haustein}}, \bibinfo
  {author} {\bibfnamefont {S.}~\bibnamefont {Katcoff}}, \ and\ \bibinfo
  {author} {\bibfnamefont {L.~K.}\ \bibnamefont {Peker}},\ }\href@noop {}
  {\bibfield  {journal} {\bibinfo  {journal} {Phys. Rev. C}\ }\textbf {\bibinfo
  {volume} {23}},\ \bibinfo {pages} {1217} (\bibinfo {year}
  {1981})}\BibitemShut {NoStop}%
\bibitem [{\citenamefont {Schmorak}\ \emph {et~al.}(1970)\citenamefont
  {Schmorak}, \citenamefont {Bemis}, \citenamefont {~}, \citenamefont {Zender},
  \citenamefont {Coffman}, \citenamefont {Ramayya},\ and\ \citenamefont
  {Hamilton}}]{Schmo70}%
  \BibitemOpen
  \bibfield  {author} {\bibinfo {author} {\bibfnamefont {M.~R.}\ \bibnamefont
  {Schmorak}}, \bibinfo {author} {\bibfnamefont {C.~E.}\ \bibnamefont {Bemis}},
  \bibinfo {author} {\bibfnamefont {J.}~\bibnamefont {~}}, \bibinfo {author}
  {\bibfnamefont {M.}~\bibnamefont {Zender}}, \bibinfo {author} {\bibfnamefont
  {F.~E.}\ \bibnamefont {Coffman}}, \bibinfo {author} {\bibfnamefont {A.~V.}\
  \bibnamefont {Ramayya}}, \ and\ \bibinfo {author} {\bibfnamefont {J.~H.}\
  \bibnamefont {Hamilton}},\ }\href@noop {} {\bibfield  {journal} {\bibinfo
  {journal} {Phys. Rev. Lett.}\ }\textbf {\bibinfo {volume} {24}},\ \bibinfo
  {pages} {1507} (\bibinfo {year} {1970})}\BibitemShut {NoStop}%
\bibitem [{\citenamefont {Thompson}\ \emph {et~al.}(1975)\citenamefont
  {Thompson}, \citenamefont {Huizenga},\ and\ \citenamefont {Elze}}]{Thomp75}%
  \BibitemOpen
  \bibfield  {author} {\bibinfo {author} {\bibfnamefont {R.~C.}\ \bibnamefont
  {Thompson}}, \bibinfo {author} {\bibfnamefont {J.~R.}\ \bibnamefont
  {Huizenga}}, \ and\ \bibinfo {author} {\bibfnamefont {T.~W.}\ \bibnamefont
  {Elze}},\ }\href@noop {} {\bibfield  {journal} {\bibinfo  {journal} {Phys.
  Rev. C}\ }\textbf {\bibinfo {volume} {12}},\ \bibinfo {pages} {1227}
  (\bibinfo {year} {1975})}\BibitemShut {NoStop}%
\bibitem [{\citenamefont {Chrien}\ \emph {et~al.}(1985)\citenamefont {Chrien},
  \citenamefont {Kopecky}, \citenamefont {Liou}, \citenamefont {Wasson},
  \citenamefont {Garg},\ and\ \citenamefont {Dritsa}}]{Chrie85}%
  \BibitemOpen
  \bibfield  {author} {\bibinfo {author} {\bibfnamefont {R.~E.}\ \bibnamefont
  {Chrien}}, \bibinfo {author} {\bibfnamefont {J.}~\bibnamefont {Kopecky}},
  \bibinfo {author} {\bibfnamefont {H.~I.}\ \bibnamefont {Liou}}, \bibinfo
  {author} {\bibfnamefont {O.~A.}\ \bibnamefont {Wasson}}, \bibinfo {author}
  {\bibfnamefont {J.~B.}\ \bibnamefont {Garg}}, \ and\ \bibinfo {author}
  {\bibfnamefont {M.}~\bibnamefont {Dritsa}},\ }\href@noop {} {\bibfield
  {journal} {\bibinfo  {journal} {Nucl. Phys. A}\ }\textbf {\bibinfo {volume}
  {436}},\ \bibinfo {pages} {205} (\bibinfo {year} {1985})}\BibitemShut
  {NoStop}%
\bibitem [{\citenamefont {Hoogduin}\ \emph {et~al.}(1996)\citenamefont
  {Hoogduin} \emph {et~al.}}]{Hoog96}%
  \BibitemOpen
  \bibfield  {author} {\bibinfo {author} {\bibfnamefont {J.~M.}\ \bibnamefont
  {Hoogduin}} \emph {et~al.},\ }\href@noop {} {\bibfield  {journal} {\bibinfo
  {journal} {Phys. Lett. B}\ }\textbf {\bibinfo {volume} {384}},\ \bibinfo
  {pages} {43} (\bibinfo {year} {1996})}\BibitemShut {NoStop}%
\bibitem [{ENSDF()}]{ENSDF}%
  \BibitemOpen
  ENSDF,\ \href@noop {} {}\bibinfo {howpublished} {NNDC Online Data Service,
  ENSDF database, \newline http://www.nndc.bnl.gov/ensdf/} (\bibinfo {year}
  {2016})\BibitemShut {NoStop}%
\bibitem [{\citenamefont {Casten}(2000)}]{Cast00}%
  \BibitemOpen
  \bibfield  {author} {\bibinfo {author} {\bibfnamefont {R.~F.}\ \bibnamefont
  {Casten}},\ }\href@noop {} {\emph {\bibinfo {title} {Nuclear Structure from a
  Simple Perspective}}}\ (\bibinfo  {publisher} {Oxford University Press Inc.,
  New York},\ \bibinfo {year} {2000})\BibitemShut {NoStop}%
\bibitem [{\citenamefont {Zamfir}\ and\ \citenamefont
  {Kusnezov}(2003)}]{Zamf03}%
  \BibitemOpen
  \bibfield  {author} {\bibinfo {author} {\bibfnamefont {N.~V.}\ \bibnamefont
  {Zamfir}}\ and\ \bibinfo {author} {\bibfnamefont {D.}~\bibnamefont
  {Kusnezov}},\ }\href@noop {} {\bibfield  {journal} {\bibinfo  {journal}
  {Phys. Rev. C}\ }\textbf {\bibinfo {volume} {67}},\ \bibinfo {pages} {014305}
  (\bibinfo {year} {2003})}\BibitemShut {NoStop}%
\bibitem [{\citenamefont {McGowan}\ \emph {et~al.}(1974)\citenamefont
  {McGowan}, \citenamefont {Bemis}, \citenamefont {Milner}, \citenamefont
  {Ford}, \citenamefont {Robinson},\ and\ \citenamefont {Stelson}}]{McG74a}%
  \BibitemOpen
  \bibfield  {author} {\bibinfo {author} {\bibfnamefont {F.~K.}\ \bibnamefont
  {McGowan}}, \bibinfo {author} {\bibfnamefont {C.~E.}\ \bibnamefont {Bemis}},
  \bibinfo {author} {\bibfnamefont {W.~T.}\ \bibnamefont {Milner}}, \bibinfo
  {author} {\bibfnamefont {J.~L.~C.}\ \bibnamefont {Ford}}, \bibinfo {author}
  {\bibfnamefont {R.~L.}\ \bibnamefont {Robinson}}, \ and\ \bibinfo {author}
  {\bibfnamefont {P.~H.}\ \bibnamefont {Stelson}},\ }\href {\doibase
  10.1103/PhysRevC.10.1146} {\bibfield  {journal} {\bibinfo  {journal} {Phys.
  Rev. C}\ }\textbf {\bibinfo {volume} {10}},\ \bibinfo {pages} {1146}
  (\bibinfo {year} {1974})}\BibitemShut {NoStop}%
\bibitem [{\citenamefont {Wollersheim}\ \emph {et~al.}(1993)\citenamefont
  {Wollersheim}, \citenamefont {Emling}, \citenamefont {Grein}, \citenamefont
  {Kulessa}, \citenamefont {Simon}, \citenamefont {Fleischmann}, \citenamefont
  {de~Boer}, \citenamefont {Hauber}, \citenamefont {Lauterbach}, \citenamefont
  {Schandera}, \citenamefont {Butler},\ and\ \citenamefont
  {Czosnyka}}]{Woll93a}%
  \BibitemOpen
  \bibfield  {author} {\bibinfo {author} {\bibfnamefont {H.}~\bibnamefont
  {Wollersheim}}, \bibinfo {author} {\bibfnamefont {H.}~\bibnamefont {Emling}},
  \bibinfo {author} {\bibfnamefont {H.}~\bibnamefont {Grein}}, \bibinfo
  {author} {\bibfnamefont {R.}~\bibnamefont {Kulessa}}, \bibinfo {author}
  {\bibfnamefont {R.}~\bibnamefont {Simon}}, \bibinfo {author} {\bibfnamefont
  {C.}~\bibnamefont {Fleischmann}}, \bibinfo {author} {\bibfnamefont
  {J.}~\bibnamefont {de~Boer}}, \bibinfo {author} {\bibfnamefont
  {E.}~\bibnamefont {Hauber}}, \bibinfo {author} {\bibfnamefont
  {C.}~\bibnamefont {Lauterbach}}, \bibinfo {author} {\bibfnamefont
  {C.}~\bibnamefont {Schandera}}, \bibinfo {author} {\bibfnamefont
  {P.}~\bibnamefont {Butler}}, \ and\ \bibinfo {author} {\bibfnamefont
  {T.}~\bibnamefont {Czosnyka}},\ }\href {\doibase
  https://doi.org/10.1016/0375-9474(93)90351-W} {\bibfield  {journal} {\bibinfo
   {journal} {Nucl. Phys. A}\ }\textbf {\bibinfo {volume} {556}},\ \bibinfo
  {pages} {261} (\bibinfo {year} {1993})}\BibitemShut {NoStop}%
\bibitem [{\citenamefont {Cejnar}\ \emph {et~al.}(2010)\citenamefont {Cejnar},
  \citenamefont {Jolie},\ and\ \citenamefont {Casten}}]{Cej10a}%
  \BibitemOpen
  \bibfield  {author} {\bibinfo {author} {\bibfnamefont {P.}~\bibnamefont
  {Cejnar}}, \bibinfo {author} {\bibfnamefont {J.}~\bibnamefont {Jolie}}, \
  and\ \bibinfo {author} {\bibfnamefont {R.~F.}\ \bibnamefont {Casten}},\
  }\href {\doibase 10.1103/RevModPhys.82.2155} {\bibfield  {journal} {\bibinfo
  {journal} {Rev. Mod. Phys.}\ }\textbf {\bibinfo {volume} {82}},\ \bibinfo
  {pages} {2155} (\bibinfo {year} {2010})}\BibitemShut {NoStop}%
\bibitem [{\citenamefont {Zhang}\ and\ \citenamefont
  {Iachello}(2017)}]{Zhang17a}%
  \BibitemOpen
  \bibfield  {author} {\bibinfo {author} {\bibfnamefont {Y.}~\bibnamefont
  {Zhang}}\ and\ \bibinfo {author} {\bibfnamefont {F.}~\bibnamefont
  {Iachello}},\ }\href@noop {} {\bibfield  {journal} {\bibinfo  {journal}
  {Phys. Rev. C}\ }\textbf {\bibinfo {volume} {95}},\ \bibinfo {pages} {034306}
  (\bibinfo {year} {2017})}\BibitemShut {NoStop}%
\bibitem [{\citenamefont {Nomura}\ \emph {et~al.}(2013)\citenamefont {Nomura},
  \citenamefont {Vretenar},\ and\ \citenamefont {Lu}}]{Nomu13a}%
  \BibitemOpen
  \bibfield  {author} {\bibinfo {author} {\bibfnamefont {K.}~\bibnamefont
  {Nomura}}, \bibinfo {author} {\bibfnamefont {D.}~\bibnamefont {Vretenar}}, \
  and\ \bibinfo {author} {\bibfnamefont {B.-N.}\ \bibnamefont {Lu}},\ }\href
  {\doibase 10.1103/PhysRevC.88.021303} {\bibfield  {journal} {\bibinfo
  {journal} {Phys. Rev. C}\ }\textbf {\bibinfo {volume} {88}},\ \bibinfo
  {pages} {021303} (\bibinfo {year} {2013})}\BibitemShut {NoStop}%
\bibitem [{\citenamefont {Heyde}\ and\ \citenamefont {Wood}(2011)}]{Heyd11a}%
  \BibitemOpen
  \bibfield  {author} {\bibinfo {author} {\bibfnamefont {K.}~\bibnamefont
  {Heyde}}\ and\ \bibinfo {author} {\bibfnamefont {J.~L.}\ \bibnamefont
  {Wood}},\ }\href {\doibase 10.1103/RevModPhys.83.1467} {\bibfield  {journal}
  {\bibinfo  {journal} {Rev. Mod. Phys.}\ }\textbf {\bibinfo {volume} {83}},\
  \bibinfo {pages} {1467} (\bibinfo {year} {2011})}\BibitemShut {NoStop}%
\bibitem [{\citenamefont {Ardisson}\ \emph {et~al.}(1994)\citenamefont
  {Ardisson}, \citenamefont {Hussonnois}, \citenamefont {LeDu}, \citenamefont
  {Trubert},\ and\ \citenamefont {Lederer}}]{Ard94a}%
  \BibitemOpen
  \bibfield  {author} {\bibinfo {author} {\bibfnamefont {G.}~\bibnamefont
  {Ardisson}}, \bibinfo {author} {\bibfnamefont {M.}~\bibnamefont
  {Hussonnois}}, \bibinfo {author} {\bibfnamefont {J.~F.}\ \bibnamefont
  {LeDu}}, \bibinfo {author} {\bibfnamefont {D.}~\bibnamefont {Trubert}}, \
  and\ \bibinfo {author} {\bibfnamefont {C.~M.}\ \bibnamefont {Lederer}},\
  }\href {\doibase 10.1103/PhysRevC.49.2963} {\bibfield  {journal} {\bibinfo
  {journal} {Phys. Rev. C}\ }\textbf {\bibinfo {volume} {49}},\ \bibinfo
  {pages} {2963} (\bibinfo {year} {1994})}\BibitemShut {NoStop}%
\bibitem [{\citenamefont {van Duppen}\ and\ \citenamefont
  {Huyse}(2000)}]{vanDu00a}%
  \BibitemOpen
  \bibfield  {author} {\bibinfo {author} {\bibfnamefont {P.}~\bibnamefont {van
  Duppen}}\ and\ \bibinfo {author} {\bibfnamefont {M.}~\bibnamefont {Huyse}},\
  }\href@noop {} {\bibfield  {journal} {\bibinfo  {journal} {Hyperfine
  Interactions}\ }\textbf {\bibinfo {volume} {129}},\ \bibinfo {pages} {149}
  (\bibinfo {year} {2000})}\BibitemShut {NoStop}%
\bibitem [{\citenamefont {Bucurescu}\ and\ \citenamefont
  {Zamfir}(2012)}]{Bucu12a}%
  \BibitemOpen
  \bibfield  {author} {\bibinfo {author} {\bibfnamefont {D.}~\bibnamefont
  {Bucurescu}}\ and\ \bibinfo {author} {\bibfnamefont {N.~V.}\ \bibnamefont
  {Zamfir}},\ }\href {\doibase 10.1103/PhysRevC.86.067306} {\bibfield
  {journal} {\bibinfo  {journal} {Phys. Rev. C}\ }\textbf {\bibinfo {volume}
  {86}},\ \bibinfo {pages} {067306} (\bibinfo {year} {2012})}\BibitemShut
  {NoStop}%
\bibitem [{\citenamefont {Bucurescu}\ and\ \citenamefont
  {Zamfir}(2013)}]{Bucu13a}%
  \BibitemOpen
  \bibfield  {author} {\bibinfo {author} {\bibfnamefont {D.}~\bibnamefont
  {Bucurescu}}\ and\ \bibinfo {author} {\bibfnamefont {N.~V.}\ \bibnamefont
  {Zamfir}},\ }\href {\doibase 10.1103/PhysRevC.87.054324} {\bibfield
  {journal} {\bibinfo  {journal} {Phys. Rev. C}\ }\textbf {\bibinfo {volume}
  {87}},\ \bibinfo {pages} {054324} (\bibinfo {year} {2013})}\BibitemShut
  {NoStop}%
\bibitem [{\citenamefont {Sheline}\ and\ \citenamefont
  {Bossinga}(1991)}]{She91a}%
  \BibitemOpen
  \bibfield  {author} {\bibinfo {author} {\bibfnamefont {R.~K.}\ \bibnamefont
  {Sheline}}\ and\ \bibinfo {author} {\bibfnamefont {B.~B.-M.}\ \bibnamefont
  {Bossinga}},\ }\href {\doibase 10.1103/PhysRevC.44.218} {\bibfield  {journal}
  {\bibinfo  {journal} {Phys. Rev. C}\ }\textbf {\bibinfo {volume} {44}},\
  \bibinfo {pages} {218} (\bibinfo {year} {1991})}\BibitemShut {NoStop}%
\bibitem [{\citenamefont {Ahmad}\ \emph {et~al.}(1975)\citenamefont {Ahmad},
  \citenamefont {Porter}, \citenamefont {Freedman}, \citenamefont {Sjoblom},
  \citenamefont {Lerner}, \citenamefont {Barnes}, \citenamefont {Milsted},\
  and\ \citenamefont {Fields}}]{Ahm75a}%
  \BibitemOpen
  \bibfield  {author} {\bibinfo {author} {\bibfnamefont {I.}~\bibnamefont
  {Ahmad}}, \bibinfo {author} {\bibfnamefont {F.~T.}\ \bibnamefont {Porter}},
  \bibinfo {author} {\bibfnamefont {M.~S.}\ \bibnamefont {Freedman}}, \bibinfo
  {author} {\bibfnamefont {R.~K.}\ \bibnamefont {Sjoblom}}, \bibinfo {author}
  {\bibfnamefont {J.}~\bibnamefont {Lerner}}, \bibinfo {author} {\bibfnamefont
  {R.~F.}\ \bibnamefont {Barnes}}, \bibinfo {author} {\bibfnamefont
  {J.}~\bibnamefont {Milsted}}, \ and\ \bibinfo {author} {\bibfnamefont
  {P.~R.}\ \bibnamefont {Fields}},\ }\href {\doibase 10.1103/PhysRevC.12.541}
  {\bibfield  {journal} {\bibinfo  {journal} {Phys. Rev. C}\ }\textbf {\bibinfo
  {volume} {12}},\ \bibinfo {pages} {541} (\bibinfo {year} {1975})}\BibitemShut
  {NoStop}%
\bibitem [{\citenamefont {Pritychenko}\ \emph {et~al.}(2016)\citenamefont
  {Pritychenko}, \citenamefont {Birch}, \citenamefont {Singh},\ and\
  \citenamefont {Horoi}}]{Prity16a}%
  \BibitemOpen
  \bibfield  {author} {\bibinfo {author} {\bibfnamefont {B.}~\bibnamefont
  {Pritychenko}}, \bibinfo {author} {\bibfnamefont {M.}~\bibnamefont {Birch}},
  \bibinfo {author} {\bibfnamefont {B.}~\bibnamefont {Singh}}, \ and\ \bibinfo
  {author} {\bibfnamefont {M.}~\bibnamefont {Horoi}},\ }\href {\doibase
  https://doi.org/10.1016/j.adt.2015.10.001} {\bibfield  {journal} {\bibinfo
  {journal} {Atomic Data and Nuclear Data Tables}\ }\textbf {\bibinfo {volume}
  {107}},\ \bibinfo {pages} {1} (\bibinfo {year} {2016})}\BibitemShut {NoStop}%
\bibitem [{\citenamefont {Garrett}\ \emph {et~al.}(2009)\citenamefont
  {Garrett}, \citenamefont {Kulp}, \citenamefont {Wood}, \citenamefont
  {Bandyopadhyay}, \citenamefont {Choudry}, \citenamefont {Dashdorj},
  \citenamefont {Lesher}, \citenamefont {McEllistrem}, \citenamefont {Mynk},
  \citenamefont {Orce},\ and\ \citenamefont {Yates}}]{Garrett09}%
  \BibitemOpen
  \bibfield  {author} {\bibinfo {author} {\bibfnamefont {P.~E.}\ \bibnamefont
  {Garrett}}, \bibinfo {author} {\bibfnamefont {W.~D.}\ \bibnamefont {Kulp}},
  \bibinfo {author} {\bibfnamefont {J.~L.}\ \bibnamefont {Wood}}, \bibinfo
  {author} {\bibfnamefont {D.}~\bibnamefont {Bandyopadhyay}}, \bibinfo {author}
  {\bibfnamefont {S.}~\bibnamefont {Choudry}}, \bibinfo {author} {\bibfnamefont
  {D.}~\bibnamefont {Dashdorj}}, \bibinfo {author} {\bibfnamefont {S.~R.}\
  \bibnamefont {Lesher}}, \bibinfo {author} {\bibfnamefont {M.~T.}\
  \bibnamefont {McEllistrem}}, \bibinfo {author} {\bibfnamefont
  {M.}~\bibnamefont {Mynk}}, \bibinfo {author} {\bibfnamefont {J.~N.}\
  \bibnamefont {Orce}}, \ and\ \bibinfo {author} {\bibfnamefont {S.~W.}\
  \bibnamefont {Yates}},\ }\href@noop {} {\bibfield  {journal} {\bibinfo
  {journal} {Phys. Rev. Lett.}\ }\textbf {\bibinfo {volume} {103}},\ \bibinfo
  {pages} {062501} (\bibinfo {year} {2009})}\BibitemShut {NoStop}%
\bibitem [{\citenamefont {Quiter}\ \emph {et~al.}(2012)\citenamefont {Quiter},
  \citenamefont {Laplace}, \citenamefont {Ludewigt}, \citenamefont {Ambers},
  \citenamefont {Goldblum}, \citenamefont {Korbly}, \citenamefont {Hicks},\
  and\ \citenamefont {Wilson}}]{Quit12a}%
  \BibitemOpen
  \bibfield  {author} {\bibinfo {author} {\bibfnamefont {B.~J.}\ \bibnamefont
  {Quiter}}, \bibinfo {author} {\bibfnamefont {T.}~\bibnamefont {Laplace}},
  \bibinfo {author} {\bibfnamefont {B.~A.}\ \bibnamefont {Ludewigt}}, \bibinfo
  {author} {\bibfnamefont {S.~D.}\ \bibnamefont {Ambers}}, \bibinfo {author}
  {\bibfnamefont {B.~L.}\ \bibnamefont {Goldblum}}, \bibinfo {author}
  {\bibfnamefont {S.}~\bibnamefont {Korbly}}, \bibinfo {author} {\bibfnamefont
  {C.}~\bibnamefont {Hicks}}, \ and\ \bibinfo {author} {\bibfnamefont
  {C.}~\bibnamefont {Wilson}},\ }\href {\doibase 10.1103/PhysRevC.86.034307}
  {\bibfield  {journal} {\bibinfo  {journal} {Phys. Rev. C}\ }\textbf {\bibinfo
  {volume} {86}},\ \bibinfo {pages} {034307} (\bibinfo {year}
  {2012})}\BibitemShut {NoStop}%
\bibitem [{\citenamefont {Hammond}\ \emph {et~al.}(2012)\citenamefont
  {Hammond}, \citenamefont {Adekola}, \citenamefont {Angell}, \citenamefont
  {Karwowski}, \citenamefont {Kwan}, \citenamefont {Rusev}, \citenamefont
  {Tonchev}, \citenamefont {Tornow}, \citenamefont {Howell},\ and\
  \citenamefont {Kelley}}]{Hamm12a}%
  \BibitemOpen
  \bibfield  {author} {\bibinfo {author} {\bibfnamefont {S.~L.}\ \bibnamefont
  {Hammond}}, \bibinfo {author} {\bibfnamefont {A.~S.}\ \bibnamefont
  {Adekola}}, \bibinfo {author} {\bibfnamefont {C.~T.}\ \bibnamefont {Angell}},
  \bibinfo {author} {\bibfnamefont {H.~J.}\ \bibnamefont {Karwowski}}, \bibinfo
  {author} {\bibfnamefont {E.}~\bibnamefont {Kwan}}, \bibinfo {author}
  {\bibfnamefont {G.}~\bibnamefont {Rusev}}, \bibinfo {author} {\bibfnamefont
  {A.~P.}\ \bibnamefont {Tonchev}}, \bibinfo {author} {\bibfnamefont
  {W.}~\bibnamefont {Tornow}}, \bibinfo {author} {\bibfnamefont {C.~R.}\
  \bibnamefont {Howell}}, \ and\ \bibinfo {author} {\bibfnamefont {J.~H.}\
  \bibnamefont {Kelley}},\ }\href {\doibase 10.1103/PhysRevC.85.044302}
  {\bibfield  {journal} {\bibinfo  {journal} {Phys. Rev. C}\ }\textbf {\bibinfo
  {volume} {85}},\ \bibinfo {pages} {044302} (\bibinfo {year}
  {2012})}\BibitemShut {NoStop}%
\bibitem [{\citenamefont {Spieker}\ \emph {et~al.}(2015)\citenamefont
  {Spieker}, \citenamefont {Pascu}, \citenamefont {Zilges},\ and\ \citenamefont
  {Iachello}}]{Spiek15a}%
  \BibitemOpen
  \bibfield  {author} {\bibinfo {author} {\bibfnamefont {M.}~\bibnamefont
  {Spieker}}, \bibinfo {author} {\bibfnamefont {S.}~\bibnamefont {Pascu}},
  \bibinfo {author} {\bibfnamefont {A.}~\bibnamefont {Zilges}}, \ and\ \bibinfo
  {author} {\bibfnamefont {F.}~\bibnamefont {Iachello}},\ }\href {\doibase
  10.1103/PhysRevLett.114.192504} {\bibfield  {journal} {\bibinfo  {journal}
  {Phys. Rev. Lett.}\ }\textbf {\bibinfo {volume} {114}},\ \bibinfo {pages}
  {192504} (\bibinfo {year} {2015})}\BibitemShut {NoStop}%
\bibitem [{\citenamefont {Spieker}\ \emph {et~al.}(2017)\citenamefont
  {Spieker}, \citenamefont {Pascu},\ and\ \citenamefont {Zilges}}]{Spiek17a}%
  \BibitemOpen
  \bibfield  {author} {\bibinfo {author} {\bibfnamefont {M.}~\bibnamefont
  {Spieker}}, \bibinfo {author} {\bibfnamefont {S.}~\bibnamefont {Pascu}}, \
  and\ \bibinfo {author} {\bibfnamefont {A.}~\bibnamefont {Zilges}},\ }\href
  {http://stacks.iop.org/1742-6596/863/i=1/a=012063} {\bibfield  {journal}
  {\bibinfo  {journal} {Journal of Physics: Conference Series}\ }\textbf
  {\bibinfo {volume} {863}},\ \bibinfo {pages} {012063} (\bibinfo {year}
  {2017})}\BibitemShut {NoStop}%
\bibitem [{\citenamefont {Ibbotson}\ \emph {et~al.}(1997)\citenamefont
  {Ibbotson}, \citenamefont {White}, \citenamefont {Czosnyka}, \citenamefont
  {Butler}, \citenamefont {Clarkson}, \citenamefont {Cline}, \citenamefont
  {Cunningham}, \citenamefont {Devlin}, \citenamefont {Helmer}, \citenamefont
  {Hoare}, \citenamefont {Hughes}, \citenamefont {Jones}, \citenamefont
  {Kavka}, \citenamefont {Kotlinski}, \citenamefont {Poynter}, \citenamefont
  {Regan}, \citenamefont {Vogt}, \citenamefont {Wadsworth}, \citenamefont
  {Watson},\ and\ \citenamefont {Wu}}]{Ibbot97a}%
  \BibitemOpen
  \bibfield  {author} {\bibinfo {author} {\bibfnamefont {R.}~\bibnamefont
  {Ibbotson}}, \bibinfo {author} {\bibfnamefont {C.}~\bibnamefont {White}},
  \bibinfo {author} {\bibfnamefont {T.}~\bibnamefont {Czosnyka}}, \bibinfo
  {author} {\bibfnamefont {P.}~\bibnamefont {Butler}}, \bibinfo {author}
  {\bibfnamefont {N.}~\bibnamefont {Clarkson}}, \bibinfo {author}
  {\bibfnamefont {D.}~\bibnamefont {Cline}}, \bibinfo {author} {\bibfnamefont
  {R.}~\bibnamefont {Cunningham}}, \bibinfo {author} {\bibfnamefont
  {M.}~\bibnamefont {Devlin}}, \bibinfo {author} {\bibfnamefont
  {K.}~\bibnamefont {Helmer}}, \bibinfo {author} {\bibfnamefont
  {T.}~\bibnamefont {Hoare}}, \bibinfo {author} {\bibfnamefont
  {J.}~\bibnamefont {Hughes}}, \bibinfo {author} {\bibfnamefont
  {G.}~\bibnamefont {Jones}}, \bibinfo {author} {\bibfnamefont
  {A.}~\bibnamefont {Kavka}}, \bibinfo {author} {\bibfnamefont
  {B.}~\bibnamefont {Kotlinski}}, \bibinfo {author} {\bibfnamefont
  {R.}~\bibnamefont {Poynter}}, \bibinfo {author} {\bibfnamefont
  {P.}~\bibnamefont {Regan}}, \bibinfo {author} {\bibfnamefont
  {E.}~\bibnamefont {Vogt}}, \bibinfo {author} {\bibfnamefont {R.}~\bibnamefont
  {Wadsworth}}, \bibinfo {author} {\bibfnamefont {D.}~\bibnamefont {Watson}}, \
  and\ \bibinfo {author} {\bibfnamefont {C.}~\bibnamefont {Wu}},\ }\href
  {\doibase http://dx.doi.org/10.1016/S0375-9474(97)00145-0} {\bibfield
  {journal} {\bibinfo  {journal} {Nucl. Phys. A}\ }\textbf {\bibinfo {volume}
  {619}},\ \bibinfo {pages} {213} (\bibinfo {year} {1997})}\BibitemShut
  {NoStop}%
\end{thebibliography}%

\end{document}